\documentclass[11pt]{article}
\usepackage{amsfonts,amsmath,amssymb,subcaption}
\usepackage{enumerate}
\usepackage{hyperref}
\usepackage{bbm}
\usepackage{nicefrac}
\usepackage[all]{xy}
\usepackage{graphicx}
\usepackage{bm}
\usepackage{makecell}
\usepackage[table]{xcolor}
\usepackage{scalerel}

\usepackage{upgreek}

\usepackage{booktabs}

\newcommand{\be}{\begin{equation}}
\newcommand{\ee}{\end{equation}}

\newcommand{\bal}{\begin{aligned}}
\newcommand{\eal}{\end{aligned}}

\newcommand{\bea}{\begin{eqnarray}}
\newcommand{\eea}{\end{eqnarray}}

\newcommand{\bes}{\begin{subequations}}
\newcommand{\ees}{\end{subequations}}

\newcommand{\cN}{{\cal N}}

\newcommand{\tr}{\mbox{tr}}

\def\sst#1{{\scriptscriptstyle #1}}

\def\0{{\sst{(0)}}}
\def\1{{\sst{(1)}}}
\def\2{{\sst{(2)}}}
\def\3{{\sst{(3)}}}
\def\4{{\sst{(4)}}}
\def\5{{\sst{(5)}}}
\def\6{{\sst{(6)}}}
\def\7{{\sst{(7)}}}
\def\8{{\sst{(8)}}}

\usepackage{multirow}
\usepackage{rotating}

\def\d{\textrm{d}}

\newcommand{\ba}{\begin{align}}
\newcommand{\ea}{\end{align}}

\newcommand{\bse}{\begin{subequations}}
\newcommand{\ese}{\end{subequations}}

\allowdisplaybreaks

\begin{document}

\makeatletter
\renewcommand{\theequation}{\thesection.\arabic{equation}}
\@addtoreset{equation}{section}
\makeatother

\begin{titlepage}

\begin{flushright}
IFT-UAM/CSIC-20-66 \\
HIP-2020-15/TH
\end{flushright}

\vspace{15pt}
   
   \begin{center}
   \baselineskip=16pt

\begin{Large}\textbf{
\textrm{Brane-jet stability of non-supersymmetric AdS vacua}
}\end{Large}

\vspace{40pt}

{\large  \textbf{Adolfo Guarino}$^{a, b}$ ,\,\, \textbf{Javier Tarr\'io}$^{c}$ \,\,and\,\,  \textbf{Oscar Varela}$^{d, e}$}

\vspace{25pt}

$^a$\,{\normalsize  
Departamento de F\'isica, Universidad de Oviedo,\\
Avda. Federico Garc\'ia Lorca 18, 33007 Oviedo, Spain.}
\\[5mm]

$^b$\,{\normalsize  
Instituto Universitario de Ciencias y Tecnolog\'ias Espaciales de Asturias (ICTEA) \\
Calle de la Independencia 13, 33004 Oviedo, Spain.}
\\[5mm]

$^c$\,{\normalsize  Department of Physics and Helsinki Institute of Physics \\ P.O.Box 64
FIN-00014, University of Helsinki, Finland.}
\\[5mm]

$^d$\,{\normalsize  
Department of Physics, Utah State University, Logan, UT 84322, USA.}
\\[5mm]

$^e$\,{\normalsize  
Departamento de F\'\i sica Te\'orica and Instituto de F\'\i sica Te\'orica UAM/CSIC, \\  Universidad Aut\'onoma de Madrid, Cantoblanco, 28049 Madrid, Spain.}
\\[10mm]

\mbox{\texttt{adolfo.guarino@uniovi.es} \, , \, \texttt{javier.tarrio@helsinki.fi} \, , \, \texttt{oscar.varela@usu.edu}}

\vspace{20pt}

\end{center}

\begin{center}
\textbf{Abstract}
\end{center}

\begin{quote}

We classify the non-supersymmetric, and perturbatively stable within $D=4$, AdS vacua of maximal $D=4$ supergravity with a dyonic ISO(7) gauging in a large sector of the supergravity. Seven such vacua are established within this sector, all of them giving rise to non-supersymmetric $\,\textrm{AdS}_{4} \times \textrm{S}^{6}\,$ type IIA backgrounds with and without non-trivial warpings and with internal fluxes. Then, we analyse the dynamics of various probe D$p$-branes in these backgrounds searching for potential brane-jet instabilities. In all these cases, such instabilities are absent.~Finally, an alternative decay channel through tunnelling is investigated, focusing on one of the seven backgrounds. We do not find instabilities either, but the analysis remains inconclusive.

\end{quote}

\vfill

\end{titlepage}

\tableofcontents

\hrulefill

\section{Motivation}

The issue of the non-perturbative (in)stability of non-supersymmetric vacua in string/M-theory has gained a renewed interest in light of the weak gravity and swampland conjectures \cite{ArkaniHamed:2006dz,Ooguri:2016pdq}. Focusing, for definiteness, in the class of anti-de Sitter (AdS) vacua that uplift from maximal supergravities in lower dimensions, all known non-supersymmetric $\,\textrm{AdS}_{5} \times \textrm{S}^{5}\,$ backgrounds of type IIB string theory that uplift \cite{Lee:2014mla,Ciceri:2014wya,Baguet:2015sma} from extrema of $D=5$ $\cN=8$ SO(6)-gauged supergravity \cite{Gunaydin:1984qu} have instabilites already at the perturbative level. This follows from \cite{Krishnan:2020sfg,Bobev:2020ttg}, where the extrema of that gauged supergravity were scanned: all the non-supersymmetric extrema have Kaluza--Klein (KK) excitations, contained within the $\,\cN=8\,$ supergravity, with mass below the Breitenlohner--Freedman (BF) bound \cite{Breitenlohner:1982bm}. Similar classification results exist \cite{Comsa:2019rcz} for $\,\textrm{AdS}_{4} \times \textrm{S}^{7}\,$ vacua of M-theory that uplift \cite{deWit:1986iy} from extrema of $D=4$ $\cN=8$ SO(8)-gauged supergravity \cite{deWit:1982ig}. In this case, and in  contrast to type IIB, there is one, and only one \cite{Comsa:2019rcz}, non-supersymmetric critical point whose KK spectrum does not contain BF-unstable modes, at least within the slice of the KK towers contained in the $\cN=8$ SO(8)-supergravity. This vacuum preserves an SO(4) subgroup of SO(8), was first obtained within the SO(8) gauged supergravity in \cite{Warner:1983du}, uplifted to M-theory in \cite{Godazgar:2014eza}, and declared to be BF-stable within the KK slice contained in the $\cN=8$ SO(8)-gauged supergravity in \cite{Fischbacher:2010ec}.

The authors of \cite{Bena:2020xxb} recently assessed the stability of the non-supersymmetric $\,\textrm{AdS}_{4} \times \textrm{S}^{7}\,$ M-theory vacuum of \cite{Godazgar:2014eza,Warner:1983du} against a class of non-perturbative decay channels that they dubbed ``brane jets" (BJs). This test, originally introduced in the similar M-theory context of \cite{Corrado:2001nv}, consists in placing a spacetime-filling probe M2-brane in the relevant $\,\textrm{AdS}_{4} \times \textrm{S}^{7}\,$ background created by a stack of M2-branes, possibly amalgamated with M5-branes. A potential is thus generated between the probe and the stack. If this potential is attractive, the probe approaches the stack and smoothly merges with it leading to a stable final state. If, on the contrary, the potential is repulsive, the probe is ejected away from the stack destabilising it. The S$^7$ warp factor of the background is crucial for this test, as only some directions along the internal S$^7$ might lead to BJ instabilities. The $\cN=2$ $\,\textrm{AdS}_{4} \times \textrm{S}^{7}\,$ solution of \cite{Corrado:2001nv,Warner:1983vz} tests negative for BJ instabilities \cite{Corrado:2001nv}, in agreement with the expectation that supersymmetric solutions should be perturbative and non-perturbatively stable. The non-supersymmetric solution of \cite{Godazgar:2014eza,Warner:1983du} turns out to exhibit BJ-instabilities along some directions within S$^7$ \cite{Bena:2020xxb}, in agreement with the general expectation of \cite{ArkaniHamed:2006dz,Ooguri:2016pdq}. Similar M5 or D4-D8 backgrounds that are non-supersymmetric but BF-stable within gauged supergravity have also been shown to be BJ-unstable \cite{Suh:2020rma}.

Motivated by these developments, in this paper we assess the BJ fate of a class of non-supersymmetric $\,\textrm{AdS}_{4} \times \textrm{S}^{6}\,$ vacua of massive type IIA string theory \cite{Romans:1985tz} that uplift \cite{Guarino:2015jca,Guarino:2015vca} from $D=4$ $\cN=8$ supergravity with a dyonic ISO(7) gauging \cite{Guarino:2015qaa}. First of all, we perform a thorough scan of vacua of the scalar potential of $\cN=8$ ISO(7) supergravity, particularised to a large subsector: the seven-chiral model that we constructed in appendix A of \cite{Guarino:2019snw}. We recover all known supersymmetric vacua, find a large number of new non-supersymmetric ones (all of them BF-unstable), and declare BF-stable within the $\cN=8$ supergravity a number of previously known non-supersymmetric vacua. We only retain the latter for our subsequent BJ analysis. It is noticeable that, while the SO(8) gauging \cite{deWit:1982ig} of $D=4$ $\cN=8$ supergravity has only one critical point \cite{Warner:1983du} which is non-supersymmetric and BF-stable within its $\cN=8$ theory \cite{Fischbacher:2010ec,Comsa:2019rcz}, the dyonic ISO(7) gauging \cite{Guarino:2015qaa,Comsa:2019rcz} has at least the seven ones contained in the seven-chiral subsector. A comprehensive scan of critical points within the full $D=4$ $\cN=8$ ISO(7) theory along the lines of \cite{Comsa:2019rcz,Krishnan:2020sfg,Bobev:2020ttg} would be timely and very interesting, but it is beyond the scope of this paper.

\begin{table}[t!]
\centering
\footnotesize{
\scalebox{0.80}{
\begin{tabular}{cccccccccc}
\hline
\\[-2.5mm]
SUSY  	& bos. sym. &  $g^{-2}  c^{\frac13} V$ &  $D=4$ & Uplift & KK spec.     & KK grav.   & BF stable & BF stable & BJ stable 
\\[0pt]
          	&      $G$           &         &       solution         &  to IIA            &  in $D=4$  & spec. in IIA & in $D=4$ & in IIA  & 
\\[2pt]
\hline
\\[-10pt]
$ \cN=3 $ &	$\textrm{SO}(3)_{\textrm{d}} \times \textrm{SO}(3)_{\textrm{R}}$ &  $-\frac{2^{16/3}}{3^{1/2}} $   &	\cite{Gallerati:2014xra} & \cite{Pang:2015vna,DeLuca:2018buk}  &\cite{Gallerati:2014xra}  & \cite{Pang:2015rwd} & $\checkmark $ & $\checkmark $  & $\checkmark $
\\[10pt]
$ \cN=2 $ & $\textrm{SU}(3) \times \textrm{U}(1)$  &   $-2^2 \, 3^{3/2} $&	\cite{Guarino:2015jca}  & \cite{Guarino:2015jca} & \cite{Guarino:2015qaa}  & \cite{Pang:2017omp} & $\checkmark $ & $\checkmark $ & $\checkmark $
\\[10pt]
$ \cN=1 $ &	G$_2$ & $- \frac{2^{28/3} \, 3^{1/2}}{5^{5/2}} $   &	\cite{Borghese:2012qm}  & \cite{Behrndt:2004km,Guarino:2015vca} & \cite{Borghese:2012qm}   & \cite{Pang:2017omp} & $\checkmark $  &  $\checkmark $ &  $\checkmark $
\\[10pt]
$ \cN=1 $ &	SU(3) &   $-\frac{2^{8} \, 3^{3/2}}{5^{5/2}} $  &	\cite{Guarino:2015qaa}  & \cite{Varela:2015uca}  & \cite{Guarino:2015qaa}   & \cite{Pang:2017omp} & $\checkmark $ & $\checkmark $  &   $\checkmark $
\\[10pt]
$ \cN=1 $ &	$\textrm{U}(1)_\textrm{R}$ &  $-25.697101$  &	\cite{Guarino:2019snw}  & ?  &  [here]   &  ? & $\checkmark $ & $\checkmark $ &   ?
\\[10pt]
$ \cN=1 $ &	$\textrm{U}(1)_\textrm{R}$ &  $-35.610235$  &	\cite{Guarino:2019snw}  & ? &  [here]  & ? & $\checkmark $ &  $\checkmark $ & ?
\\[10pt]
\hline
\\[-2.5mm]
$ \cN=0 $ &	G$_2$   &  $-\frac{2^{16/3}}{3^{1/2}}$ &	\cite{Borghese:2012qm} & \cite{Lust:2008zd,Guarino:2015vca} & \cite{Borghese:2012qm}   & \cite{Pang:2017omp} & $\checkmark $ & ?  & $\checkmark $
\\[10pt]
$ \cN=0 $ &	SU(3)   & $-23.413628$  &	\cite{Guarino:2015qaa}  & \cite{Varela:2015uca} &  \cite{Guarino:2015qaa}  & \cite{Pang:2017omp} & $\checkmark $ &  ?  & $\checkmark $
\\[10pt]
$ \cN=0 $ &	SU(3)  &  $-23.456779$ &	\cite{Guarino:2015qaa}    & \cite{Varela:2015uca} &  \cite{Guarino:2015qaa}  & \cite{Pang:2017omp} & $\checkmark $ & ?  & $\checkmark $
\\[10pt]
$ \cN=0 $ &	$\textrm{SO}(3)_{\textrm{d}} \times \textrm{SO}(3)_{\textrm{R}}$ &  $-23.512690$  &	\cite{Guarino:2015qaa} & \cite{DeLuca:2018buk}  &  \cite{Guarino:2015qaa}  & ? & $\checkmark $  &  ? & $\checkmark $
\\[10pt]
$ \cN=0 $ &	$\textrm{U}(1)_{\textrm{d}} \times \textrm{SO}(3)_{\textrm{R}}$  & $-23.456053$ &	\cite{Guarino:2019jef} & \cite{Guarino:2019snw}$^*$ &  [here]  & ? & $\checkmark $ & ?  & $\checkmark $
\\[10pt]
$ \cN=0 $ &	$\textrm{U}(1)_{\textrm{d}} \times \textrm{SO}(3)_{\textrm{R}}$ & $-23.458780$ & 	\cite{Guarino:2019jef} & \cite{Guarino:2019snw}$^*$ &   [here]  & ? & $\checkmark $ & ?  & $\checkmark $
\\[10pt]
$ \cN=0 $ &	$\textrm{SO}(3)_{\textrm{R}}$ & $-23.456098$ &  	\cite{Guarino:2019jef} & \cite{Guarino:2019snw}$^*$ &   [here]  & ? & $\checkmark $ &  ? & $\checkmark $
\\[5pt]
\hline
\end{tabular}}
\caption{\footnotesize{All known supersymmetric, and non-supersymmetric but BF-stable within $D=4$, $\,\textrm{AdS}_{4} \times \textrm{S}^{6}\,$ solutions of massive IIA supergravity that uplift from critical points of $D=4$ $\cN=8$ dyonic ISO(7) supergravity. The reference marked with $^*$ contains only partial results on the IIA uplift. Non-available data are denoted with ?}}\normalsize \label{Table:KnownBFStableCritical points}}
\end{table}

The upshot of this analysis of BF-stable vacua within $D=4$ $\cN=8$ ISO(7) supergravity is summarised in table \ref{Table:KnownBFStableCritical points}. The residual $\cN < 8$ supersymmetry, bosonic symmetry group $G \subset \textrm{SO(7)}$ and the cosmological constant (in units involving $c \equiv m/g$, with $g>0$ the electric and $m >0$ the magnetic couplings of the $\cN=8$ supergravity) are recorded. The groups SU(3) and G$_2$ therein are the unique such subgroups of SO(7), while $\textrm{SO}(3)_{\textrm{d}}$ and $\textrm{SO}(3)_{\textrm{R}}$ are embedded into SO(7) as in (\ref{Embedding_SO3_R}). Finally, $\textrm{U}(1)_{\textrm{d}}$ and $\textrm{U}(1)_{\textrm{R}}$ are the Cartan subgroups of $\textrm{SO}(3)_{\textrm{d}}$ and $\textrm{SO}(3)_{\textrm{R}}$, respectively. For convenience, the table also includes pointers to the literature indicating the references where each critical point was first found in the $D=4$ $\cN=8$ supergravity and uplifted to the resulting $\,\textrm{AdS}_{4} \times \textrm{S}^{6}\,$ type IIA solution, possibly recovering solutions like \cite{Behrndt:2004km,Lust:2008zd} first found by other methods. Perturbative and non-perturbative stability is guaranteed for the supersymmetric solutions but, notwithstanding the arguments of \cite{ArkaniHamed:2006dz,Ooguri:2016pdq}, stability of the non-supersymmetric solutions should be addressed on a case-by-case basis. 

Perturbative instabilities should manifest themselves as KK modes about the $\,\textrm{AdS}_{4} \times \textrm{S}^{6}\,$ solutions with mass below the BF bound on AdS$_4$. Two sectors of these KK towers are easily accessible: the slice containing the KK modes with all spins $s \leq 2$ that are also contained in $D=4$ $\cN=8$ ISO(7) supergravity, and the massive $s=2$ KK gravitons. The latter have been computed in the cases indicated in the table and, by gauge symmetry, are not expected to induce instabilities. The scalar, $s=0$, and vector, $s=1$, KK spectra in the $D=4$ $\cN=8$ slice have been computed in the references indicated in the table or in this paper if noted as [here], see appendix \ref{app_vacua}. Out of all the $D=4$ vacua that we find, only those supersymmetric and non-supersymmetric but free from BF-instabilities in the $D=4$ $\cN=8$ slice have been reported in table \ref{Table:KnownBFStableCritical points}. Obtaining the full KK spectra about these $\,\textrm{AdS}_{4} \times \textrm{S}^{6}\,$ solutions and determining whether the non-supersymmetric ones contain BF-unstable modes outside the $D=4$ $\cN=8$ slice remains an open problem.

The main objective of the paper is to test the non-supersymmetric vacua summarised in table \ref{Table:KnownBFStableCritical points} for non-perturbative BJ instabilities, along the lines of \cite{Bena:2020xxb,Corrado:2001nv}. For this purpose, in section \ref{sec:D2-brane} we place spacetime-filling D2-brane probes on each of these non-supersymmetric $\,\textrm{AdS}_{4} \times \textrm{S}^{6}\,$ vacua. Remarkably, we find no BJ instabilities. This is perhaps not so surprising for the non-supersymmetric G$_2$-invariant AdS critical point \cite{Borghese:2012qm}, as its $\,\textrm{AdS}_{4} \times \textrm{S}^{6}\,$ uplift has a trivial warp factor. However,  all the other solutions do involve non-trivial warp factors and yet they are BJ-stable under spacetime-filling D2-brane probes. BJ instabilities might still occur associated to other D$p$-branes, with $p=4,6,8$, wrapped around (contractible) cycles of the internal S$^6$. These instabilities could be expected on the grounds that all of these solutions are supported by internal fluxes. In section \ref{sec:Dp-branes} we address this question for the simplest of these solutions, the one with G$_2$ invariance. Again, rather surprisingly, we find no BJ instabilities. 

Of course, the absence of BJ instabilities does not contradict the statements of \cite{ArkaniHamed:2006dz,Ooguri:2016pdq}: these non-supersymmetric solutions might still decay in some other way, for instance tunnelling  into a stable vacuum. In the supergravity, this decay would be signalled by the existence of a domain-wall connecting this solution to a different one, possibly supersymmetric. In section \ref{sec:domain-walls} we search for this type of domain-walls in the effective $D=4$ $\cN=8$ supergravity, focusing again on the simplest solution: the one with G$_2$ symmetry. We find no conclusive evidence for the existence of such domain-walls, suggesting that the non-supersymmetric G$_2$ solution is also stable against this decay channel. However, we do not claim comprehensiveness of this analysis, which deserves further future investigation.

For later reference, we conclude by collecting some relevant expressions related to the D$p$-brane action. Following the Einstein frame conventions of \cite{Cederwall:1996ri}, an extended D$p$-brane ($p=0,2,4,6,8$) with tension $\,T_{p}\,$ and charge $\,Q_{p}\,$ in the presence of non-vanishing Romans mass $\,\hat{F}_{0}\,$  is described by the action \cite{Cederwall:1996ri,Green:1996bh}  (see also \cite{Araujo:2016jlx})
\begin{equation}
\label{SDp}
\begin{array}{lll}
S_{\textrm{D}_p} &=& S_{\textrm{D}_p}^{\textrm{(DBI)}} + S_{\textrm{D}_p}^{\textrm{(WZ)}}  + S_{\textrm{D}_{p}}^{(\hat{F}_{0})} \\[4mm]
&=& - \, T_{p} \,  \displaystyle\int d^{p+1}\xi \,  e^{\frac{p-3}{4} \hat{\phi}} \, \sqrt{-\textrm{det}\left[ \hat{g}_{ij}  + e^{-\frac{1}{2} \hat{\phi}} \left( 2 \pi \alpha' F_{ij} - \hat{B}_{ij} \right) \right]  }  \\[4mm]
& & + \, Q_{p} \,  \displaystyle\int \left.  \left( \displaystyle\sum \hat{C}_{n} \wedge e^{-\hat{B}_2}\right) \wedge  e^{2 \pi \alpha' F} \right|_{p+1}  \\[4mm]
& & - \, Q_{p} \,  \hat{F}_{0} \, \displaystyle \int  \sum_{r=0} \left.  (2 \pi \alpha')^r \, \frac{A \wedge F^r}{(r+1)!} \right|_{p+1} \; .
\end{array}
\end{equation}
The first piece here is the Dirac--Born--Infeld (DBI) term involving the pull-backs of the type II metric $\hat{g}_{ij}$, dilaton $\hat \phi$, and B-field $\hat{B}_{ij}$, as well as the field strength $F_{ij}$ of the Born-Infeld field $A_i$ defined on the worldvolume. The latter is parameterised by the $\,(p+1)\,$ coordinates $\,\xi^{i}\,$. The second piece is the Wess--Zumino (WZ) term involving the Ramond--Ramond (RR) potentials $\hat{C}_{n}$. Note that a non-vanishing $\,\hat{B}_{2}\,$ induces a set of $\,(p - 2 r)\,$ RR charges with $\,r=1,\ldots, \frac{p}{2}\,$ on a D$p$-brane via the $\,e^{-\hat{B}_2}\,$ factor in the WZ term. The last piece is the Chern--Simons form contribution to the WZ term for $\hat{F}_{0}$ \cite{Green:1996bh}. In order for the configuration to be supersymmetric, the tension-charge relation $\,T_{p} = \mp Q_{p}\,$  must hold, with the upper sign for the D$p$-brane and the lower sign for the anti-D$p$-brane in our conventions. We will restrict our study to the case of unmagnetised D$p$-branes whereby
\begin{equation}
\label{unmagnetised_F}
F_{ij}=0 \ .
\end{equation}
The first two terms in (\ref{SDp}) simplify accordingly and the third term vanishes.

\section{D$2$-brane-jet stability}
\label{sec:D2-brane}

\subsection{Spacetime-filling D2-brane probes on $\,\textrm{AdS}_{4} \times \textrm{S}^{6}\,$} \label{sec:Uplifting}

The non-supersymmetric type IIA $\,\textrm{AdS}_{4} \times \textrm{S}^{6}\,$ solutions of interest have been given on a case-by-case basis in the references indicated in table \ref{Table:KnownBFStableCritical points}. As argued in appendix \ref{app:SO3_R_sector}, all the non-supersymmetric solutions in the table, as well as the supersymmetric ones with the exception of the two $\cN=1$ $\textrm{U}(1)_\textrm{R}$-invariant solutions, are encompassed by the formalism of \cite{Guarino:2019snw,Guarino:2019jef}. Specifically, all the relevant $\,\textrm{AdS}_{4} \times \textrm{S}^{6}\,$ solutions can be parameterised in terms of eight real constants $\varphi$, $\chi$, $\phi_i$, $b_i$, $i=1,2,3$, corresponding to $D=4$ supergravity scalars: see appendices \ref{app_vacua} and \ref{app:SO3_R_sector} for details. The specific values that these constants attain at each of the specific solutions of table \ref{Table:KnownBFStableCritical points}, namely, the corresponding $D=4$ scalar vacuum expectation values (vevs), can be found in table \ref{Table:ScalarVevs} of appendix \ref{app:SO3_R_sector}. Along with these constants, the solutions also depend on the $\mathbb{R}^7$ coordinates $\mu^I$, $I=1, \ldots, 7$, constrained to lie on the S$^6$ locus 
\begin{equation} \label{eq:S6constraint}
\delta_{IJ} \, \mu^I \mu^J = 1 \; .
\end{equation}
These backgrounds are created by a stack of D2-branes in the presence of other D$p$-branes. 

The ten-dimensional Einstein frame metric, the dilaton and the RR three-form potential for all these solutions are given by 
\begin{eqnarray} 
\label{10Dmetric_SO(3)_R}
& d\hat{s}^2_{10} = \Delta_1^{1/8} \, \Delta_2^{1/4} \,  \big( ds_{\textrm{AdS}_{4}}^2 + g^{-2} \, \Delta_2^{-1} \, ds_{\textrm{S}^6}^2 \big) \; , \qquad 
e^{\hat \phi} = \Delta_1^{3/4} \Delta_2^{-1/2} \; , \nonumber \\[8pt]
& \hat{C}_3 = C \,  e^{3 A(r)} \,  dx^{0} \wedge dx^{1} \wedge dx^{2} + \ldots  
\end{eqnarray}
The quantities $\Delta_1$, $\Delta_2$ and $C$ here depend both on the constants $\varphi$, $\chi$, $\phi_i$, $b_i$, and on the S$^6$ angles $\mu^I$. Their explicit expressions \cite{Guarino:2019snw,Guarino:2019jef} are reviewed in appendix \ref{app:SO3_R_sector}. In (\ref{10Dmetric_SO(3)_R}), we take $ds_{\textrm{AdS}_{4}}^2$ to be the line element 
\begin{equation}
\label{4D_metric_SO(3)_R}
ds^2_{\textrm{AdS}_{4}} = e^{2 A(r)} \, \eta_{\alpha \beta} \, dx^{\alpha} \, dx^{\beta} + dr^2
\hspace{3mm}  \hspace{8mm} (\alpha=0,1,2) \ ,
\end{equation}
of radius $\,L\,$ AdS$_4$ in the Poincar\'e patch, so that here and in (\ref{10Dmetric_SO(3)_R}), $\,A(r)=\frac{r}{L}\,$ and $x^\alpha$, $\alpha =0,1,2$, are the Poincar\'e coordinates. The AdS$_4$ radius is specified in terms of the function $V$ (the $D=4$ scalar potential) of $\varphi$, $\chi$, $\phi_i$, $b_i$ given in (\ref{V_SO3}) of appendix \ref{app:SO3_R_sector}. Finally, $g$ in (\ref{10Dmetric_SO(3)_R}) is an additional constant (the $D=4$ $\cN=8$ electric coupling) that sets an overall scale, and $ds_{\textrm{S}^6}^2$ in (\ref{10Dmetric_SO(3)_R}) is a family of metrics on a topological S$^6$ that depend on the $D=4$ scalars $\varphi$, $\chi$, $\phi_i$, $b_i$ and on the S$^6$ coordinates $\mu^I$. The explicit expression of this metric can be found in \cite{Guarino:2019snw}, but it will not be needed for the analysis of this section. The legs of the RR three-form $\hat{C}_3$ along the internal $S^6$ (denoted with ellipses in (\ref{10Dmetric_SO(3)_R})), the RR one-form and the B-field will not be needed either.

We now move on to place a probe D2-brane on this family of backgrounds. We choose to put the probe parallel to the AdS$_{4}$ boundary, \textit{i.e.} along $\,\mathbb{R}^{1,2}\,$, so that the worldvolume coordinates are $\,{\xi^{i}=x^{i}}\,$, $\,i=0,1,2\,$. In this case, and with the simplifying assumption (\ref{unmagnetised_F}), the D$2$-brane action that follows from (\ref{SDp}) reads
\begin{equation}
\label{SD2_SO(3)_R}
S_{\textrm{D}_2} = - T_{2} \,  \displaystyle\int d^{3}\xi \,  e^{-\frac{1}{4} \hat{\phi}}  \, \sqrt{-\textrm{det}(\hat{g}_{ij}})  -  T_{2}  \displaystyle\int  \hat{C}_{3}   \ .
\end{equation}
Pulling back the background (\ref{10Dmetric_SO(3)_R}), (\ref{4D_metric_SO(3)_R}) on the D2-brane worldvolume 
\begin{equation}
\label{vol_3}
\hat{\textrm{vol}}_{3}=  e^{3 A(r)} \,  \Delta^{-\frac{3}{2}} \,  dx^{0} \wedge dx^{1} \wedge dx^{2} \ ,
\end{equation}
the action (\ref{SD2_SO(3)_R}) simplifies into
\begin{equation}
\label{SD2_final_SO(3)_R}
S_{\textrm{D}_2} = - T_{2} \,  \displaystyle \int  \hat{\textrm{vol}}_{3} \left[  e^{-\frac{1}{4} \hat{\phi}}  +  \Delta^{\frac{3}{2}} \, C \right] \ ,
\end{equation}
where we have introduced the usual shorthand notation for the metric warp factor in (\ref{10Dmetric_SO(3)_R}),
\begin{equation} \label{eq:Deltam1}
\Delta^{-1} \equiv \Delta_1^{1/8} \, \Delta_2^{1/4}  \; .
\end{equation}
The action (\ref{SD2_final_SO(3)_R}) implies an effective radial potential density\footnote{This potential shouldn't be confused with the scalar potential of $D=4$ supergravity, which we denote with the same symbol $V$.}
\begin{equation}
\label{V(r)_D2_SO(3)_R}
V(r) = T_{2} \, e^{3 A(r)} \left[  e^{-\frac{1}{4} \hat{\phi}} \,  \Delta^{-\frac{3}{2}} +  C \right] \ ,
\end{equation}
experienced by the D$2$-brane probe in the class of backgrounds (\ref{10Dmetric_SO(3)_R}). The force exerted by the background on the probe is therefore computed from
\begin{equation}
\frac{dV(r)}{dr} = T_{2} \, e^{3 A(r)} \, \frac{3}{L} \, \left[ e^{-\frac{1}{4} \hat{\phi}} \,   \Delta^{-\frac{3}{2}} +  C  \right]  \ .
\end{equation}
This force is thus attractive or repulsive depending on the sign of the quantity
\begin{equation}
\label{Theta_D2_SO(3)_R}
\Theta \equiv    e^{-\frac{1}{4} \hat{\phi}} \, \Delta^{-\frac{3}{2}}  +  C  \ ,
\end{equation}
analogue to that introduced in \cite{Bena:2020xxb} in an M2-brane context (with the same relative sign once a different sign convention on the Freund--Rubin term is taken into account). For each solution in table \ref{Table:KnownBFStableCritical points} with the $D=4$ scalar vevs $\varphi$, $\chi$, $\phi_i$, $b_i$ fixed as in table \ref{Table:ScalarVevs}, $\Theta$ is a function of the S$^6$ angles $\mu^I$. If $\,\Theta \geq 0\,$ everywhere on S$^6$, the resulting force attracts the probe D$2$-brane towards the stack of branes located at $\,r \rightarrow -\infty\,$ that creates the background geometry. In this case, the IIA background is stable with respect to this decay channel. On the contrary, if $\,\Theta < 0\,$ on certain directions along the S$^6$, the resulting force pushes the probe D$2$-brane towards $\,r \rightarrow \infty\,$ along those directions, and the massive IIA background suffers from a BJ-instability analogue to \cite{Bena:2020xxb,Suh:2020rma}. Let us now determine the behaviour of (\ref{Theta_D2_SO(3)_R}) on a case-by-case basis.

\subsection{Solutions with G$_2$ symmetry} \label{sec:D2G2Text}

As explained in appendix \ref{sec:D2G2}, the coefficient $\Theta$ becomes constant (independent of the S$^6$ angles), for the solutions with G$_2$-invariance. Specifically, we obtain
\begin{eqnarray} \label{ThetaG2N=1}
&& {\cal N} = 1 \, , \, \mathrm{G}_2 \, :  \quad \Theta = 2^{\frac{31}{6}} \cdot 15^{-\frac{7}{4}} ( 4- \sqrt{15}  ) \, c^{-\frac{1}{6}}  \approx 0.0399 \, c^{-\frac{1}{6}}  \; , %
\\[5pt]
\label{ThetaG2N=0}
&& {\cal N} = 0 \, , \, \mathrm{G}_2 \, :  \quad  \Theta = 2^{\frac{19}{6}} \cdot 3^{-\frac{7}{4}} ( \sqrt{2} - 1 )  \, c^{-\frac{1}{6}}  \approx 0.5439 \, c^{-\frac{1}{6}}  \; , 
\end{eqnarray}
positive in both cases (recall that $c \equiv m/g >0$). Both G$_2$-invariant configurations are thus BJ-stable. This was expected for the supersymmetric solution, but comes a bit as a surprise for the non-supersymmetric one. In retrospect, perhaps the ${\cal N} = 0 $, $\mathrm{G}_2$ result is not so surprising either, as the warping becomes constant and in the previous examples 
\cite{Bena:2020xxb,Suh:2020rma} the BJ instabilities tend to come associated with non-trivial warpings (not always, though: the ${D=11}$ SO$(7)_-$ solution is unwarped and BJ-stable \cite{Bena:2020xxb}; however, it is BF-unstable). We have also evaluated (\ref{Theta_D2_SO(3)_R}) for the BF-unstable SO(7) point \cite{DallAgata:2011aa}, which also lies in the sector with at least G$_2$-invariance and thus also leads to trivial warping in IIA (see (4.8) of \cite{Varela:2015uca}). It turns out to also be BJ-unstable with $\Theta = 3^{-1} \cdot 5^{\frac{1}{12}} ( 3- \sqrt{10}  ) \, c^{-\frac{1}{6}} $. More importantly, all other non-supersymmetric solutions in table \ref{Table:KnownBFStableCritical points} involve non-trivial warpings and, as we will now next see, they are also free from D2-BJ-instabilities.

\subsection{Solutions with at least SU(3) symmetry} \label{sec:D2SU3Text}

\begin{figure}[t]
\begin{center}
\hspace{6mm}
\includegraphics[width=0.4\textwidth]{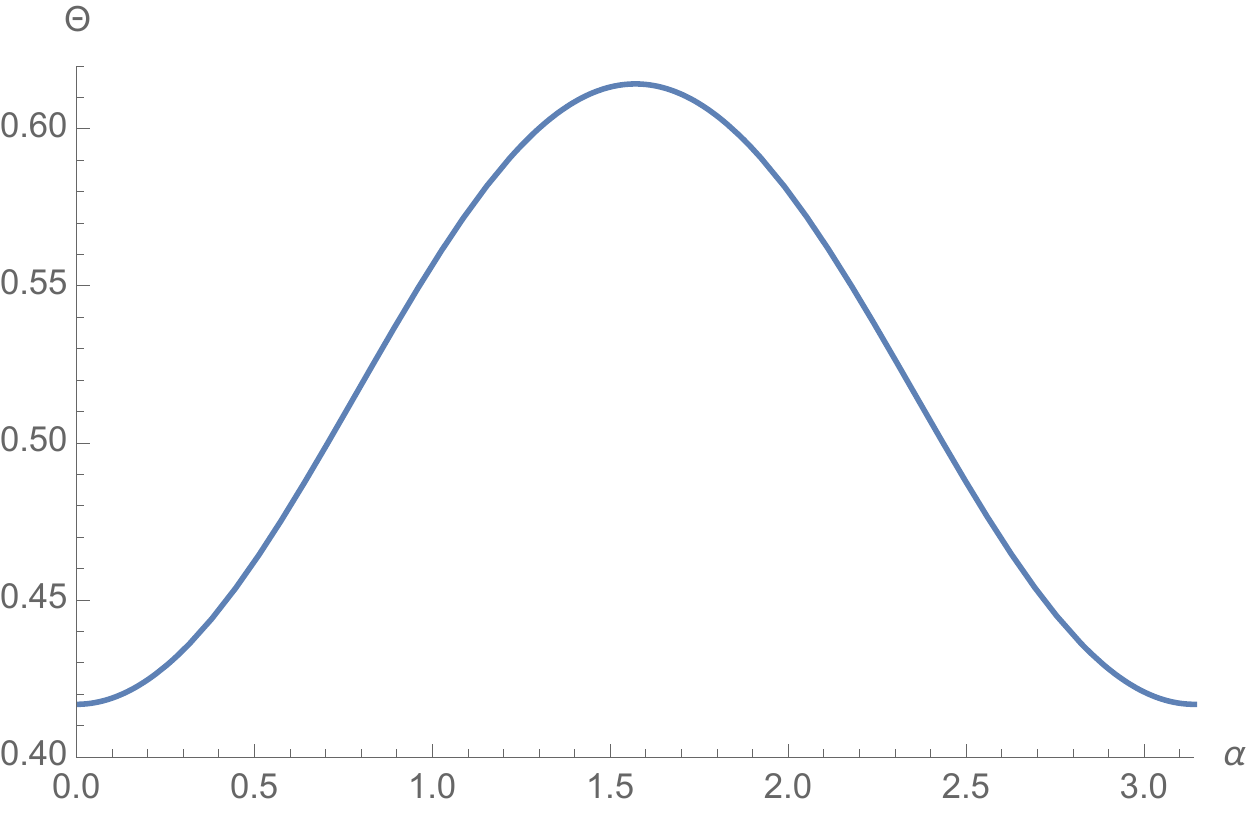}\hspace{2cm}
\includegraphics[width=0.4\textwidth]{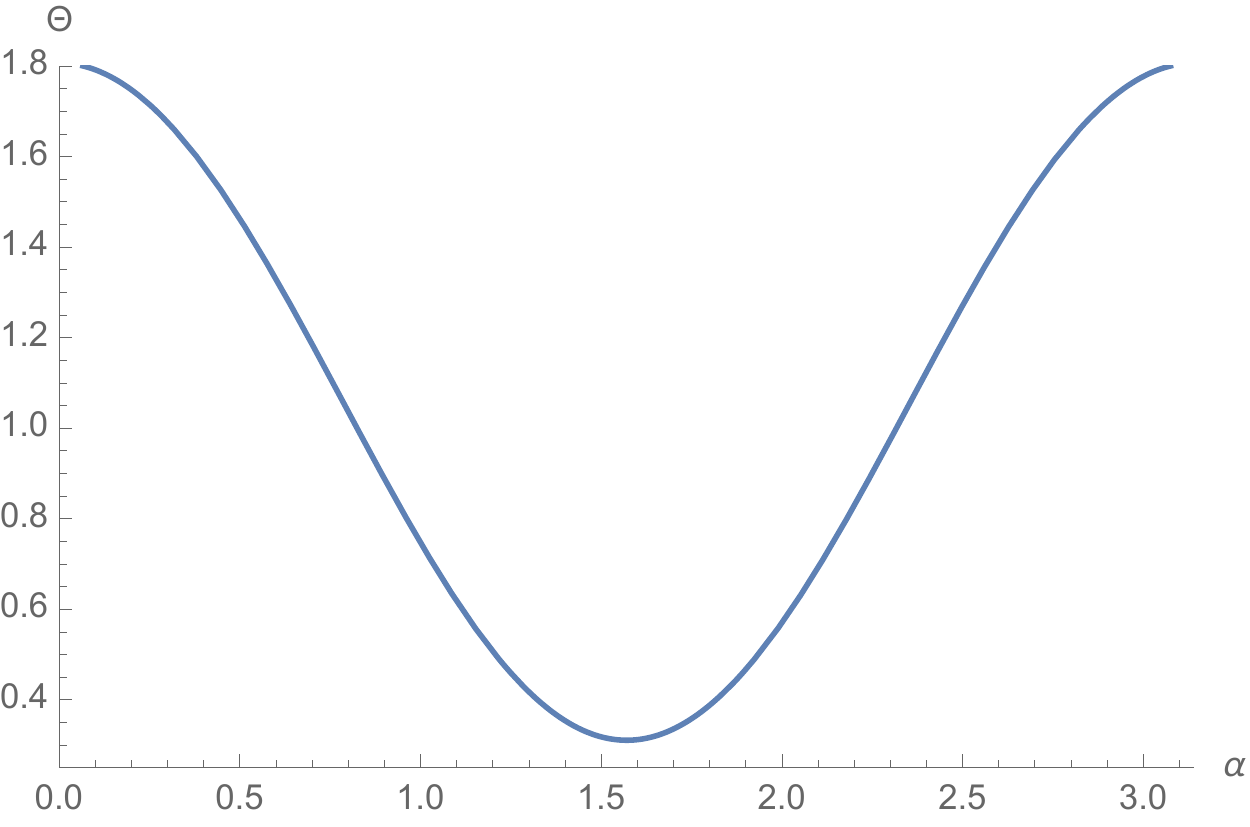}
\caption{\footnotesize{The coefficient $\Theta (\alpha)$ in (\ref{Theta_D2_SO(3)_R}) for spacetime-filling probe D2-branes on the $\cN=0$ SU(3)-invariant vacua with cosmological constants $V = -23.4136 \, g^{2}  c^{-\frac13}  $ (left) and $V = -23.4567 \, g^{2}  c^{-\frac13} $ (right).}\normalsize}\label{fig:SU3N=0Theta}
\end{center}
\end{figure}

The IIA backgrounds with at least SU(3) symmetry but not G$_2$, are cohomogeneity-one, and the coefficient $\Theta$ develops a dependence in the S$^6$ angle $\alpha$ described in appendix \ref{sec:D2SU3}. Bringing (\ref{Delta1})--(\ref{V_SO3}) with (\ref{eq:SU3locus}), (\ref{eq:musSU3}) to the expression (\ref{Theta_D2_SO(3)_R}), and further particularising to the $D=4$ scalar vevs contained in table  \ref{Table:ScalarVevs}, we compute:
\begin{eqnarray} \label{ThetaN=2SU3U1}
 {\cal N} = 2 \, , \, \mathrm{SU}(3) \times \textrm{U}(1) \, & : &  \quad \Theta (\alpha) = 2^{\frac{3}{2}} \cdot 3^{-\frac{3}{4}} \, c^{-\frac{1}{6}} \, \cos^2 \alpha  \; , %
\\[5pt]
\label{ThetaN=1SU3}
 {\cal N} = 1 \, , \, \mathrm{SU}(3) \, & : &  \quad  \Theta (\alpha) = 16 \cdot 3^{-\frac{3}{4}} \cdot 5^{-\frac{7}{4}} \, c^{-\frac{1}{6}} \, \big(  6 -\sqrt{10}+ 2 \cos 2\alpha  \big)  \; , 
\\[5pt]
\label{ThetaN=0SU3p1}
 {\cal N} = 0 \, , \, \mathrm{SU}(3) \, & : &  \quad  \Theta (\alpha) \approx  c^{-\frac{1}{6}} \, \big( 0.515522  -  0.098760 \,  \cos 2\alpha \big)  \; , 
\\[5pt]
\label{ThetaN=0SU3p2}
 {\cal N} = 0 \, , \, \mathrm{SU}(3) \, & : &  \quad  \Theta (\alpha) \approx  c^{-\frac{1}{6}} \, \big( 1.058613 + 0.747962 \,  \cos 2\alpha   \big)  \; .
\end{eqnarray}
The last two cases correspond to the relevant solutions in the order they appear in table~\ref{Table:KnownBFStableCritical points}. The functions (\ref{ThetaN=2SU3U1}) and (\ref{ThetaN=1SU3}) are non-zero in the entire domain (\ref{alpharange}) of $\alpha$, leading to stability of these supersymmetric solutions against D2 BJs. This is analogue to the M2 BJ-stability of the supersymmetric solution discussed in \cite{Corrado:2001nv}. More surprisingly, the functions (\ref{ThetaN=0SU3p1}) and (\ref{ThetaN=0SU3p2}) are also non-negative for all $\alpha$ (see figure \ref{fig:SU3N=0Theta}), leading as well to D2-BJ stability of the corresponding warped, non-supersymmetric $\,\textrm{AdS}_{4} \times \textrm{S}^{6}\,$ solutions. This is in contrast to the non-supersymmetric warped $\,\textrm{AdS}_{4} \times \textrm{S}^{7}\,$ solutions analysed in \cite{Bena:2020xxb,Suh:2020rma}, which are BJ unstable w.r.t.~the corresponding spacetime-filling probe branes.

Equations (\ref{ThetaN=2SU3U1})--(\ref{ThetaN=0SU3p2}) show that all solutions in the SU(3) sector that are BF-stable, are also D2-BJ-stable. For completeness, we have tested the D2-BJ-stability of the SO(6) solution ((4.7) of \cite{Varela:2015uca}), which also belongs to the class of solutions with at least SU(3) symmetry but is known to be BF-unstable \cite{DallAgata:2011aa}. This solution is also BJ-unstable, as it has $\Theta (\alpha) = 2^{-\frac{13}{12}} \cdot 3^{-1} \, c^{-\frac{1}{6}} \, \big( 9 \cos 2\alpha  + 7 \big)$, which dips below zero in a subinterval of (\ref{alpharange}).

\subsection{Solutions with an explicit factor of $\textrm{SO}(3)_{\textrm{R}}$} \label{sec:SO3RText}

\begin{figure}[t]
\begin{center}
\hspace{6mm}
\includegraphics[width=0.4\textwidth]{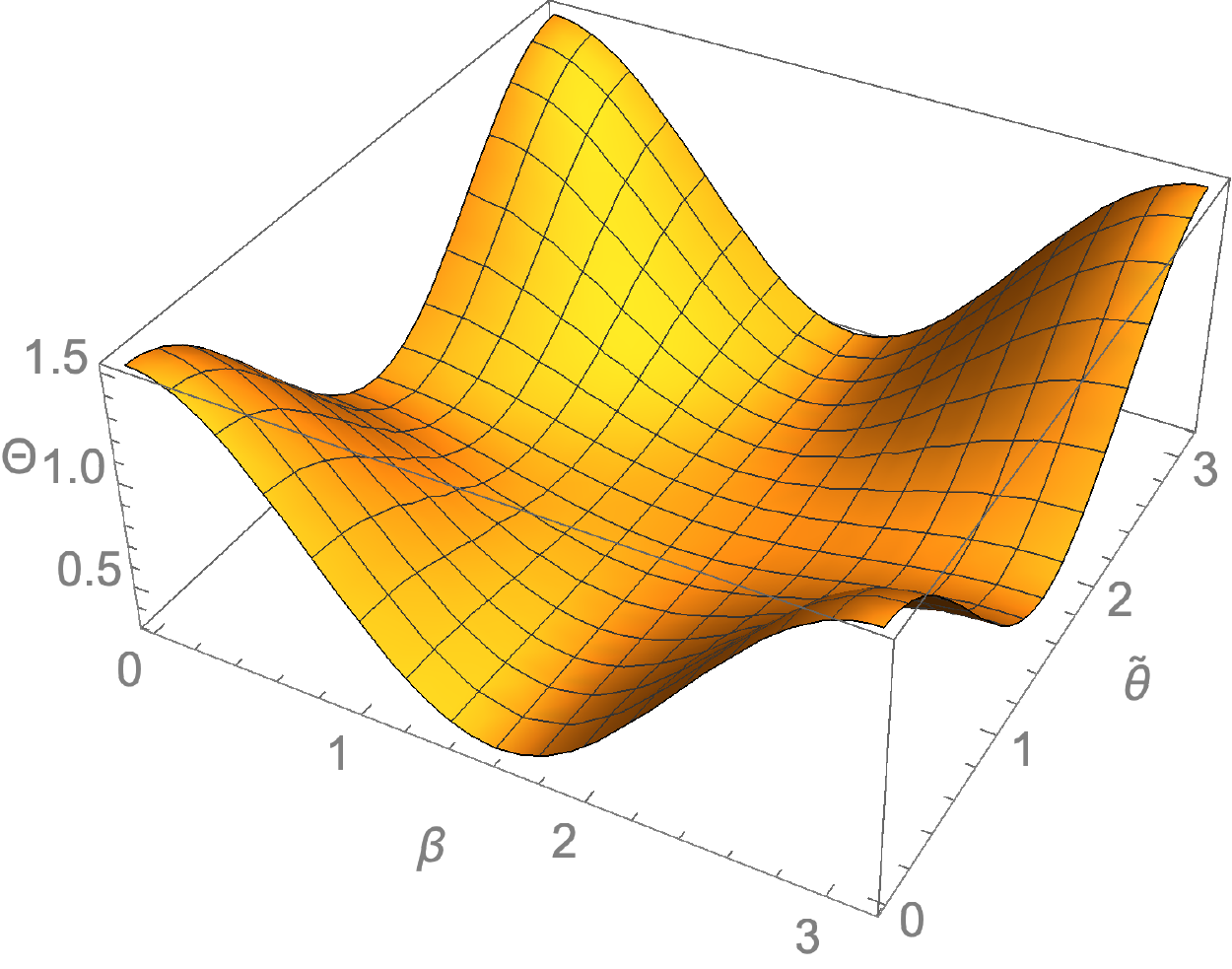}\hspace{2cm}
\includegraphics[width=0.4\textwidth]{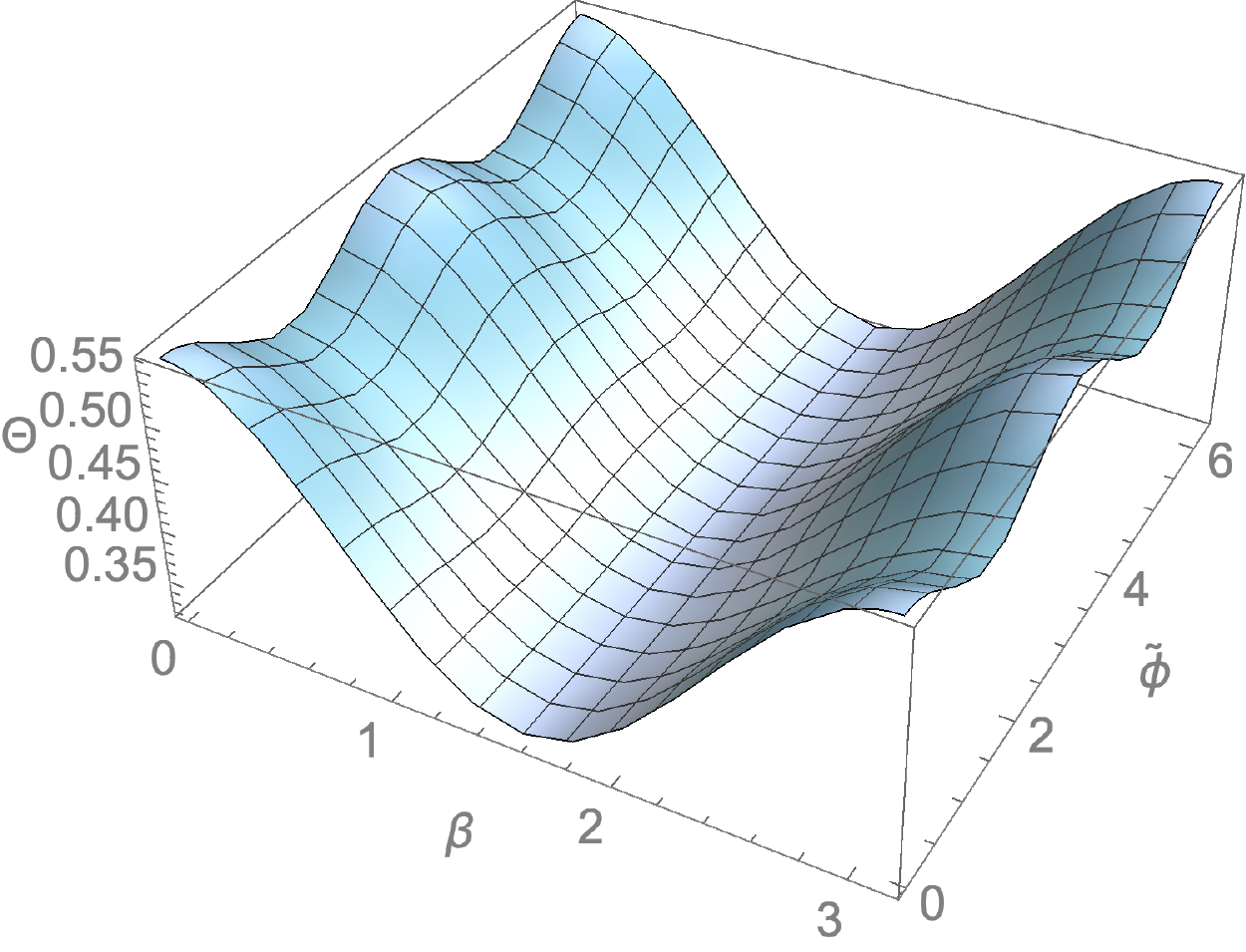}
\caption{\footnotesize{Left: The function $\Theta(\beta , \tilde{\theta})$ for the $\cN=0$ $\textrm{U}(1)_{\textrm{d}} \times \textrm{SO}(3)_{\textrm{R}} $ solution with cosmological constant $V =  -23.4561 \, g^{2}  c^{-\frac13} $. The behaviours of $\Theta(\beta , \tilde{\theta})$ for the solution with the same symmetry and $V = -23.4588 \, g^{2}  c^{-\frac13} $, and  of 
$\Theta(\beta , \tilde{\theta}, \tilde{\phi})$ at fixed $\tilde{\phi}$ for the $\cN=0$ $\textrm{SO}(3)_{\textrm{R}} $ solution,
are qualitatively similar. Right: The function $\Theta(\beta , \tilde{\theta} = \frac{\pi}{2}, \tilde{\phi})$ for the $\cN=0$ $\textrm{SO}(3)_{\textrm{R}} $ solution.}\normalsize}\label{fig:SO3N=0ThetaSolutions}
\end{center}
\end{figure}

Finally, we turn to assess the D2-BJ-stability of the solutions in table \ref{Table:KnownBFStableCritical points} whose residual symmetry exhibits an explicit factor of $\textrm{SO}(3)_{\textrm{R}}$ as defined in equation (\ref{Embedding_SO3_R}). These solutions also involve non-trivial warpings and, as reviewed in appendix \ref{sec:SO3R}, they are co-homogeneity one, two or three depending on whether their symmetry is enhanced to  include $\textrm{SO}(3)_{\textrm{d}}$, $\textrm{U}(1)_{\textrm{d}}$, or no additional factors, on top of $\textrm{SO}(3)_{\textrm{R}}$. Accordingly, in each of these three cases the function $\Theta$ develops a dependence on the S$^6$ coordinates $\beta$, or $(\beta , \tilde{\theta})$, or $(\beta , \tilde{\theta}, \tilde{\phi})$, introduced in appendix \ref{sec:SO3R}. Similarly as before, the calculation entails particularising (\ref{Theta_D2_SO(3)_R}) to (\ref{Delta1})--(\ref{V_SO3}) with, now, (\ref{musSO3R}), (\ref{eq:mutildes}), and fixing the $D=4$ scalar vevs as in table  \ref{Table:ScalarVevs}. For simplicity, we only record here the result for the cohomogeneity-one, $\textrm{SO}(3)_{\textrm{d}} \times \textrm{SO}(3)_{\textrm{R}} $-invariant solutions:
\begin{eqnarray} \label{ThetaN=3SO4}
\hspace{-5mm} {\cal N} = 3 & : & \Theta (\beta) = 2^{\frac{7}{6}} \cdot 3^{-\frac{7}{4}} \, c^{-\frac{1}{6}} \, \big( -4 +\sqrt{ 37 + 24 \cos 2 \beta +3 \cos 4\beta } \big)   \; , %
\\[5pt]
\label{ThetaN=0SO4}
\hspace{-5mm} {\cal N} = 0   & : & \Theta (\beta) \approx  c^{-\frac{1}{6}} \, \big( -1.3197 +\sqrt{ 3.7667 + 1.4641 \cos 2 \beta +0.0951 \cos 4\beta } \big)  \; . 
\end{eqnarray}
Both these functions are non-negative on the entire range (\ref{betarange}) of $\beta$: they have a global minimum at $\beta = \frac{\pi}{2} $, with $ \Theta (\frac{\pi}{2} ) =0 $ for (\ref{ThetaN=3SO4}) and $ \Theta (\frac{\pi}{2} ) \approx  0.2288$ for (\ref{ThetaN=0SO4}). Their plots are qualitatively similar to the plot for the $\cN=0$ SU(3) solution on the right panel of figure~\ref{fig:SU3N=0Theta}. Both solutions are thus BJ-stable with respect to spacetime-filling probe D2-branes. The $\cN=3$ case is again analogue to the supersymmetric cases considered in \cite{Corrado:2001nv} and above. The non-supersymmetric case is also BJ-stable, like the cases above and unlike \cite{Bena:2020xxb,Suh:2020rma}. 

We will omit the explicit form of $\Theta$ for the higher cohomogeneity cases, as the resulting expressions are not particularly enlightening, and simply refer to the plots in figure \ref{fig:SO3N=0ThetaSolutions}. The behaviour of the function $\Theta(\beta , \tilde{\theta})$ is qualitatively very similar for both non-supersymmetric solutions in table \ref{Table:KnownBFStableCritical points} with $\textrm{U}(1)_{\textrm{d}} \times \textrm{SO}(3)_{\textrm{R}} $ symmetry. In both cases, $\beta = \frac{\pi}{2}$ defines a line of global minima independent of $\tilde{\theta}$, with $\Theta(\frac{\pi}{2}, \tilde{\theta} ) \approx 0.3162$ and $\Theta(\frac{\pi}{2}, \tilde{\theta} ) \approx 0.2965$ for the solutions with cosmological constants $V = -23.4561 \, g^{2}  c^{-\frac13} $ and $V = -23.4588 \, g^{2}  c^{-\frac13} $, respectively. In both cases the function is positive at its global minimum, leading to D2-BJ stability for these solutions. Similarly, for the non-supersymmetric $\textrm{SO}(3)_{\textrm{R}} $-invariant solution, $\beta = \frac{\pi}{2}$ defines a plane of global minima of $\Theta(\beta , \tilde{\theta}, \tilde{\phi})$ independent of $\tilde{\theta}$ and $\tilde{\phi}$. This minimum is positive, $\Theta(\frac{\pi}{2} , \tilde{\theta}, \tilde{\phi}) \approx 0.3195$, leading to D2-BJ stability for the $\textrm{SO}(3)_{\textrm{R}} $ solution as well. 

We have also tested the BJ-stability of the remaining non-supersymmetric solution of ISO(7) supergravity with residual $\textrm{U}(1)_{\textrm{d}} \times \textrm{SO}(3)_{\textrm{R}} $ symmetry: the solution labelled {\it ii}) in \cite{Guarino:2019jef}. This solution is BF-unstable (see appendix \ref{app_vacua} below) yet, curiously, it is D2-BJ-instability-free. Finally, note that the two $\cN=1$ solutions in table \ref{Table:KnownBFStableCritical points} with symmetry $\textrm{U}(1)_{\textrm{R}}$ are excluded from our analysis. Being supersymmetric, they are also expected to be BJ-stable.

\section{D$p$-brane-jet stability of the  $\,\textrm{G}_{2}$-invariant vacua}
\label{sec:Dp-branes}

The presence of internal fluxes in the non-supersymmetric solutions listed in table \ref{Table:KnownBFStableCritical points} might lead to BJ instabilities associated to D$p$-brane probes with $p >0$ wrapped on (contractible) cycles of the internal S$^6$. In this section, we systematically compute the effective potential density $\,V(r)\,$ for such D$4$, D$6$ and D$8$ brane probes, focusing in the simplest backgrounds: the two of them with G$_2$ symmetry. We find that the $\cN=1$ vacuum is stable under these class of perturbations, as expected. More interestingly, the $\cN=0$ solutions is also stable.

Our starting point is the class of $\textrm{G}_{2}$-invariant backgrounds of massive IIA supergravity presented in \cite{Guarino:2015vca}, parameterised by the two scalars $\,(\varphi,\chi)\,$ in (\ref{eq:G2locus}), and obtained upon uplift of the $\textrm{G}_{2}$-invariant sector of the $\textrm{ISO}(7)$ maximal supergravity \cite{Guarino:2015qaa}. These backgrounds are given by
\begin{eqnarray} 
\label{10D_G2-sector}
& d \hat{s}_{10}^2 = \Delta^{-1} \,  ds^2_{\textrm{AdS}_{4}}  +  g^{-2} \, \Delta^{\frac{1}{3}} \, ds^2_{\textrm{S}^6} \; , \qquad 
e^{\hat \phi} =  e^{4 \varphi}  \, \Delta^{2}  \; , \qquad 
\hat{B}_2 = B \,  {\cal J} \; , \nonumber \\[8pt]
& \hat{C}_3 = C \,  e^{3 A(r)} \,  dx^{0} \wedge dx^{1} \wedge dx^{2} + g^{-3} \, \chi \, \textrm{Im} \, \Upomega \; , \qquad 
\hat{C}_1 = 0  \; ,
\end{eqnarray} 
in terms of the $\,\textrm{SU}(3)$-invariant two-form $\,\mathcal{J}\,$ and three-form $\,\Upomega\,$ specifying the homogeneous nearly-K\"ahler structure on $\,\textrm{S}^6\,$ (see appendix~\ref{app:Geometric_structures}). The scalar-dependent quantities $\,\Delta\,$, $\,B\,$ and $\,C\,$ take the expressions
\begin{equation}
\label{DeltaBC_func}
\begin{array}{rll}
\Delta & = & e^{-\frac{3}{4} \varphi} \big( 1+e^{2 \varphi} \chi^2 \big)^{-\frac{3}{4}} \ , \\[2mm]
B  & = & g^{-2} \, e^{2 \varphi} \, \chi \, \big( 1+e^{2 \varphi} \chi^2 \big)^{-1} \ , \\[2mm]
C  &=& - \frac{L}{3} \, \left[   g \, e^\varphi \big( 1+ e^{2\varphi} \chi^2 \big)^2 \big( 5 - 7 e^{2\varphi} \chi^2  \big)   + \,  m \, e^{7\varphi} \chi^3 \right] \ ,
\end{array}
\end{equation}
before being evaluated at the scalar vevs for the $\,\mathcal{N}=1\,$ and $\,\mathcal{N}=0\,$ $\,\textrm{G}_{2}$-invariant vacua of table~\ref{Table:ScalarVevs}. Once such vevs are inserted into (\ref{10D_G2-sector})-(\ref{DeltaBC_func}), the ten-dimensional $\textrm{G}_{2}$-invariant backgrounds, as expressed in \cite{Varela:2015uca}, are obtained.

We will now consider higher-dimensional D$p$-branes ($p>2$) wrapping specific (contractible) cycles within the $\,\textrm{S}^6\,$. To this end, it is convenient to describe the round S$^6$ as the sine-cone over S$^5$, 
\begin{equation}
\label{S6=S5xS1_metric}
ds^2_{\textrm{S}^6} = d\alpha^2 + \sin^2\alpha  \, ds_{\textrm{S}^5}^2 = d\alpha^2 + \sin^2\alpha \left(   ds^{2}_{\mathbb{CP}_{2}} + \boldsymbol{\eta} \right) \ ,
\end{equation}
and refer its homogeneous nearly-K\"ahler structure to the Sasaki-Einstein structure on S$^5$, see appendix \ref{app:Geometric_structures}. In (\ref{S6=S5xS1_metric}), $\alpha$ is the S$^6$ angle with range (\ref{alpharange}) introduced in (\ref{eq:musSU3}) that also appears in section \ref{sec:D2SU3Text} above. The volume form associated to (\ref{S6=S5xS1_metric}) is
\begin{equation}
\textrm{vol}_{6}(\textrm{S}^6) = \sin^5\alpha \,\,\, d\alpha \wedge  \textrm{vol}_{4}(\mathbb{CP}_{2}) \wedge \boldsymbol{\eta} \ .
\end{equation}
Here and in (\ref{S6=S5xS1_metric}), $\,\boldsymbol{\eta} \equiv d\psi + \boldsymbol{A}_{1}\,$ is a one-form along the S$^5$ Hopf fibre, with $\psi$ an angle ranging as $\,\psi \in [0 , 2\pi ]\,$ and $\,\boldsymbol{A}_{1}\,$ a local one-form potential for the K\"ahler form $ \, \boldsymbol{J}\,$ on $\mathbb{CP}_{2}$, normalised as $\,d\boldsymbol{A}_{1} = 2 \, \boldsymbol{J}\,$. See appendix~\ref{app:Geometric_structures} for further details.

\subsection{D$4$-brane wrapping internal two-cycles}
\label{Sec:D4_brane}

We start by considering a probe D$4$-brane parallel to the AdS$_{4}$ boundary, \textit{i.e.} along $\,\mathbb{R}^{1,2}\,$, and wrapping $\,\textrm{S}^{2} \subset \mathbb{CP}_{2} \subset \textrm{S}^6\,$.\footnote{The one-form basis elements on $\,\textrm{S}^{2} \, $ are identified with $\,\{\sigma_{1} \,,\, \sigma_{2}\}\,$  (see appendix~\ref{app:Geometric_structures}). Then the $\,{\textrm{S}^2 \sim \mathbb{CP}_{1}}\,$ geometry in (\ref{CP1_metric}) emerges from the $\,\mathbb{CP}_{2}\,$ metric in (\ref{CP2_metric}) upon setting the angle $\,\gamma=\frac{\pi}{2}\,$, which consistently reduces the K\"ahler form $\,\boldsymbol{J}\,$ (\ref{SU(2)_structure_forms}) on $\,\mathbb{CP}_{2}\,$ to $\,\boldsymbol{J}_{\mathbb{CP}_{1}}= -\textrm{vol}_{2}(\textrm{S}^2)\,$, (\ref{J_S2}), on $\,\mathbb{CP}_{1}\,$.} From (\ref{10D_G2-sector}) and (\ref{S6=S5xS1_metric}), combined with (\ref{CP2_metric}) and (\ref{CP1_metric}), the worldvolume of the D$4$-brane reads
\begin{equation}
\label{vol_5}
\hat{\textrm{vol}}_{5} =  e^{3 A(r)} \, \Delta^{-\frac{7}{6}} \,  g^{-2} \, \sin^2 \alpha  \,  \left(\frac{\sin\gamma}{2}\right)^2 \,\,  dx^{0} \wedge dx^{1} \wedge dx^{2} \wedge \textrm{vol}_{2}(\textrm{S}^2) \ ,
\end{equation}
and the action (\ref{SDp})-(\ref{unmagnetised_F}) simplifies into
\begin{equation}
\label{SD4}
S_{\textrm{D}_4} = - T_{4} \, \displaystyle\int d^{5}\xi \,   e^{\frac{1}{4} \hat{\phi}}  \,  \sqrt{-\textrm{det}(\hat{g}_{ij}  -   e^{-\frac{1}{2} \hat{\phi}} \, \hat{B}_{ij})} - T_{4}  \displaystyle\int -  \hat{C}_{3} \wedge \hat{B}_2  \, + \, \hat{C}_{5}  \ .
\end{equation}
One therefore encounters non-trivial WZ terms of the form $\, \hat{C}_{3} \wedge \hat{B}_2 \,$ and $\,\hat{C}_{5} \,$ (see (\ref{WZ_D4_1}) and (\ref{WZ_D4_2})) when pulling back the background (\ref{10D_G2-sector}) into the D4-brane worldvolume (\ref{vol_5}).

Inserting (\ref{WZ_D4_1}) and (\ref{WZ_D4_2}) into (\ref{SD4}), and computing the DBI term using (\ref{10D_G2-sector}), the effective action for the probe D$4$-brane takes the form
\begin{equation}
\label{SD4_2}
\begin{array}{lll}
S_{\textrm{D}_4} &=& - \, T_{4} \displaystyle\int  \hat{\textrm{vol}}_{5} \, \Delta^{-\frac{1}{3}}  \left[   e^{\frac{1}{4} \hat{\phi}}  \,  \sqrt{ \Delta^{\frac{2}{3}} +  e^{-\hat{\phi}} \, (g^2 \, B \,  \cos\alpha)^2 } \right.  \\[4mm]
&&   \left.   + \, \dfrac{L}{3}  \, e^{\frac{1}{2} \hat{\phi}}  \, \Delta^{-\frac{5}{6}}  \,  g \, \left( m \, g^3 \,  B^2 - 4 \, \chi \right) \, \cos\alpha  \right]   \ ,
\end{array}
\end{equation}
with $\,e^{\hat{\phi}}\,$ and the quantities $\,\Delta\,$ and $\,B\,$ given in (\ref{10D_G2-sector}) and (\ref{DeltaBC_func}).  The D$4$-brane wrapped over $\,\textrm{S}^2 \subset \mathbb{CP}_{2}\,$ then then experiences an effective potential density
\begin{equation}
\label{V_D4}
V(r)  =  T_{4} \,\,  e^{3 A(r)} \, g^{-2} \, \left(\dfrac{\sin\gamma}{2}\right)^2  \, \sin^2\alpha \,\,  \Theta(\alpha)  \ ,
\end{equation}
specified by the function
\begin{equation}
\label{Theta_D4}
\Theta(\alpha)  =  \Delta^{-\frac{3}{2}} \,  e^{\frac{1}{4} \hat{\phi}}  \left[ \sqrt{ \Delta^{\frac{2}{3}} + e^{-\hat{\phi}} \, (g^2 \, B \, \cos\alpha)^2 } +    \frac{L}{3}  \, e^{\frac{1}{4} \hat{\phi}}  \, \Delta^{-\frac{5}{6}}  \,  g  \, \left( m \,  g^3 \, B^2 - 4 \, \chi \right) \, \cos\alpha  \right] \ .
\end{equation}
This function depends on the coordinate $\,\alpha\,$, and therefore BJ instabilities could occur at certain values of this angle. However, an explicit evaluation of (\ref{Theta_D4}) shows that 
\begin{eqnarray}
\label{Theta_G2_D4}
{\cal N} = 1 \, , \, \mathrm{G}_2  & : &  \Theta (\alpha) = 2^{\frac{7}{2}} \cdot 15^{-\frac{7}{4}} \, c^{-\frac{1}{2}} \, \left( \, 4 \sqrt{15 + \cos^2\alpha}  +  \sqrt{15} \,  \cos\alpha \, \right)  \, > \, 0 \; ,   \\[5pt]
{\cal N} = 0 \, , \, \mathrm{G}_2  &  : &  \Theta (\alpha) = 2^{\frac{5}{2}} \cdot 3^{-\frac{7}{4}} \, c^{-\frac{1}{2}} \, \left( \, \sqrt{7 + \cos(2 \alpha)}   -  \cos\alpha   \, \right)  \, > \, 0  \; , 
\end{eqnarray}
namely, the net force is always attractive towards $\,r=0\,$. As a result, there is no BJ instability in this probe D$4$-brane channel. This is a remarkable fact as the $\,\hat{B}_{2}\,$ potential in (\ref{10D_G2-sector}) has an internal component on $\,\textrm{S}^{2}\,$, as can be seen from (\ref{SU(2)/SU(3)_correspondence}) and (\ref{J_S2}), that could have triggered such an instability. 

Together with the probe D$4$-brane wrapping $\,\textrm{S}^{2} \subset \mathbb{CP}_{2} \subset \textrm{S}^6\,$ above, there are other D$4$-branes that also require an analysis of BJ instabilities. The reason is that the WZ term in (\ref{SD4}) has non-vanishing components along $\,\mathcal{J}\,$ as required by the $\,\textrm{G}_{2}\,$ invariance of the backgrounds (\ref{10D_G2-sector}). Then, by virtue of (\ref{SU(2)/SU(3)_correspondence}) and (\ref{SU(2)_structure_forms}), there are seven independent such WZ components which stand a chance of inducing BJ instabilities on the probe D$4$-branes wrapping the respective two-cycles. The computation of the effective potential density for the six new D$4$-branes parallels the one performed above so we omit all the details here. The final outcome is summarised as follows:
\begin{itemize}
\item[$\circ$] D$4$-branes on two-cycles $\,\left\lbrace \, d\gamma \wedge \sigma_{1} \,\, , \,\, d\gamma \wedge \sigma_{2} \,\, , \,\, \sigma_{3} \wedge \sigma_{1} \,\, , \,\, \sigma_{3} \wedge \sigma_{2} \, \right\rbrace\,$ $\,\,\Rightarrow\,\,$  $\,\Theta(\alpha,\psi) > 0\ .$ 
\item[$\circ$] D$4$-branes on two-cycles $\,\left\lbrace \, \textrm{S}^2 \,\, (\textrm{see} (\ref{Theta_D4})) \,\, , \,\, d\gamma \wedge \sigma_{3} \, \right\rbrace\,$ $\,\,\Rightarrow\,\,$  $\,\Theta(\alpha) > 0\ .$
\item[$\circ$] D$4$-brane on two-cycle $\,d\alpha \wedge \boldsymbol{\eta}\,$ $\,\,\Rightarrow\,\,$  $\,\Theta = \textrm{constant} > 0\ .$
\end{itemize}
As a result, we find no BJ instabilities for any of the seven D$4$-branes when placed in the class of $\,\textrm{G}_{2}$-invariant backgrounds in (\ref{10D_G2-sector}).

\subsection{D$6$-brane wrapping internal four-cycles}
\label{Sec:D6_brane}

We consider now a probe D$6$-brane parallel to the AdS$_{4}$ boundary, \textit{i.e.} along $\,\mathbb{R}^{1,2}\,$, and wrapping $\,\mathbb{CP}_{2} \subset \textrm{S}^6\,$. The D$6$-brane coordinates are identified with $\,{\xi^{i}=x^{i}}\,$ ($\,i=0,1,2\,$) together with those on $\,\mathbb{CP}_{2}\subset \textrm{S}^6\,$.\footnote{The set of one-form basis elements on $\,\mathbb{CP}_{2}\,$ consists of $\,\{d\gamma \,, \,\sigma_{1} \,,\, \sigma_{2}\,, \,\sigma_{3} \}\,$ (see appendix~\ref{app:Geometric_structures}).} From (\ref{10D_G2-sector}) and (\ref{S6=S5xS1_metric}), the D$6$-brane worldvolume is given by
\begin{equation}
\label{vol_7}
\hat{\textrm{vol}}_{7} =  e^{3 A(r)} \, \Delta^{-\frac{5}{6}} \,  g^{-4} \, \sin^4 \alpha  \,\,  dx^{0} \wedge dx^{1} \wedge dx^{2} \wedge \textrm{vol}_{4}(\mathbb{CP}_{2}) \ ,
\end{equation}
and the action (\ref{SDp})-(\ref{unmagnetised_F}) reduces to
\begin{equation}
\label{SD6}
S_{\textrm{D}_6} = - T_{6} \, \displaystyle\int d^{7}\xi \,   e^{\frac{3}{4} \hat{\phi}}  \,  \sqrt{-\textrm{det}(\hat{g}_{ij} -  e^{-\frac{1}{2} \hat{\phi}} \,  \hat{B}_{ij})}  -  T_{6} \displaystyle\int  \, \tfrac{1}{2} \, \hat{C}_{3} \wedge \hat{B}_2{}^2  -  \hat{C}_{5} \wedge \hat{B}_2   +  \hat{C}_{7}   \ .
\end{equation}
Three non-trivial WZ terms (see (\ref{WZ_D6_1}) and (\ref{WZ_D6_2})) contribute now to the D$6$-brane action (\ref{SD6}) when pulling back the background (\ref{10D_G2-sector}) into the D$6$-brane worldvolume (\ref{vol_7}).

Using (\ref{10D_G2-sector}), a straightforward computation of the DBI and WZ terms yields an action for a probe D$6$-brane of the form
\begin{equation}
\label{SD6_2}
S_{\textrm{D}_6} =  - \, T_{6} \,  \displaystyle\int \hat{\textrm{vol}}_{7}  \, \Delta^{-\frac{2}{3}} \left[ e^{\frac{3}{4} \hat{\phi}} \, \left( \Delta^{\frac{2}{3}} +  e^{-\hat{\phi}} \,  B^2 \, g^4 \right)   +  m \, \frac{L}{3} \, e^{\frac{3}{2} \hat{\phi}} \,  \Delta^{-\frac{1}{6}} \, g^2  \,  B     \right] \ ,
\end{equation}
with $\,e^{\hat{\phi}}\,$, and $\,\Delta\,$ and $\,B\,$, given in (\ref{10D_G2-sector}) and (\ref{DeltaBC_func}). As a result, the effective potential density for the probe D$6$-brane simplifies into
\begin{equation}
\label{V_D6}
V(r) = T_{6} \,\,  e^{3A(r)} \, g^{-4} \,\,  \sin^4\alpha \,\, \Theta \ ,
\end{equation}
with
\begin{equation}
\label{Theta_D6}
\Theta  =  \Delta^{-\frac{3}{2}} \left[ \,  e^{\frac{3}{4} \hat{\phi}} \, \left( \Delta^{\frac{2}{3}} + e^{-\hat{\phi}}  \,  B^2 \, g^4 \right) +  m \, \dfrac{L}{3} \, e^{\frac{3}{2} \hat{\phi}} \,  \Delta^{-\frac{1}{6}} \, g^2  \,  B    \,  \right] \ .
\end{equation}
Note that $\,\Theta\,$ in (\ref{Theta_D6}) does not depend on the $\,\textrm{S}^{6}\,$ angles. A direct evaluation of this coefficient at the two $\textrm{G}_{2}$-invariant vacua gives
\begin{eqnarray}
\label{Theta_G2_D6}
{\cal N} = 1 \, , \, \mathrm{G}_2  & : &  \quad \Theta  = 2^{\frac{11}{6}}  \cdot 3^{-1}  \cdot 15^{-\frac{7}{4}} \, c^{-\frac{5}{6}} \, \left( \,  192 -  5 \, \sqrt{15}  \, \right)  \, > \, 0 \; ,   \\[5pt]
{\cal N} = 0 \, , \, \mathrm{G}_2  &  : &   \quad \Theta = 2^{\frac{4}{3}} \cdot 3^{-\frac{7}{4}} \, c^{-\frac{5}{6}} \, \left( \,  8 + \sqrt{2}  \, \right)  \, > \, 0  \; .
\end{eqnarray}
Therefore, a probe D$6$-brane wrapping $\,\mathbb{CP}_{2} \subset \textrm{S}^6\,$ does not suffer from a BJ instability despite the fact that the $\,\hat{F}_{4}\,$ flux in (\ref{10D_G2-sector_field_strengths}) has an internal component on $\,\mathbb{CP}_{2}\,$.

Finally, there are other D$6$-branes that require a careful analysis of BJ instabilities. Again, the reason is that the WZ term in (\ref{SD6}) has non-vanishing components along $\,\mathcal{J} \wedge \mathcal{J}\,$ due to the $\,\textrm{G}_{2}\,$ invariance of the backgrounds (\ref{10D_G2-sector}). Combining (\ref{SU(2)/SU(3)_correspondence}) and (\ref{SU(2)_structure_forms}), one finds seven independent such WZ components which can potentially induce BJ instabilities on the probe D$6$-branes wrapping the respective four-cycles. Omitting the details on the computation of the effective potential densities, the final outcome is summarised as follows:
\begin{itemize}
\item[$\circ$] D$6$-branes on four-cycles
\begin{equation}
\begin{array}{ll}
&  \left\lbrace \,
d\alpha \wedge \sigma_{2} \wedge \sigma_{3} \wedge  \boldsymbol{\eta}  
\,\, , \,\, 
d\alpha \wedge \sigma_{1} \wedge \sigma_{3} \wedge  \boldsymbol{\eta}  
\,\, , \,\, 
d\alpha \wedge d\gamma \wedge \sigma_{2} \wedge \boldsymbol{\eta}  
\,\, , \,\, 
d\alpha \wedge d\gamma \wedge \sigma_{1} \wedge \boldsymbol{\eta}  
\, \right\rbrace  \\[2mm]
& \Rightarrow \,\,  \Theta(\alpha,\psi) > 0 \ .
\end{array} \nonumber
\end{equation}
\item[$\circ$] D$6$-branes on four-cycles $\,\left\lbrace \, d\alpha \wedge \sigma_{1} \wedge \sigma_{2} \wedge  \boldsymbol{\eta}   \,\, , \,\, d\alpha \wedge d\gamma \wedge \sigma_{3} \wedge  \boldsymbol{\eta}   \, \right\rbrace\,$ $\,\,\Rightarrow\,\,$  $\,\Theta(\alpha) > 0\ .$
\item[$\circ$] D$6$-brane on the four-cycle $\,\mathbb{CP}_{2}\,$ $\,\,\Rightarrow\,\,$  $\,\Theta = \textrm{cst} > 0\,$ \,\,\,\, (\textrm{see} (\ref{Theta_D6}))  .
\end{itemize}
Therefore, we do not observe BJ instabilities for any of the seven D$6$-branes when placed in the class of $\,\textrm{G}_{2}$-invariant backgrounds in (\ref{10D_G2-sector}).

\subsection{D$8$-brane wrapping $\, \textrm{S}^6\,$}
\label{Sec:D8_brane}

Lastly we consider a probe D$8$-brane parallel to the AdS$_{4}$ boundary, \textit{i.e.} along $\,\mathbb{R}^{1,2}\,$, and wrapping the whole internal $\,\textrm{S}^6\,$. The D$8$-brane coordinates are identified with $\,{\xi^{i}=x^{i}}\,$ ($\,i=0,1,2\,$) together with those on $\,\textrm{S}^6\,$. From (\ref{10D_G2-sector}), the worldvolume of the D$8$-brane is
\begin{equation}
\label{vol_9}
\hat{\textrm{vol}}_{9} =  e^{3 A(r)} \, \Delta^{-\frac{1}{2}} \,  g^{-6} \,\,  dx^{0} \wedge dx^{1} \wedge dx^{2} \wedge \textrm{vol}_{6}(\textrm{S}^{6}) \ ,
\end{equation}
and the action (\ref{SDp})-(\ref{unmagnetised_F}) becomes
\begin{equation}
\label{SD8}
\begin{array}{lll}
S_{\textrm{D}_8} &=& - \, T_{8} \, \displaystyle\int d^{9}\xi \,   e^{\frac{5}{4} \hat{\phi}}  \,  \sqrt{-\textrm{det}(\hat{g}_{ij} - e^{-\frac{1}{2} \hat{\phi}} \,  \hat{B}_{ij})}  \\[2mm]
&  & - \, T_{8} \displaystyle\int  -  \tfrac{1}{6} \, \hat{C}_{3} \wedge \hat{B}_2{}^3  +  \tfrac{1}{2} \, \hat{C}_{5} \wedge \hat{B}_2{}^{2}   -  \hat{C}_{7} \wedge \hat{B}_2   +  \hat{C}_9   \ .
\end{array}
\end{equation}
The low co-dimensionality of the D$8$-brane translates into the appearance of four non-trivial WZ terms in the action (\ref{SD8}) (see (\ref{WZ_D8_1}) and (\ref{WZ_D8_2})) when pulling back the background (\ref{10D_G2-sector}) into the D$8$-brane worldvolume (\ref{vol_9}).

The computation of the WZ terms, together with the gravitational DBI term using (\ref{10D_G2-sector}), proceed uneventfully and yields a probe D$8$-brane action
\begin{equation}
\label{SD8_2}
S_{\textrm{D}_8} =  - T_{8} \,  \displaystyle\int \hat{\textrm{vol}}_{9}  \, \Delta^{-1} \left[ \,  e^{\frac{5}{4} \hat{\phi}} \, \left( \Delta^{\frac{2}{3}} +  e^{-\hat{\phi}} \,  B^2 \, g^4 \right)^{\frac{3}{2}}  
+ \frac{L}{3} \, m \,  e^{\frac{5}{2} \hat{\phi}} \, \Delta^{\frac{1}{2}} \,  \right]  \ ,
\end{equation}
with $\,e^{\hat{\phi}}\,$ and the quantities $\,\Delta\,$ and $\,B\,$ given in (\ref{10D_G2-sector}) and (\ref{DeltaBC_func}). The effective potential density for the probe D$8$-brane then takes the form
\begin{equation}
\label{V_D8}
V(r) = T_{8} \,\,  e^{3A(r)} \, g^{-6} \,\, \Theta \ ,
\end{equation}
with
\begin{equation}
\label{Theta_D8}
\Theta  =  \Delta^{-\frac{3}{2}} \left[ \,    e^{\frac{5}{4} \hat{\phi}} \, \left( \Delta^{\frac{2}{3}} +   e^{-\hat{\phi}} \,  B^2 \, g^4 \right)^{\frac{3}{2}} + \dfrac{L}{3} \, m \,  e^{\frac{5}{2} \hat{\phi}} \, \Delta^{\frac{1}{2}}  \, \right] \ .
\end{equation}
Once again $\,\Theta\,$ turns out to be independent of the $\,\textrm{S}^{6}\,$ angles. Finally, a direct evaluation of the coefficient (\ref{Theta_D8}) shows that
\begin{eqnarray}
\label{Theta_G2_D8}
{\cal N} = 1 \, , \, \mathrm{G}_2  & : &  \quad \Theta  = 2^{\frac{1}{6}}  \cdot 15^{-\frac{7}{4}} \, c^{-\frac{7}{6}} \, \left( \,  256 + 25 \, \sqrt{15} \, \right)  \, > \, 0 \; ,   \\[5pt]
{\cal N} = 0 \, , \, \mathrm{G}_2  &  : &   \quad \Theta = 2^{\frac{7}{6}} \cdot 3^{-\frac{7}{4}} \, c^{-\frac{7}{6}} \, \left( \,  3 + 8 \, \sqrt{2}   \, \right)  \, > \, 0  \; .
\end{eqnarray}
Therefore, as in all the previous cases, a probe D$8$-brane wrapping the internal $\,\textrm{S}^6\,$ is also free from a BJ instability.

\section{(Meta)stability and decay through domain-walls}
\label{sec:domain-walls}

In previous sections we have established the absence of BJ instabilities for spacetime-filling D2-branes in all non-supersymmetric, yet perturbatively BF stable, vacua listed in table \ref{Table:KnownBFStableCritical points}, and of more general BJ instabilities for the vacuum with $\,\text{G}_2\,$ symmetry. However, generic swampland arguments \cite{ArkaniHamed:2006dz,Ooguri:2016pdq} suggest that non-supersymmetric $\,\text{AdS}_4\,$ vacua should present some type of instability. Thus, we dedicate this section to the exploration of an alternative decay channel for the non-supersymmetric vacuum with $\,\text{G}_2\,$ symmetry in table~\ref{Table:KnownBFStableCritical points}. More concretely, we look at its potential quantum tunnelling into a different vacuum with strictly lower potential energy.

An unstable vacuum in a bulk theory of gravity can decay to a true, stable vacuum by bubble nucleation under certain conditions \cite{Coleman:1980aw}. The nucleation is quantum in nature and has a decay rate (per unit volume per unit time) given in the semi-classical approximation by the expression
\be\label{eq.probability}
\frac{\Gamma}{V} = A\, e^{-(S_\text{DW} - S_\text{false})/\hbar} \ ,
\ee
where $\,S_\text{false}\,$ is the Euclidean on-shell action evaluated at the non-supersymmetric $\,\text{AdS}_4\,$ vacuum, and $\,S_\text{DW}\,$ refers to the same quantity evaluated at the domain-wall (DW) solution that interpolates between the non-supersymmetric and supersymmetric $\,\text{AdS}_4\,$ vacua. 
The computation of the coefficient $\,A\,$ in (\ref{eq.probability}) involves the evaluation of a functional determinant that depends on the details of the model \cite{Callan:1977pt}.
The expression \eqref{eq.probability} for the nucleation decay rate assumes the existence of a DW solution, with a (in this case Lorentzian) 4D metric
\be\label{eq.DWansatz}
\d s_4^2 = e^{2A(r)} \, \eta_{\alpha\beta} \, \d x^\alpha \, \d x^\beta + \d r^2 \qquad\quad (\alpha=0,1,2) \ ,
\ee
with $\,A(r)\sim r/L_\pm\,$  in the limit $\,r\to \pm \infty\,$, where $\,L_+\,$ ($\,L_-\,$) is the radius of the unstable (stable)  $\,\text{AdS}_4\,$ spacetime in the Poincar\'e patch, as in \eqref{4D_metric_SO(3)_R}. 
We focus on DWs that asymptote to the $\,\mathcal{N}=0\,$, $\,\text{G}_2\,$ vacuum at $\,r\to+\infty\,$ (ultraviolet, UV). Then, due to the condition $\,V_-<V_+\,$ (with $\,V_\pm\,$ being the gravitational potential evaluated at the scalars vevs that determine the $\,\text{AdS}_4\,$ vacua with radius $\,L_\pm\,$), only a limited number of $\,\text{AdS}_4\,$ vacua can be asymptotically approached when $\,r\to-\infty\,$ (infrared, IR), in agreement with the holographic c-theorem of \cite{Freedman:1999gp} and the F-theorem of \cite{Jafferis:2011zi}. Note that the $\,\cN=3\,$ vacuum of table \ref{Table:KnownBFStableCritical points} is degenerate in energy with the $\,\cN=0\,$, $\,\text{G}_2\,$ vacuum, and it is thus excluded from our analysis.

In this section we investigate DW solutions that asymptote to (some of) the $\,\textrm{AdS}_{4}\,$ vacua of table~\ref{Table:KnownBFStableCritical points}, in the IR, with a value of the potential $\,V\,$ lower than the corresponding value at the $\,\mathcal{N}=0\,$, $\,\text{G}_2\,$ vacuum. To keep the discussion simple, we will focus on the $\,\text{SU}(3)\,$ invariant sector presented in \cite{Guarino:2015qaa}, whose action we bring to \eqref{eq.SU3invariantaction} below. Our aim here is to qualitatively describe the procedure without entering technical details, since these have been extensively explained elsewhere. The interested reader can find detailed accounts in the early work \cite{Freedman:1999gp}, in \cite{Guarino:2016ynd} for numerical integrations in the $\,\text{SU}(3)\,$ invariant model \eqref{eq.SU3invariantaction}, or in \cite{Guarino:2019snw} for calculations in other sectors of the $\,\text{ISO}(7)\,$ maximal supergravity. We have not found a DW configuration connecting the $\,\cN=0\,$, $\,\text{G}_2\,$ vacuum in the ultraviolet (UV) to any of the $\,\text{SU}(3)\,$ symmetric vacua of table~\ref{Table:KnownBFStableCritical points}, either with $\,\mathcal{N}=0\,$ or $\,\mathcal{N}=1\,$ supersymmetry, in the IR. However, it is worth emphasising that we have restricted our search to the $\,\text{SU}(3)\,$ invariant sector of the theory.

\subsection{$\textrm{SU}(3)$ invariant sector: bosonic action}

The bosonic sector of the $\,\text{SU}(3)\,$ invariant model in \cite{Guarino:2015qaa} consists of four real scalars and the metric field. The dynamics is dictated by the Einstein-scalar action
\be\label{eq.SU3invariantaction}
S = \frac{1}{2\kappa^2} \int \d^4x \,  \sqrt{-g} \left( R - \frac{3}{2} \left( (\partial \varphi)^2 + e^{2\varphi} (\partial \chi)^2 \right) - 2 \left( (\partial \phi)^2 + e^{2\phi} (\partial \rho)^2 \right)  - V \right) \ ,
\ee
which includes a scalar potential
\be\bal
V & = \frac{e^{4\phi+3\varphi}}{2} \Big[ m - g\, \chi\, \left(\chi^2 + 6\rho^2 \right)   \Big]^2 - 6\, g^2 e^{\varphi} \Bigg[ \left( \frac{1}{2} - e^{2\phi} \rho^2 \right)^2 + \left( 1-\frac{1}{2}  e^{2\phi} \chi^2 \right)^2  \\
& \qquad\qquad\qquad  + e^{4\phi} \rho^2 \chi^2 + \frac{\chi^2}{4} \left( e^{4\phi-2\varphi} +12 e^{2\phi + 2 \varphi} \rho^2 \right) - \frac{13}{4}  + \frac{1}{2} e^{4\phi-4\varphi} - e^{2\phi-2\varphi} \Bigg]\ .
\eal\ee
Here $\,g\,$ and $\,m\,$ are the electric and magnetic couplings of $\,\text{ISO}(7)\,$ maximal supergravity, being related to the inverse $\,\textrm{S}^6\,$ radius and the Romans mass parameter, respectively. We will set $\,g=m=1\,$ in this section without loss of generality.
To compare with the parameterisation in table~\ref{Table:ScalarVevs} of appendix~\ref{app:SO3_R_sector}, one must identify $\,\rho=2\,\zeta\,$ in \eqref{eq:SU3locus}.

Contrary to the case in \cite{Guarino:2015qaa}, where supersymmetric DW solutions of this model were investigated, we are interested here in solutions involving the $\,\cN=0\,$, $\,\text{G}_2\,$ vacuum. This requires that we consider the second order Euler--Lagrange equations of motion for the scalars and the metric. This gives rise to a coupled set of five second order differential equations: one for each scalar field and one for the DW function $\,A(r)\,$ in the ansatz~\eqref{eq.DWansatz}. This system is supplemented with a first order differential equation, given by the Hamiltonian constraint of the gravitational system. This implies that deviations around any AdS$_{4}$ vacuum will be parameterised by nine constants of integration, although one of them can always be reabsorbed in a shift of the radial coordinate $\,r\,$.

\subsection{Numerical analysis}

DW solutions that asymptote to the $\,\cN=0\,$, $\,\text{G}_2\,$ vacuum in the UV ($\,r\to+\infty\,$) are described by deviations around such a solution (see  table~\ref{Table:ScalarVevs}) that involve four different powers of the radius:
\be
e^{-\frac{3-\sqrt{33}}{2} \frac{r}{L}} \ , \quad e^{-\frac{3-\sqrt{5}}{2} \frac{r}{L}} \ , \quad e^{-\frac{3+\sqrt{5}}{2} \frac{r}{L}} \ , \quad e^{-\frac{3+\sqrt{33}}{2} \frac{r}{L}} \ ,
\ee
each one being parameterised by two constants of integration, and with $\,L=3^{3/4}2^{-13/6}\,$ (recall that we are setting $\,g=m=1\,$).
Regularity in the UV immediately forbids the $\,e^{-\frac{3-\sqrt{33}}{2} \frac{r}{L}} \,$ radial dependence, leading to a problem with six parameters. In a way, this parameter space can be thought of as the phase space that will give rise to the coefficient $\,A\,$ in \eqref{eq.probability} upon evaluation of a functional determinant \cite{Callan:1977pt}. In order to find DW solutions in this parameter space one has to resort to numerical methods, which turn out to be plagued with technical difficulties, rendering this a daunting task. However, it is common lore that the numeric integration of the equations of motion is simpler if one constructs the DW starting from the IR end-point instead, which we will do in the following.

\subsubsection*{Flowing to the $\,\cN=0\,$, $\,\text{SU}(3)\,$ vacua in the IR}

To present the logic in a simpler manner, we will focus momentarily on one specific candidate for the IR asymptotics of the DW: the $\,\cN=0\,$, $\,\text{SU}(3)\,$ vacuum of table~\ref{Table:KnownBFStableCritical points} with scalar potential $\,V=-23.413628\,$.\footnote{Although one would not expect this $\,\text{AdS}_4\,$ vacuum to be stable, for the same token as the $\,\mathcal{N}=0\,$, $\,\text{G}_2\,$ one, this still provides a potential decay channel for the latter. If such a channel exists then one still has to consider the decay of the $\,\mathcal{N}=0\,$, $\,\text{SU}(3)\,$ vacuum, which cannot occur via D$2$-brane-jet instabilities, as showed in section~\ref{sec:D2SU3Text}.} The radial deviation from this vacuum in the IR ($\,r \to -\infty\,$) is codified into the exponentials
\be
\label{exponentials_N0&SU3}
e^{1.409 \frac{r}{L}} \ , \quad e^{1.359 \frac{r}{L}} \ , \quad e^{0.345 \frac{r}{L}} \ , \quad e^{-0.517 \frac{r}{L}} \ ,\quad
e^{-2.483 \frac{r}{L}} \ , \quad e^{-3.345 \frac{r}{L}} \ , \quad e^{-4.359 \frac{r}{L}} \ , \quad e^{-4.409 \frac{r}{L}} \ ,
\ee
each one being parameterised by a single constant of integration, and with $\,L=0.50576\,$. Regularity in the IR selects the first three exponentials in (\ref{exponentials_N0&SU3}) so that a regular expansion around the AdS$_{4}$ vacuum determines a three-dimensional parameter space
\be\bal
\chi & = 0.45466 \left( 1 - \alpha_1 \, e^{1.409 \frac{r}{L}}  + \alpha_2 \,e^{1.359 \frac{r}{L}} + \alpha_3 \,e^{0.345 \frac{r}{L}}  + \cdots \right) \ , \\
\rho & = 0.33481 \left( 1 - 0.6655 \,\alpha_1 \,e^{1.409 \frac{r}{L}}  + 0.7297\, \alpha_2 \,e^{1.359 \frac{r}{L}} - 4.3692 \,\alpha_3\, e^{0.345 \frac{r}{L}}  + \cdots \right) \ , \\
e^\phi & = 1.66344 \left( 1 - 0.6390\, \alpha_1 \,e^{1.409 \frac{r}{L}}  -0.4399 \,\alpha_2 \,e^{1.359 \frac{r}{L}} + 1.9454 \,\alpha_3 \,e^{0.345 \frac{r}{L}}  + \cdots \right) \ , \\
e^\varphi & = 1.19382 \left( 1 - 0.2044\, \alpha_1 \,e^{1.409 \frac{r}{L}}  + 0.5903 \, \alpha_2 \,e^{1.359 \frac{r}{L}} -3.6678\, \alpha_3\, e^{0.345 \frac{r}{L}}  + \cdots \right) \ .
\eal\ee

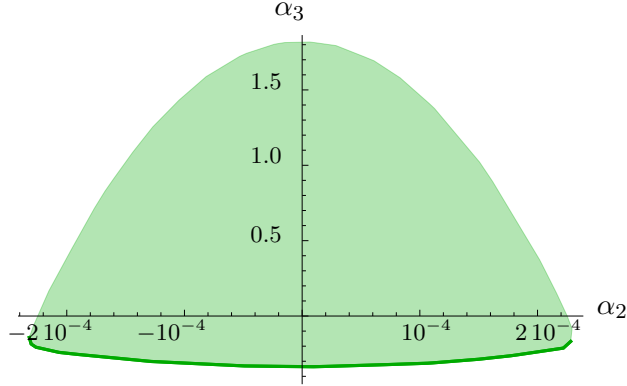
\begin{figure}[t]
\begin{center}
\includegraphics[width=0.5\textwidth]{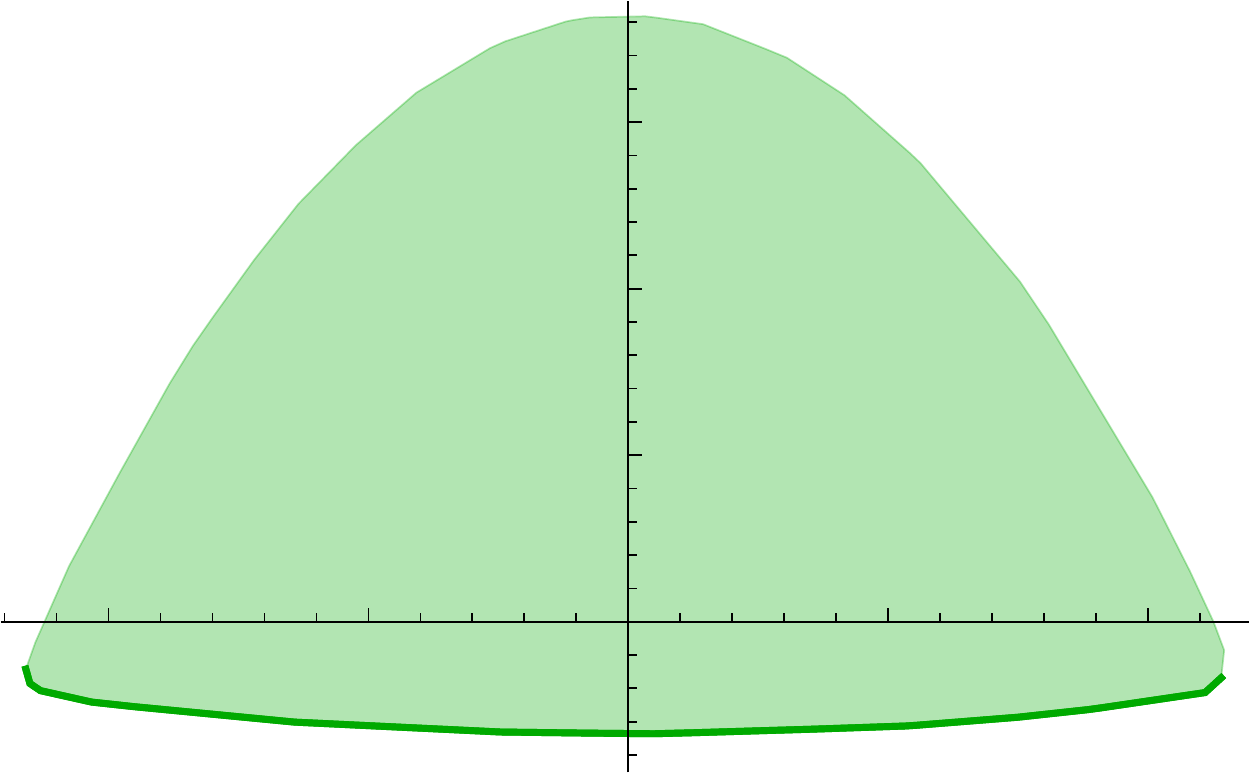}
\put(-126,54){\footnotesize$0.5$}
\put(-126,84){\footnotesize$1.0$}
\put(-126,112){\footnotesize$1.5$}
\put(-69,17){\footnotesize$10^{-4}$}
\put(-26,17){\footnotesize$2\,10^{-4}$}
\put(-169,17){\footnotesize$-10^{-4}$}
\put(-218,17){\footnotesize$-2\,10^{-4}$}
\put(5,27){{$\alpha_2$}}
\put(-117,140){{$\alpha_3$}}
\caption{Region within the two-dimensional parameter space $\,(\alpha_{2},\alpha_{3})\,$ describing regular deviations around the $\,\cN=0\,$, $\,\text{SU}(3)\,$ vacuum of table~\ref{Table:KnownBFStableCritical points} with potential $\,V=-23.413628\,$ in the IR.}
\label{fig.spacecraft}
\end{center}
\end{figure}

An exploratory analysis of solutions shows that the constant of integration  $\,\alpha_1\,$ must always be non-zero for non-trivial solutions to exist. Then, by virtue of a shift of the radial coordinate, it can be set to any convenient value, such that its magnitude does not carry physical significance. This allows us to set $\,\alpha_1=1\,$ and leaves us with a two-dimensional parameter space spanned by $\,(\alpha_{2},\alpha_{3})\,$.
At the origin of this two-dimensional parameter space, namely $\,(\alpha_{2},\alpha_{3})=(0,0)\,$, the solution at the UV does not approach an $\,\text{AdS}_4\,$ vacuum. Instead, the scalars and DW function $\,A(r)\,$ approach asymptotically a scaling behavior dictated by the presence of D2-branes in the setup (see e.g.~\cite{Guarino:2016ynd}), which are the objects that dominate generically the stress-energy tensor near the boundary. The presence of the Romans mass and the condition that the DW ends on an $\,\text{AdS}_4\,$ geometry in the IR imply that this scaling behavior is never a full solution to the equations of motion, but just an asymptotic solution. By exploring the space of parameters $\,(\alpha_{2},\alpha_{3})\,$ we find that there is a closed region where the UV asymptotics of the solutions are governed by the presence of D2-branes, denoted by the shaded area in figure~\ref{fig.spacecraft}.

For values of the parameters $\,(\alpha_{2},\alpha_{3})\,$ outside the shaded area in figure~\ref{fig.spacecraft} we find that at least one scalar diverges at a finite value of the radius, rendering these integrations unphysical. Most importantly, in the highlighted parts of the border between DWs that approach the D2-brane behaviour in the UV and unphysical ones (the thick green line at the bottom of figure~\ref{fig.spacecraft}), new phenomenology appears. For this uni-parametric family of solutions the D2-branes are not the only dominant sources in the stress-energy tensor near the UV. Instead, all the branes present in the setup contribute and change the dynamics accordingly. Geometrically, the metric approaches an $\,\text{AdS}_4\,$ solution in the UV, with all scalars approaching constant values. By examining the radius of the geometry, and the magnitude of the scalar vevs, we conclude that the asymptotic UV  solution corresponds to the $\,\cN=0\,$, $\,\text{SO}(7)\,$ vacuum with $V=-19.614907$  of \cite{DallAgata:2011aa}. For values of the parameters close to, but not exactly, the ones in the one-parameter family of solutions that asymptote to an $\,\text{AdS}_4\,$ metric, the DWs asymptote to the D2-branes geometry, as generically in the shaded area of the figure. However there is a range of the radial coordinate $\,r\,$ where the scalar fields and DW function $\,A(r)\,$ are very well approximated by their values at the $\,\cN=0\,$, $\,\text{SO}(7)\,$ vacuum, with very small gradients. The closer one gets to the critical values in parameter space, the longer (in radial coordinate) the $\,\cN=0\,$, $\,\text{SO}(7)\,$ vacuum is realised before the DW continues towards the D2-brane behaviour.

Importantly, within the precision of our numerics, we have not found DW solutions other than the ones exposed above and summarised in figure~\ref{fig.spacecraft}. This suggests that, within the $\,\text{SU}(3)\,$ invariant sector of the $\,\text{ISO}(7)\,$ supergravity, a DW connecting the $\,\cN=0\,$, $\,\text{G}_2\,$ vacuum in the UV with the $\,\cN=0\,$, $\,\text{SU}(3)\,$ vacuum in the IR does not exist. Thus the probability of quantum tunnelling between the two vacua must vanish. Qualitatively, the same results hold if we perform our numerical integration starting from the second $\,\cN=0\,$, $\,\text{SU}(3)\,$ vacuum of table~\ref{Table:KnownBFStableCritical points} with $\,V = -23.456779\,$.

\subsubsection*{Flowing to the $\,\cN=1\,$, $\,\text{SU}(3)\,$ vacuum in the IR}

We move to explore now the more interesting case of vacuum decay into the $\,\cN=1\,$, $\,\text{SU}(3)\,$ vacuum. We conduct a similar analysis to the one described above. In this case, deviations from the $\,\text{AdS}_4\,$ vacuum in the IR are characterised by radial dependences of the form
\be
\label{exponentials_N1&SU3}
e^{-(1-\sqrt{6}) \frac{r}{L}} \ , \quad e^{-(2-\sqrt{6}) \frac{r}{L}} \ , \quad e^{-(1+\sqrt{6}) \frac{r}{L}} \ , \quad e^{-(2+\sqrt{6}) \frac{r}{L}} \ ,
\ee
each one being parameterised by two constants of integration, and with $\,L=5^{5/4}3^{-1/4}2^{-9/4}\,$. Regularity of the flows in the IR selects the first two radial dependences in (\ref{exponentials_N1&SU3}), so that four parameters are needed to describe the $\,r\to-\infty\,$ behaviour
\be\bal
\chi & = \frac{1}{4} \left( 1 + \left( \frac{3(4+3\sqrt{6})}{4} \beta_2 + \frac{5(4-\sqrt{6}}{4} \beta_3  \right) e^{-(1-\sqrt{6}) \frac{r}{L}}  + \beta_1 \, e^{-(2-\sqrt{6}) \frac{r}{L}}  + \cdots \right) \ , \\
\rho & = \frac{-\sqrt{3}}{4} \left( 1+  \beta_2\, e^{-(1-\sqrt{6}) \frac{r}{L}}  + \left( -\frac{3(4+3\sqrt{6})}{16}\beta_1 -\frac{5(4+\sqrt{6}}{16} \beta_4  \right)  e^{-(2-\sqrt{6}) \frac{r}{L}}   + \cdots \right) \ , \\
e^\phi & = \frac{4}{\sqrt{5}} \left( 1 -  \beta_3\, e^{-(1-\sqrt{6}) \frac{r}{L}}  +  \left( \frac{4-\sqrt{6}}{16}\beta_1 + \frac{3-3\sqrt{6}-4)}{16}\beta_4  \right) e^{-(2-\sqrt{6}) \frac{r}{L}}    + \cdots \right) \ , \\
e^\varphi & = \frac{4}{\sqrt{15}} \left( 1  - \left(  \frac{4+\sqrt{6}}{4}\beta_2 + \frac{3\sqrt{6}-4}{4}\beta_3  \right)  e^{-(1-\sqrt{6}) \frac{r}{L}} - \beta_4\, e^{-(2-\sqrt{6}) \frac{r}{L}}   + \cdots \right) \ .
\eal\ee

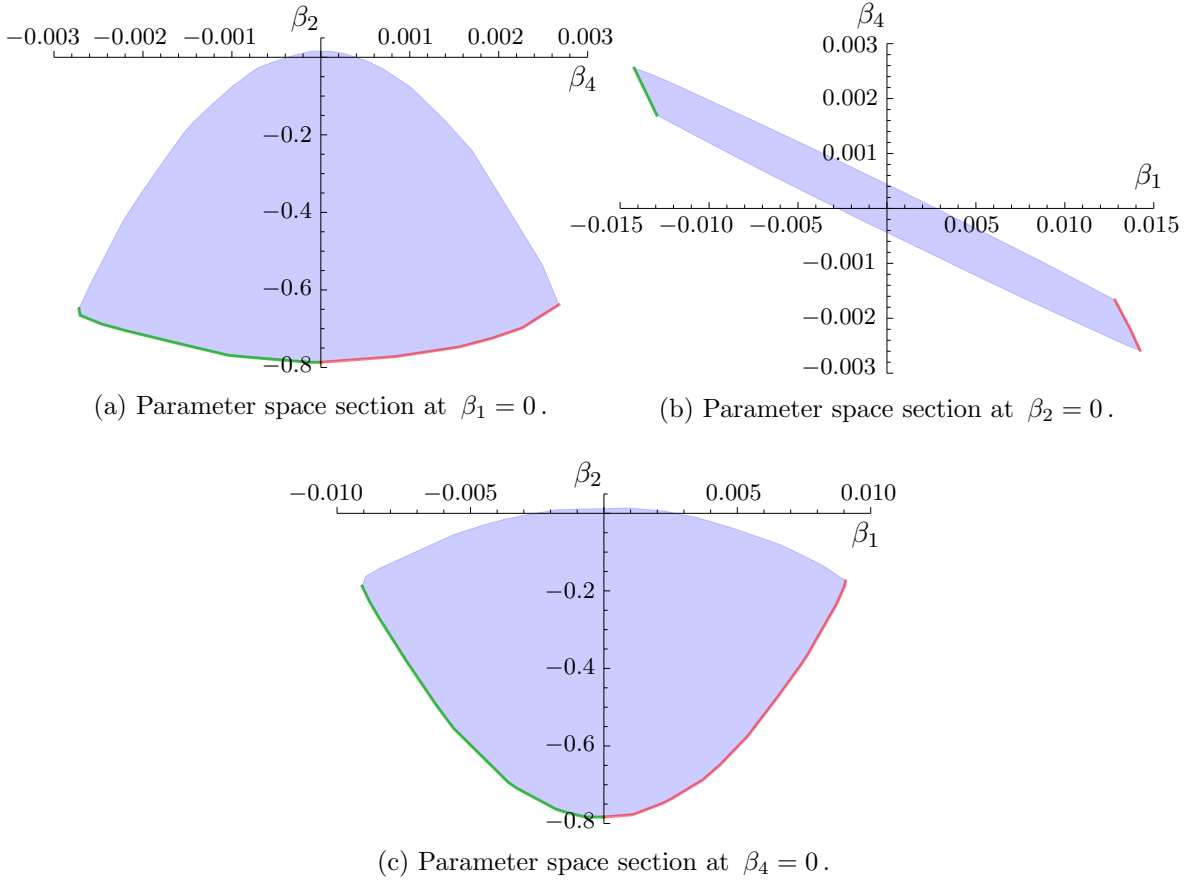
\begin{figure}[t]
\begin{center}
\begin{subfigure}{.5\textwidth}
  \centering
  \includegraphics[width=0.95\textwidth]{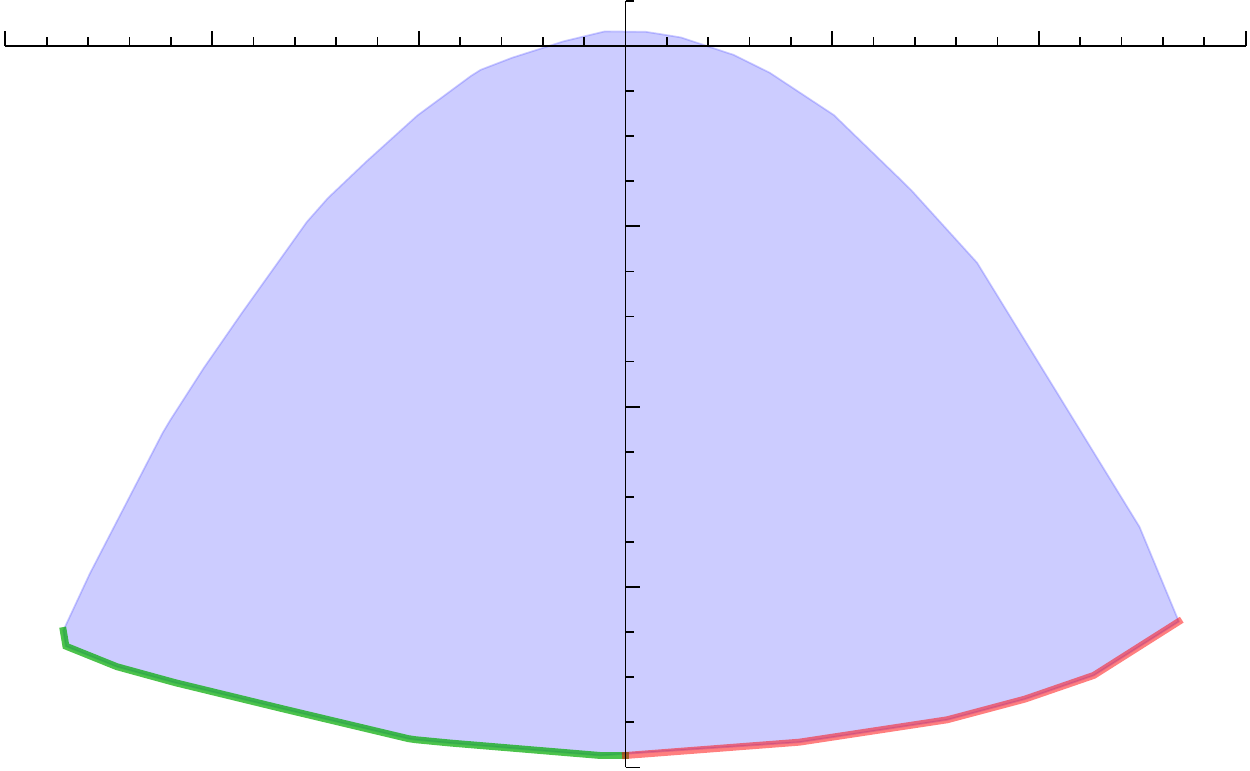}
  \put(-8,107){{$\beta_4$}}
  \put(-113,130){{$\beta_2$}}
  \put(-124,86){\footnotesize$-0.2$}
  \put(-124,57){\footnotesize$-0.4$}
  \put(-124,28){\footnotesize$-0.6$}
  \put(-124,-1){\footnotesize$-0.8$}
  \put(-154,123){\footnotesize{$-0.001$}}
  \put(-188,123){\footnotesize{$-0.002$}}
  \put(-221,123){\footnotesize{$-0.003$}}
  \put(-80,123){\footnotesize{$0.001$}}
  \put(-46,123){\footnotesize{$0.002$}}
  \put(-12,123){\footnotesize{$0.003$}}
  \caption{Parameter space section at $\,\beta_1=0\,$.}
\end{subfigure}%
\begin{subfigure}{.5\textwidth}
  \centering
  \includegraphics[width=0.95\textwidth]{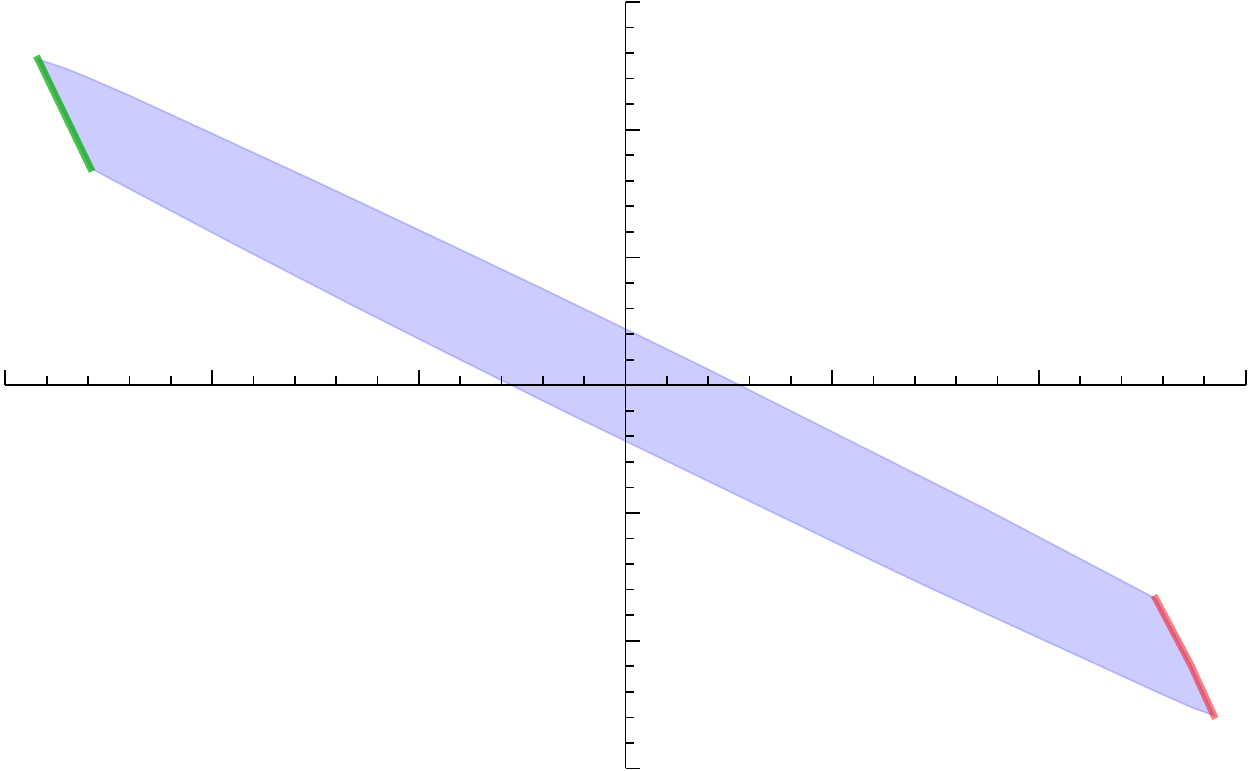}
  \put(-8,72){{$\beta_1$}}
  \put(-113,133){{$\beta_4$}}
  \put(-126,122){\footnotesize$0.003$}
  \put(-126,102){\footnotesize$0.002$}
  \put(-126,81){\footnotesize$0.001$}
  \put(-134,40){\footnotesize$-0.001$}
  \put(-134,19){\footnotesize$-0.002$}
  \put(-134,0){\footnotesize$-0.003$}
  \put(-154,54){\footnotesize{$-0.005$}}
  \put(-188,54){\footnotesize{$-0.010$}}
  \put(-222,54){\footnotesize{$-0.015$}}
  \put(-80,54){\footnotesize{$0.005$}}
  \put(-45,54){\footnotesize{$0.010$}}
  \put(-12,54){\footnotesize{$0.015$}}
  \caption{Parameter space section at $\,\beta_2=0\,$.}
\end{subfigure}
\par\bigskip
\begin{subfigure}{.5\textwidth}
  \centering
  \includegraphics[width=0.95\textwidth]{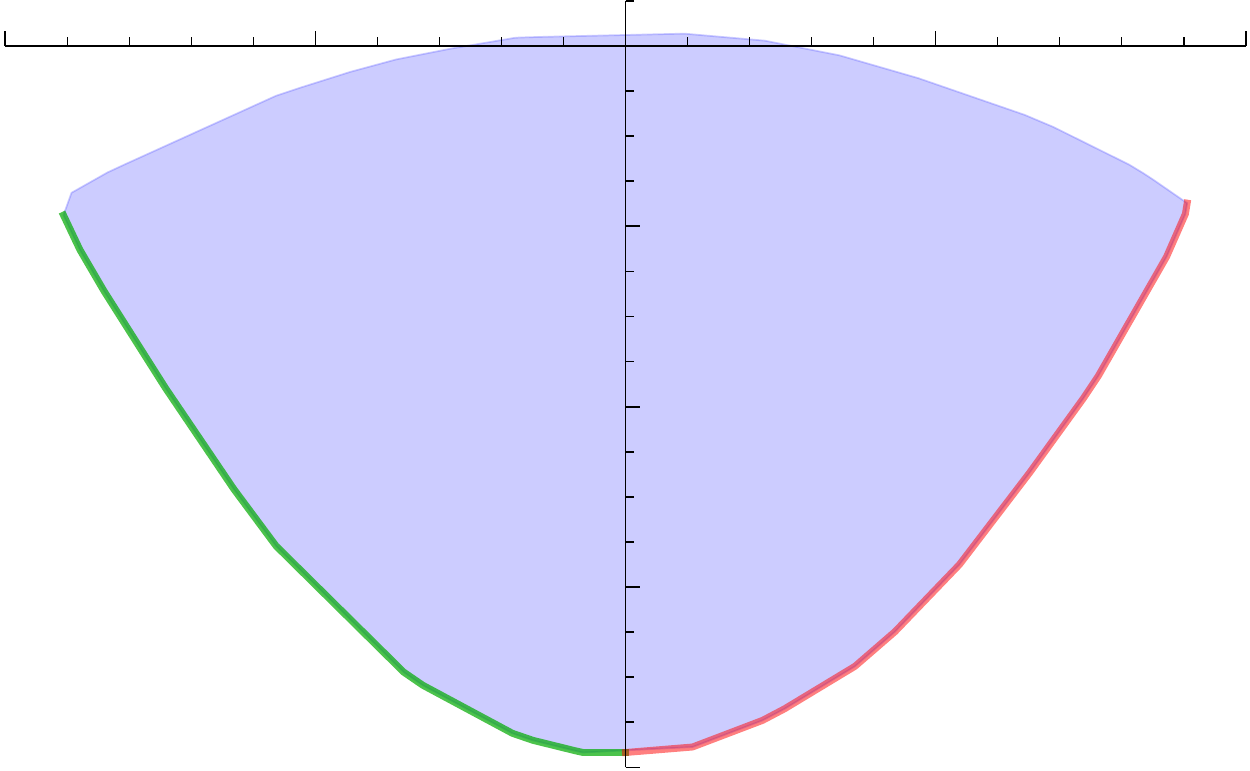}
  \put(-8,107){{$\beta_1$}}
  \put(-113,130){{$\beta_2$}}
  \put(-124,86){\footnotesize$-0.2$}
  \put(-124,57){\footnotesize$-0.4$}
  \put(-124,28){\footnotesize$-0.6$}
  \put(-124,-1){\footnotesize$-0.8$}
  \put(-170,123){\footnotesize{$-0.005$}}
  \put(-221,123){\footnotesize{$-0.010$}}
  \put(-63,123){\footnotesize{$0.005$}}
  \put(-12,123){\footnotesize{$0.010$}}
  \caption{Parameter space section at $\,\beta_4=0\,$.}
\end{subfigure}
\caption{Three slices of the three-dimensional parameter space $\,(\beta_1,\beta_2,\beta_4)\,$ describing regular deviations around the $\,\cN = 1\,$, $\,\text{SU}(3)\,$ vacuum of table~\ref{Table:KnownBFStableCritical points} in the IR.}
\label{fig.spacecraft3}
\end{center}
\end{figure}

An exploratory analysis shows this time that the parameter $\,\beta_3\,$ must be turned on for a regular DW to exist in the UV, and we can set $\,\beta_3=1\,$ without loss of generality. This leaves us with a large three-dimensional parameter space spanned by $\,(\beta_1,\beta_2,\beta_4)\,$ to scan, and we have done it by repeating the procedure explained above for several slices along one of the parameters. We present three such slices in figure~\ref{fig.spacecraft3}.

We observe a connected region in the parameter space where the DW generically approaches the behaviour dictated by dominating D2-branes in the UV. Outside this region there are diverging scalars that render the solutions unphysical. At the border between these two regions there exist lower-dimensional families of DWs that approach an $\,\text{AdS}_4\,$ vacuum in the UV, but none of them correspond with the $\,\cN=0\,$, $\,\text{G}_2\,$ vacuum we are after. Instead, we only find DWs that asymptote to the $\,\mathcal{N}=1\,$, $\,\text{G}_2\,$ vacuum in the UV of the type discussed in \cite{Guarino:2016ynd}. The way these solutions are approached falls into two categories, denoted by two different highlighted areas in figure~\ref{fig.spacecraft3}. 
\begin{itemize}
\item[$\circ$] Close to the critical region with $\,\text{AdS}_4\,$ UV asymptotics denoted by the green line in figure~\ref{fig.spacecraft3}, the DW solutions approach the D2-brane behaviour by first spending an arbitrarily large (the closer one gets to the critical values) range of radial coordinate near the $\,\cN=1\,$, $\,\text{G}_2\,$ vacuum.

\item[$\circ$]  For DW solutions close to the critical region with $\,\text{AdS}_4\,$ UV asymptotics denoted by the red line in figure~\ref{fig.spacecraft3}, the solutions first spend an arbitrary large range of radial coordinate near the $\,\cN=1\,$, $\,\text{G}_2\,$ vacuum and then, closer to the UV, the radial profile spends an arbitrary large range of radial coordinate near the $\,\cN=0\,$, $\,\text{SO}(7)\,$ vacuum. After that, the DWs approach the D2-brane behaviour in the far UV.
\end{itemize}

Therefore, due to the collected evidence against the existence of DWs connecting the $\,\cN=0\,$, $\,\text{G}_2\,$ vacuum in the UV with other $\,\text{AdS}_4\,$ vacua in the IR, we conclude that the probability of decaying by quantum tunnelling for the $\,\cN=0\,$, $\,\text{G}_2\,$ vacuum must be highly suppressed, at least withing the $\,\textrm{SU}(3)\,$ invariant sector of the $\,\textrm{ISO}(7)\,$ maximal supergravity.

\section{Discussion} 
\label{sec:Discussion}

We have determined the non-supersymmetric, but BF-stable within $D=4$, AdS vacua of maximal supergravity in four dimensions with a dyonic ISO(7) gauging \cite{Guarino:2015qaa} that lie within the seven $\cN=1$ chiral subsector \cite{Guarino:2019snw} of the supergravity. There are seven such solutions, summarised in table \ref{Table:KnownBFStableCritical points} of the introduction along the six known supersymmetric ones. By the results of \cite{Guarino:2015jca,Guarino:2015vca}, all these vacua give rise to $\,\textrm{AdS}_{4} \times \textrm{S}^{6}\,$ backgrounds of massive type IIA string theory. Then, we have tested these solutions for BJ-instabilities against spacetime-filling D2-branes, following the analogue spacetime-filling M2-brane analysis of \cite{Bena:2020xxb,Corrado:2001nv}. All the type IIA solutions test negative for this type of BJ-instabilities. This is to be contrasted to the $\,\textrm{AdS}_{4} \times \textrm{S}^{7}\,$ M-theory case, where the only non-supersymmetric, but BF-stable within the $D=4$ supergravity, solution \cite{Warner:1983du,Godazgar:2014eza} is BJ-unstable \cite{Bena:2020xxb}. We have also performed a thorough stability test of the simplest of these non-supersymmetric vacua, the one with G$_2$ symmetry, against BJs of D$p$-branes, $p >2$, wrapped around different contractible cycles of the internal S$^6$. We find no BJ instabilities either. It would be interesting to extend the wrapped-D$p$-BJ-stability test to the other non-supersymmetric solutions in table \ref{Table:KnownBFStableCritical points}. 

Of course, these non-supersymmetric $\,\textrm{AdS}_{4} \times \textrm{S}^{6}\,$ solutions could still decay through other channels. In fact, one would expect that they did, based on the general arguments of \cite{ArkaniHamed:2006dz,Ooguri:2016pdq}. One possibility is that their full KK spectrum does contain BF-unstable modes. After all, the $D=4$ $\cN=8$ supergravity captures only a small slice of the full KK towers, and these might still contain modes below the BF bound outside that slice. It would be possible to address this question with the technology put forth in \cite{Malek:2019eaz}. Another possibility is that these vacua tunnelled into other ones, becoming eventually stable at the end of a decay chain. Such decay would be signalled by the existence of a DW solution connecting the non-supersymmetric AdS solution to a different one --an iterative process that would stop when a stable, supersymmetric vacuum be reached. In this paper we have partially assessed this possibility for the non-supersymmetric G$_2$-invariant vacuum. 

We have not found a DW connecting the  $\,\cN=0\,$ $\,\text{G}_2\,$ solution in the UV to any $\,\text{SU}(3)\,$-symmetric one in the IR, within the $\,\text{SU}(3)\,$ invariant sector of $\,\text{ISO}(7)\,$ supergravity, and irrespectively of the amount of supersymmetry present in the latter. Our analysis is numeric in nature and is therefore subject to an inherent degree of uncertainty, but the evidences suggest that there is no such DW. Therefore no quantum tunnelling process seems possible for this solution. Some caveats of this interpretation are the following: 

\begin{enumerate}

\item It is possible that the DW does after all exist in the $\,\text{SU}(3)\,$ invariant sector \eqref{eq.SU3invariantaction} but our procedure was not able to find it. In this case, given the extreme amount of fine-tuning of parameters that such solution may require, we consider that the decay rate factor $\,A\,$ in \eqref{eq.probability} would be highly suppressed in this channel, rendering the $\,\cN=0\,$ $\,\text{G}_2\,$ solution metastable.

\item A second possibility is that the decay is protected in the presence of $\,\text{SU}(3)\,$  invariance, such that a DW solution exists with lower, or none, symmetry. Considering for concreteness the model with seven-chiral fields introduced in \cite{Guarino:2019snw}, the three supersymmetric candidates present there would have 4-, 6- and 9-dimensional parameter spaces\footnote{These dimensions can be obtained by counting the number of scalars with non-negative mass squared, minus one. This counting arises from considerations of regularity in the IR for the radial behavior $\,e^{-\Delta \frac{r}{L}}\,$, and the holographic relation
\be
\Delta(\Delta-3) = M^2\,L^2 \ .
\ee}
to be explored, with each dimension corresponding respectively to the last three supersymmetric points of table~\ref{Table:KnownBFStableCritical points}. In the full $\cN=8$ supergravity, the dimensions of the parameter spaces would be 35, 40 and 55 (see appendix~\ref{app_vacua} for the corresponding spectra). The search of this type of DW solutions in such high-dimensional parameter spaces is certainly beyond the purposes of this work.

\item A third possibility is that there is indeed no DW connecting the $\cN=0$ $\,\text{G}_2\,$ $\,\text{AdS}_4\,$ solution to any other vacuum of $D=4$ $\cN=8$ ISO(7) supergravity in the IR. 

\end{enumerate}

In any case, the swampland logic \cite{ArkaniHamed:2006dz,Ooguri:2016pdq} would still require a decay channel to exist for the non-supersymmetric $\,\textrm{AdS}_{4} \times \textrm{S}^{6}\,$ vacua of type IIA string theory that we have considered in this work. Such an explicit decay channel for these solutions remains to be found.

\section*{Acknowledgements}

We are grateful to Ant\'on Faedo and Patrick Meessen for discussions. AG is partially supported by the Spanish government grant PGC2018-096894-B-100 and by the Principado de Asturias through the grant FC-GRUPIN-IDI/2018/000174.  JT is supported  by the European Research Council, grant no. 725369.  OV is supported by the NSF grant PHY-1720364 and, partially, by grants SEV-2016-0597 and PGC2018-095976-B-C21 from MCIU/AEI/FEDER, UE.

\appendix

 
 \addtocontents{toc}{\protect\setcounter{tocdepth}{1}}

\section{All known vacua of $\,D=4\,$ $\,\mathcal{N}=8\,$ ISO(7) supergravity}
\label{app_vacua}

In appendix A of \cite{Guarino:2019snw} we constructed an $\cN=1$ discrete, $(\mathbb{Z}_2)^3$--invariant truncation of $D=4$ $\cN=8$ dyonic ISO(7) supergravity \cite{Guarino:2015qaa} that retains seven real scalars, seven pseudoscalars, and no vector fields. In $\cN=1$ language, this truncation retains the largest number, seven, of $\cN=1$ chiral fields contained in the $\cN=8$ scalar manifold $\textrm{E}_{7(7)}/\textrm{SU}(8)$, and is analogue to the $(\mathbb{Z}_2)^3$ truncation constructed for other $D=4$ $\cN=8$ gaugings in \cite{Aldazabal:2006up,Aldazabal:2008zza,Aldazabal:2010ef,Bobev:2019dik}. The real K\"ahler potential $K$ and the holomorphic superpotential $W$ for this model can be found in equations (A.8) and (A.9) of \cite{Guarino:2019snw}. The scalar potential $V$ follows from $K$ and $W$ through the standard $\cN=1$ formulae (2.7), (2.8) of \cite{Guarino:2019snw}. 

In \cite{Guarino:2019snw}, we scanned the superpotential $W$ of this model and found two previously unknown $\cN=1$ points. For the present project, we have scanned the potential $V$ in this sector for vacua, and we

\begin{enumerate}

\item \label{item1} recover all known supersymmetric points summarised in table \ref{Table:KnownBFStableCritical points} of the text;

\item \label{item2} recover the non-supersymmetric points previously known to be BF-stable within the full $\cN=8$ ISO(7) theory: one point with G$_2$ symmetry \cite{Borghese:2012qm}, two points with SU(3) symmetry \cite{Guarino:2015qaa}, and one point with $\textrm{SO}(3)_{\textrm{d}} \times \textrm{SO}(3)_{\textrm{R}}$ symmetry \cite{Guarino:2015qaa};

\item \label{item3} recover the non-supersymmetric points labelled as {\it ii})--{\it v}) in \cite{Guarino:2019jef}, which were found in that reference to be free of BF instabilities within the subsector considered therein;

\item \label{item4} recover the non-supersymmetric points previously known to have BF-instabilities: the SO(7) and SO(6) points of \cite{DallAgata:2011aa}, and the $\textrm{U}(1) \times \textrm{SO}(3)_{\textrm{R}}$-invariant point labelled as {\it i}) in \cite{Guarino:2019jef}; and 

\item \label{item5} find $43$ new critical points, all of them non-supersymmetric.

\end{enumerate}

\begin{table}[t!]
\rowcolors{2}{gray!25}{white}
\tiny{
\begin{tabular}{|c|c|}
\hline
$g^{-2} c^{\frac13}\,V$ & $M^2 L^2$ \\
\hline
$-23.9529$ &
   $-2.396$,\, $-2.017$,\, $-1.940$,\, $-1.871$,\, $-1.578$,\, $-1.318$,\, $-1.193$,\, $-0.987$,\, $-0.867$,\, $1.015$,\, $2.208$,\, $2.823$,\, $5.609$,\, $6.743$ \\
$ -24.0968$ &
   $-2.435$,\, $-2.143$,\, $-1.921$,\, $-1.889$,\, $-1.760$,\, $-1.659$,\, $-1.161$,\, $-1.118$,\, $-0.558$,\, $0.808$,\, $2.691$,\, $3.269$,\, $5.586$,\, $6.722$ \\
$ -24.1499$ &
   $-2.316$,\, $-2.200$,\, $-2.122$,\, $-1.791$,\, $-1.689$,\, $-1.525$,\, $-1.266$,\, $-1.217$,\, $0.705$,\, $1.032$,\, $1.568$,\, $3.853$,\, $5.715$,\, $6.822$ \\
$ -24.3184$ &
   $-2.915$,\, $-2.746$,\, $-2.703$,\, $-2.462$,\, $-1.744$,\, $-1.336$,\, $-1.095$,\, $-0.505$,\, $-0.459$,\, $2.059$,\, $3.667$,\, $4.214$,\, $6.680$,\, $7.188$ \\
$ -24.5332$ &
   $-2.694$,\, $-2.631$,\, $-2.416$,\, $-2.200$,\, $-1.870$,\, $-1.466$,\, $-1.049$,\, $-0.925$,\, $0.488$,\, $1.629$,\, $2.678$,\, $4.459$,\, $6.033$,\, $6.863$ \\
$ -24.6910$ & $-3.000$,\, $-2.729$,\, $-2.571$,\, $-2.269$ ($\times 2$),\, $-1.183$,\, $-0.430$ ($\times 2$),\, $1.273$ ($\times 2$),\, $2.891$,\, $3.964$,\, $6.993$,\, $7.538 $ \\
$ -24.6930$ &
   $-2.804$,\, $-2.459$,\, $-2.291$,\, $-2.176$,\, $-1.638$,\, $-1.455$,\, $-1.240$,\, $-0.767$,\, $0.614$,\, $1.350$,\, $2.951$,\, $4.837$,\, $5.761$,\, $7.072$ \\
$ -25.1119$ &
   $-3.244$,\, $-3.139$ ($\times 2$),\, $-3.000$,\, $-1.569$,\, $-0.954$,\, $-0.438$ ($\times 2$),\, $1.799$ ($\times 2$),\, $3.815$,\, $4.211$,\, $6.753$,\, $8.710$ \\
$ -25.6934$ &
   $-2.445$,\, $-2.233$,\, $-2.156$,\, $-1.818$,\, $-1.737$,\, $-1.580$,\, $-1.404$,\, $-0.136$,\, $0.842$,\, $2.376$,\, $3.116$,\, $4.428$,\, $8.090$,\, $8.186$ \\
$ -25.9471$ &
   $-2.919$,\, $-2.309$ ($\times 2$),\, $-1.556$,\, $-1.484$ ($\times 2$),\, $-1.152$,\, $0.944$ ($\times 2$),\, $1.200$,\, $4.973$,\, $5.124$ ($\times 2$),\, $8.358$ \\
$ -27.1014$ &
   $-3.046$,\, $-2.540$,\, $-2.477$,\, $-1.730$,\, $-1.503$,\, $-1.419$,\, $-0.56$,\, $0.663$,\, $2.290$,\, $3.495$,\, $5.021$,\, $6.216$,\, $7.624$,\, $8.402$ \\
$ -27.1368$ &
   $-2.322$,\, $-1.981$,\, $-1.973$,\, $-1.814$,\, $-1.670$,\, $-1.591$,\, $-0.617$,\, $1.256$,\, $1.915$,\, $2.307$,\, $3.366$,\, $5.504$,\, $6.748$,\, $7.539$ \\
$ -27.4182$ &
   $-2.507$,\, $-2.317$,\, $-2.216$,\, $-1.975$,\, $-1.752$,\, $-1.519$,\, $-1.149$,\, $0.858$,\, $1.854$,\, $4.166$,\, $6.007$,\, $6.702$,\, $8.330$,\, $9.625$ \\
$ -27.6100$ &
   $-3.323$,\, $-2.074$ ($\times 2$),\, $-1.873$,\, $-1.193$ ($\times 2$),\, $-0.269$,\, $1.724$ ($\times 2$),\, $3.790$,\, $4.040$,\, $4.611$ ($\times 2$),\, $7.379$ \\
$ -28.0504$ &
   $-2.484$ ($\times 2$),\, $-2.344$,\, $-1.928$,\, $-1.455$ ($\times 2$),\, $1.045$,\, $2.246$ ($\times 2$),\, $3.204$,\, $6.193$ ($\times 2$),\, $6.383$,\, $9.002$ \\
$ -28.1978$ &
   $-3.404$,\, $-2.971$,\, $-2.709$,\, $-1.400$,\, $-1.373$,\, $-0.432$,\, $-0.355$,\, $1.436$,\, $3.430$,\, $3.564$,\, $4.051$,\, $5.469$,\, $7.139$,\, $8.004$ \\
$ -28.4866$ &
   $-2.868$,\, $-2.534$,\, $-2.225$,\, $-1.832$,\, $-1.575$,\, $-0.844$,\, $-0.430$,\, $0.886$,\, $1.458$,\, $2.970$,\, $6.165$,\, $7.085$,\, $9.312$,\, $14.329$ \\
$ -29.3540$ &
   $-2.804$,\, $-2.756$,\, $-2.049$,\, $-1.487$,\, $-1.250$,\, $-0.614$,\, $1.072$,\, $1.464$,\, $2.099$,\, $3.782$,\, $5.550$,\, $7.071$,\, $7.894$,\, $10.504$ \\
$ -29.7825$ & $-3.000$,\, $-2.281$,\, $-1.502$,\, $-1.432$ ($\times 2$),\, $0.838$,\, $1.055$ ($\times 2$),\, $3.054$,\, $3.358$ ($\times 2$),\, $7.199$,\, $8.307$,\, $9.703$  \\
$ -30.2420$ & $-2.655$,\, $-2.410$,\, $-2.086$,\, $-1.506$,\, $-1.38$,\, $-0.390$,\, $1.347$,\, $1.957$,\, $2.222$,\, $4.143$,\, $5.493$,\, $6.788$,\, $7.827$,\, $9.731$  \\
$ -30.4178$ &
   $-2.488$,\, $-2.285$,\, $-2.029$,\, $-1.560$,\, $-1.448$,\, $0.345$,\, $1.139$,\, $2.640$,\, $3.238$,\, $4.464$,\, $5.967$,\, $6.598$,\, $8.285$,\, $9.654$ \\
$ -30.4349$ &
   $-2.607$,\, $-2.240$,\, $-2.099$,\, $-1.816$,\, $-0.956$,\, $-0.895$,\, $0.601$,\, $1.836$,\, $4.341$,\, $5.396$,\, $6.157$,\, $8.114$,\, $8.218$,\, $10.512$ \\
$ -30.8806$ &
   $-3.093$,\, $-2.842$,\, $-2.147$,\, $-1.819$,\, $-1.129$,\, $-0.333$,\, $0.466$,\, $1.955$,\, $4.646$,\, $4.949$,\, $6.473$,\, $7.525$,\, $9.845$,\, $11.923$ \\
$ -31.0405$ &
   $-3.649$,\, $-2.196$,\, $-2.194$,\, $-1.458$,\, $-1.008$,\, $-0.246$,\, $0.941$,\, $1.930$,\, $2.762$,\, $5.218$,\, $6.444$,\, $6.850$,\, $7.106$,\, $8.163$ \\
$ -31.1950$ &
   $-3.701$,\, $-2.228$,\, $-1.889$,\, $-1.178$,\, $-1.112$,\, $-0.889$,\, $0.219$,\, $3.611$,\, $4.170$,\, $4.527$,\, $6.033$,\, $6.610$,\, $6.896$,\, $10.938$ \\
$ -31.3341$ &
   $-3.726$,\, $-2.211$,\, $-1.809$,\, $-1.292$,\, $-0.360$,\, $-0.013$,\, $0.843$,\, $1.696$,\, $3.122$,\, $5.609$,\, $5.915$,\, $6.860$ ($\times 2$),\, $9.369$ \\
$ -32.2195$ &
   $-3.397$,\, $-2.093$,\, $-1.971$,\, $-1.929$,\, $-1.337$,\, $0.569$,\, $0.935$,\, $2.846$,\, $4.439$,\, $5.685$,\, $7.215$,\, $7.315$,\, $8.954$,\, $12.456$ \\
$ -32.8169$ & $-3.717$,\, $-1.661$,\, $-1.554$,\, $-1.140$,\, $-1.000$,\, $0.258$,\, $2.524$,\, $2.587$,\, $3.897$,\, $4.716$,\, $5.723$,\, $6.913$,\, $8.595$,\, $8.933$  \\
$ -33.0425$ &
   $-3.881$,\, $-1.814$,\, $-1.244$,\, $-1.013$,\, $-0.614$,\, $0.445$,\, $2.695$,\, $2.856$,\, $2.979$,\, $4.181$,\, $5.875$,\, $6.802$,\, $8.554$,\, $9.239$ \\
$ -33.1815$ & $-3.865$,\, $-1.550$,\, $-1.509$,\, $-0.094$,\, $0.030$,\, $0.799$,\, $1.293$,\, $2.998$,\, $3.679$,\, $3.773$,\, $5.885$,\, $6.709$,\, $8.245$,\, $8.804$  \\
$ -33.5995$ &
   $-3.139$,\, $-2.463$,\, $-2.134$,\, $-1.240$,\, $-1.027$,\, $-0.502$,\, $1.359$,\, $3.813$,\, $4.670$,\, $5.354$,\, $7.494$,\, $7.765$,\, $8.597$,\, $9.814$ \\
$ -34.7574$ &
   $-3.144$,\, $-2.315$,\, $-1.611$,\, $-1.273$,\, $-0.769$,\, $0.319$,\, $2.625$,\, $4.449$,\, $5.386$,\, $6.038$,\, $6.262$,\, $8.261$,\, $10.515$,\, $10.990$ \\
$ -34.7680$ &
   $-2.801$,\, $-2.149$,\, $-1.782$,\, $-0.676$,\, $-0.414$,\, $0.746$,\, $2.360$,\, $3.715$,\, $5.325$,\, $6.716$,\, $7.356$,\, $8.018$,\, $8.162$,\, $11.988$ \\
$ -34.7968$ &
   $-3.531$,\, $-3.009$,\, $-2.045$,\, $-1.031$,\, $-0.511$,\, $1.030$,\, $2.412$,\, $3.047$,\, $4.161$,\, $4.309$,\, $7.523$,\, $7.807$,\, $9.557$,\, $11.599$ \\
$ -35.0016$ & $-2.900$,\, $-2.287$,\, $-1.630$,\, $-0.556$,\, $0.421$,\, $1.001$,\, $1.258$,\, $4.261$,\, $5.329$,\, $6.261$,\, $7.849$,\, $8.414$,\, $8.603$,\, $11.746$  \\
$ -35.6102$ &
   $-2.272$,\, $-2.202$,\, $-1.458$,\, $-1.184$,\, $-0.044$,\, $0.093$,\, $2.319$,\, $3.805$,\, $4.027$,\, $4.143$,\, $4.654$,\, $7.581$,\, $9.802$,\, $10.865$ \\
$ -35.6102$ &
   $-2.266$,\, $-2.237$,\, $-1.435$,\, $-1.192$,\, $-0.015$,\, $0.138$,\, $2.277$,\, $3.876$,\, $4.015$,\, $4.136$,\, $4.604$,\, $7.560$,\, $9.816$,\, $10.844$ \\
$ -36.0555$ &
   $-2.660$,\, $-2.527$,\, $-1.551$,\, $-1.046$,\, $-0.129$,\, $0.676$,\, $2.603$,\, $3.115$,\, $4.852$,\, $5.618$,\, $7.527$,\, $8.510$,\, $9.803$,\, $10.35$ \\
$ -36.0727$ &
   $-2.758$,\, $-2.211$,\, $-1.476$,\, $-1.141$,\, $0.113$,\, $1.305$,\, $2.062$,\, $2.828$,\, $4.938$,\, $5.549$,\, $7.510$,\, $8.837$,\, $9.622$,\, $10.117$ \\
$ -38.5392$ &
   $-3.473$,\, $-3.263$,\, $-1.578$,\, $-0.594$,\, $0.561$,\, $1.248$,\, $2.234$,\, $4.693$,\, $7.003$,\, $7.666$,\, $8.221$,\, $8.714$,\, $11.401$,\, $14.475$ \\
$ -38.7246$ &
   $-2.630$,\, $-1.935$,\, $-1.519$,\, $-1.378$,\, $-0.614$,\, $1.722$,\, $2.693$,\, $4.782$,\, $5.253$,\, $6.976$,\, $7.872$,\, $9.411$,\, $10.101$,\, $12.432$ \\
$ -39.7613$ &
   $-3.462$,\, $-2.968$,\, $-1.405$,\, $-0.955$,\, $0.439$,\, $1.591$,\, $2.468$,\, $5.149$,\, $6.919$,\, $7.451$,\, $8.146$,\, $9.024$,\, $9.718$,\, $13.626$ \\
$ -40.6009$ &
   $-3.035$,\, $-2.837$,\, $-1.968$,\, $-0.772$,\, $0.479$,\, $4.156$,\, $4.501$,\, $5.787$,\, $7.227$,\, $8.340$,\, $9.278$,\, $9.843$,\, $14.123$,\, $17.872$ \\
 \hline
\end{tabular}
\caption{ \footnotesize{Scalar masses within the seven-chiral model for the $43$ new non-supersymmetric AdS$_{4}$ vacua found in this paper, labelled by the value of the potential. All of them present BF instabilities.}\normalsize \label{Table:new_non-susy} } }
\end{table}

Next, we scan for BF instabilities for the critical points in items \ref{item3} and \ref{item5} above. In order to do this, we first compute the scalar masses around these points within the seven-chiral sector. This analysis is enough to find BF instabilities in point {\it ii}) of \cite{Guarino:2019jef} and in all $43$ new critical points (see table~\ref{Table:new_non-susy}). Finally, we scan for BF instabilities for the remaining points in item \ref{item3} by computing their spectra within the full $\cN=8$ theory using the mass formulae of \cite{LeDiffon:2011wt}. The points {\it iii})--{\it v}) in \cite{Guarino:2019jef} are found to be BF-stable within the full $\cN=8$ theory. Together with those in item \ref{item2}, these are the critical points that have been brought to table \ref{Table:KnownBFStableCritical points} of the introduction: the points with symmetry $\textrm{U}(1)_{\textrm{d}} \times \textrm{SO}(3)_{\textrm{R}}$ are {\it iii}) and {\it v}) of \cite{Guarino:2019jef}, and the point with symmetry $\textrm{SO}(3)_{\textrm{R}}$ is {\it iv}) of \cite{Guarino:2019jef}. This analysis brings the knowledge of vacua of $D=4$ $\cN=8$ dyonic ISO(7) supergravity to the following state-of-the-art: there exist, at least,
\begin{itemize}
\item 6 supersymmetric points, recorded in table \ref{Table:KnownBFStableCritical points}, with bosonic spectra within $D=4$ $\cN=8$ ISO(7) supergravity reviewed below;
\item 7 non-supersymmetric points, but BF-stable within the $D=4$ $\cN=8$ supergravity, recorded in table \ref{Table:KnownBFStableCritical points}; their bosonic spectra within the $D=4$ supergravity is given below; and
\item 47 non-supersymmetric points, which are BF-unstable already within the $D=4$ $\cN=8$ supergravity; the spectra of the previously known ones can be found in the references above, while for the new ones the spectra within the seven-chiral model of \cite{Guarino:2019snw} is given in table \ref{Table:new_non-susy} above. 
\end{itemize}
Further comments on the symmetries of the critical points in table \ref{Table:KnownBFStableCritical points} will be made in appendix \ref{app:SO3_R_sector}. We emphasise that the present classification is not exhaustive within the full $\cN=8$ theory: a complete classification along the lines of \cite{Comsa:2019rcz,Krishnan:2020sfg,Bobev:2020ttg} should still be made. We also note that the present perturbative analysis only sees a small cross section of the full KK towers about the corresponding type IIA $\,\textrm{AdS}_{4} \times \textrm{S}^{6}\,$ solutions: nothing excludes BF instabilites in KK modes not contained within $D=4$ $\cN=8$ supergravity. 

For convenience, we now summarise the bosonic spectra within $\cN=8$ supergravity of all the solutions of table \ref{Table:KnownBFStableCritical points}. Please refer to the table for the original references. Together with a massless graviton, the vector and scalar mass spectra are given subsequently.

\subsubsection*{$\mathcal{N}=3$ vacuum with $\,\textrm{SO}(3)_{\textrm{d}} \times \textrm{SO}(3)_{\textrm{R}}\,$ symmetry}

The set of normalised vector masses is given by
\begin{equation}
\label{N=3&SO4_spectrum_vec}
\begin{array}{ccrrrrrrrr}
M^2 \, L^2 &=& ( 3 \pm \sqrt{3}) \,\,(\times 3) & , &  \tfrac{15}{4} \,\,(\times 4) & , & \tfrac{3}{4} \,\,(\times 12) & , & 0 \,\,(\times 6) & , 
\end{array}
\end{equation}
whereas the normalised scalar masses read
\begin{equation}
\label{N=3&SO4_spectrum_scal}
\begin{array}{ccrrrrrrrr}
M^2 \, L^2 &=& 3(1 \pm \sqrt{3}) \,\,(\times 1) & , &  (1 \pm \sqrt{3}) \,\,(\times 6) & , & -\tfrac{9}{4} \,\,(\times 4) & , &  -2 \,\,(\times 18) & , \\
& & -\tfrac{5}{4} \,\,(\times 12) & , & 0 \,\,(\times 22) & .
\end{array}
\end{equation}

\subsubsection*{$\mathcal{N}=2$ vacuum with $\,\textrm{SU}(3) \times \textrm{U}(1)\,$ symmetry}

The set of normalised vector masses is given by
\begin{equation}
\label{N=2&U3_spectrum_vec}
\begin{array}{ccrrrrrrrr}
M^2 \, L^2 &=& 4 \,\,(\times 1) & , &  \tfrac{28}{9} \,\,(\times 6) & , & \tfrac{4}{9} \,\,(\times 12) & , & 0 \,\,(\times 9) & ,
\end{array}
\end{equation}
whereas the normalised scalar masses read
\begin{equation}
\label{N=2&U3_spectrum_scal}
\begin{array}{ccrrrrrrrr}
M^2 \, L^2 &=& (3 \pm \sqrt{17}) \,\,(\times 1) & , &   -\tfrac{20}{9} \,\,(\times 12) & , & -2 \,\,(\times 16) & , & -\tfrac{14}{9} \,\,(\times 18) & , \\
& & 2 \,\,(\times 3) & , & 0 \,\,(\times 19) & .
\end{array}
\end{equation}

\subsubsection*{$\mathcal{N}=1$ vacuum with $\,\textrm{G}_{2}\,$ symmetry}

The set of normalised vector masses is given by
\begin{equation}
\label{N=1&G2_spectrum_vec}
\begin{array}{ccrrrrrrrr}
M^2 \, L^2 &=& \tfrac{1}{2} ( 3 \pm \sqrt{6}) \,\,(\times 7) & , &  0 \,\,(\times 14) & ,
\end{array}
\end{equation}
whereas the normalised scalar masses read
\begin{equation}
\label{N=1&G2_spectrum_scal}
\begin{array}{ccrrrrrrrr}
M^2 \, L^2 &=& (4 \pm \sqrt{6}) \,\,(\times 1) & , &  -\tfrac{1}{6} (11 \pm \sqrt{6} ) \,\,(\times 27) & , & 0 \,\,(\times 14) & .
\end{array}
\end{equation}

\subsubsection*{$\mathcal{N}=1$ vacuum with $\,\textrm{SU}(3)\,$ symmetry}

The set of normalised vector masses is given by
\begin{equation}
\label{N=1&SU3_spectrum_vec}
\begin{array}{ccrrrrrrrr}
M^2 \, L^2 &=& 6 \,\,(\times 1) & , &  \tfrac{28}{9} \,\,(\times 6) & , & \tfrac{25}{9} \,\,(\times 6) & , & 2 \,\,(\times 1) & , \\
& & \tfrac{4}{9} \,\,(\times 6) & , & 0 \,\,(\times 8) & ,
\end{array}
\end{equation}
whereas the normalised scalar masses read
\begin{equation}
\label{N=1&SU3_spectrum_scal}
\begin{array}{ccrrrrrrrr}
M^2 \, L^2 &=& (4 \pm \sqrt{6}) \,\,(\times 2) & , &   -\tfrac{20}{9} \,\,(\times 12) & , & -2 \,\,(\times 8) & , & -\tfrac{8}{9} \,\,(\times 12) & , \\
& &\tfrac{7}{9} \,\,(\times 6) & , & 0 \,\,(\times 28) & .
\end{array}
\end{equation}

\subsubsection*{$\mathcal{N}=1$ vacuum with $\,\textrm{U}(1)_{\textrm{R}}\,$ symmetry and $\, V= -25.697 \,\, g^{2} \left(\tfrac{m}{g}\right) ^{-\frac13}\,$}

The set of normalised vector masses is given by
\begin{equation}
\label{N=1&U1_1_spectrum_vec}
\begin{array}{ccrrrrrrrr}
M^2 \, L^2 &=& 6.030 \,\,(\times 1) & , &  5.719 \,\,(\times 2) & , & 4.749 \,\,(\times 2) & , & 4.660 \,\,(\times 1) & , \\
&  & 4.459 \,\,(\times 1) & , & 3.636 \,\,(\times 2) & , & 2.892 \,\,(\times 2)  & , & 2.018 \,\,(\times 1) & , \\ 
&  & 1.914 \,\,(\times 2)  & , & 1.229 \,\,(\times 1) & , & 1.119 \,\,(\times 1) & , & 0.711 \,\,(\times 1) & , \\
&  & 0.693 \,\,(\times 2)  & , & 0.616 \,\,(\times 1)  & , & 0.359 \,\,(\times 1)  & , & 0.347 \,\,(\times 2)  & , \\
&  & 0.251 \,\,(\times 2)  & , & 0.177 \,\,(\times 2)  & , & 0 \,\,(\times 1)  & , 
\end{array}
\end{equation}
whereas the normalised scalar masses read
\begin{equation}
\label{N=1&U1_1_spectrum_scal}
\begin{array}{ccrrrrrrrr}
M^2 \, L^2 &=& 
 8.164  \,\,(\times 1) & , &
 8.099  \,\,(\times 2) & , &
 4.222  \,\,(\times 1) & , &
 3.719  \,\,(\times 2) & ,  \\
&  &  
 3.022  \,\,(\times 1) & , &
 2.749 \,\,(\times 2) & , &
 2.710  \,\,(\times 1) & , &
 2.665 \,\,(\times 2) & ,  \\
&  &  
 -2.240 \,\,(\times 2) & , &
 -2.207  \,\,(\times 1) & , &
 -2.150  \,\,(\times 1) & , &
 -2.107 \,\,(\times 2) & ,  \\
&  &  
 -2.099  \,\,(\times 1) & , &
 -2.040 \,\,(\times 2) & , &
 -1.956 \,\,(\times 2) & , &
 -1.877  \,\,(\times 1) & ,  \\
&  &  -1.863 \,\,(\times 2) & , &
 -1.823 \,\,(\times 2) & , &
 -1.783  \,\,(\times 1) & , &
 -1.749 \,\,(\times 2) & ,  \\
&  &  
 -1.700  \,\,(\times 1) & , &
 -1.641  \,\,(\times 1) & , &
 -1.623  \,\,(\times 1) & , &
 -1.443 \,\,(\times 2) & ,  \\
&  &  
 -1.384  \,\,(\times 1) & , &
 -1.289  \,\,(\times 1) & , &
  0.784  \,\,(\times 1) & , &
 -0.570  \,\,(\times 1) & ,  \\
&  &  0.134  \,\,(\times 1) & , &
 -0.086 \,\,(\times 2) & , &
  0 \,\,(\times 27) & .
\end{array}
\end{equation}

\subsubsection*{$\mathcal{N}=1$ vacuum with $\,\textrm{U}(1)_{\textrm{R}}\,$ symmetry and $\, V= -35.610 \,\, g^{2} \left(\tfrac{m}{g}\right) ^{-\frac13}\,$}

The set of normalised vector masses is given by
\begin{equation}
\label{N=1&U1_2_spectrum_vec}
\begin{array}{ccrrrrrrrr}
M^2 \, L^2 &=&  
 7.199 \,\,(\times 1) & , &
 6.908 \,\,(\times 1) & , &
 6.765 \,\,(\times 1) & , &
 6.245 \,\,(\times 1) & ,  \\
&  & 
 6.055 \,\,(\times 2) & , &
 5.939 \,\,(\times 2) & , &
 5.613 \,\,(\times 1) & , &
 4.995 \,\,(\times 2) & ,  \\
&  & 
 4.714 \,\,(\times 2) & , &
 4.474 \,\,(\times 1) & , &
 4.158 \,\,(\times 2) & , &
 2.740 \,\,(\times 1) & ,  \\
&  & 
 2.557 \,\,(\times 1) & , &
 1.415 \,\,(\times 2) & , &
 1.127 \,\,(\times 1) & , &
 0.959 \,\,(\times 2) & ,  \\
&  & 
 0.654 \,\,(\times 2) & , &
 0.385 \,\,(\times 2) & , &
 0 \,\,(\times 1) & ,
\end{array}
\end{equation}
whereas the normalised scalar masses read
\begin{equation}
\label{N=1&U1_2_spectrum_scal}
\begin{array}{ccrrrrrrrr}
M^2 \, L^2 &=&  
10.856 \,\,(\times 1) & , &
 10.142 \,\,(\times 1) & , &
 9.809 \,\,(\times 1) & , &
 8.315 \,\,(\times 2) & ,  \\
 &  &
 7.571 \,\,(\times 1) & , &
 4.765 \,\,(\times 1) & , &
 4.615 \,\,(\times 1) & , &
 4.245 \,\,(\times 1) & ,  \\
 &  &
 4.113 \,\,(\times 2) & , &
 4.101 \,\,(\times 1) & , &
 4.055 \,\,(\times 2) & , &
 4.025 \,\,(\times 1) & ,  \\
 &  &
 3.939 \,\,(\times 2) & , &
 3.864 \,\,(\times 1) & , &
 3.613 \,\,(\times 1) & , &
 2.814 \,\,(\times 2) & ,  \\
 &  &
 2.714 \,\,(\times 2) & , &
 2.303 \,\,(\times 1) & , &
 -2.249 \,\,(\times 2) & , &
 -2.247 \,\,(\times 1) & ,  \\
 &  &
 -2.239 \,\,(\times 1) & , &
 -1.961 \,\,(\times 2) & , &
 -1.615 \,\,(\times 2) & , &
 -1.447 \,\,(\times 1) & ,  \\
 &  &
 -1.346 \,\,(\times 2) & , &
 -1.188 \,\,(\times 2) & , &
 -1.132 \,\,(\times 1) & , &
 0.115 \,\,(\times 2) & ,  \\
 &  &
 0.068 \,\,(\times 2) & , &
 0.015 \,\,(\times 1) & , &
 0 \,\,(\times 27) & .
\end{array}
\end{equation}

\subsubsection*{$\mathcal{N}=0$ vacuum with $\,\textrm{G}_{2}\,$ symmetry}

The set of normalised vector masses is given by
\begin{equation}
\label{N=0&G2_spectrum_vec}
\begin{array}{ccrrrrrrrr}
M^2 \, L^2 &=& 3 \,\,(\times 14) & , &  0 \,\,(\times 14) & ,
\end{array}
\end{equation}
whereas the normalised scalar masses read
\begin{equation}
\label{N=0&G2_spectrum_scal}
\begin{array}{ccrrrrrrrr}
M^2 \, L^2 &=& 6 \,\,(\times 2) & , &   -1 \,\,(\times 54) & , &  0 \,\,(\times 14) & .
\end{array}
\end{equation}

\subsubsection*{$\mathcal{N}=0$ vacuum with $\,\textrm{SU}(3)\,$ symmetry and $\, V= -23.414 \,\, g^{2} \left(\tfrac{m}{g}\right) ^{-\frac13}\,$}

The set of normalised vector masses is given by
\begin{equation}
\label{N=0&SU3_2_spectrum_vec}
\begin{array}{ccrrrrrrrr}
M^2 \, L^2 &=& 4.373 \,\,(\times 1) & , &  2.490 \,\,(\times 1) & , & 3.200 \,\,(\times 6) & , & 2.791 \,\,(\times 6) & , \\
& & 0.111 \,\,(\times 6) & , & 0 \,\,(\times 8) & ,
\end{array}
\end{equation}
whereas the normalised scalar masses read
\begin{equation}
\label{N=0&SU3_2_spectrum_scal}
\begin{array}{ccrrrrrrrr}
M^2 \, L^2 &=& 6.230 \,\,(\times 1) & , &  5.905 \,\,(\times 1) & , & 1.130 \,\,(\times 1) & , & -1.264 \,\,(\times 1) & , \\
& & -1.582 \,\,(\times 12) & , & -0.954 \,\,(\times 12) & , & -1.396 \,\,(\times 8) & , & -0.309 \,\,(\times 8) & , \\
& & -1.082 \,\,(\times 6) & , & 0 \,\,(\times 20) & .
\end{array}
\end{equation}

\subsubsection*{$\mathcal{N}=0$ vacuum with $\,\textrm{SU}(3)\,$ symmetry and $\, V= -23.457 \,\, g^{2} \left(\tfrac{m}{g}\right) ^{-\frac13}\,$}

The set of normalised vector masses is given by
\begin{equation}
\label{N=0&SU3_1_spectrum_vec}
\begin{array}{ccrrrrrrrr}
M^2 \, L^2 &=& 4.677 \,\,(\times 1) & , &  2.136 \,\,(\times 1) & , & 3.184 \,\,(\times 6) & , & 2.715 \,\,(\times 6) & , \\
& & 0.150 \,\,(\times 6) & , & 0 \,\,(\times 8) & ,
\end{array}
\end{equation}
whereas the normalised scalar masses read
\begin{equation}
\label{N=0&SU3_1_spectrum_scal}
\begin{array}{ccrrrrrrrr}
M^2 \, L^2 &=& 6.214 \,\,(\times 1) & , &  5.925 \,\,(\times 1) & , & 1.145 \,\,(\times 1) & , & -1.284 \,\,(\times 1) & , \\
& & -1.707 \,\,(\times 12) & , & -0.860 \,\,(\times 12) & , & -1.623 \,\,(\times 8) & , & -0.159 \,\,(\times 8) & , \\
& & -1.061 \,\,(\times 6) & , & 0 \,\,(\times 20) & .
\end{array}
\end{equation}

\subsubsection*{$\mathcal{N}=0$ vacuum with $\,\textrm{SO}(3)_{\textrm{d}} \times \textrm{SO}(3)_{\textrm{R}}\,$ symmetry}

The set of normalised vector masses is given by
\begin{equation}
\label{N=0&SO4_spectrum_vec}
\begin{array}{ccrrrrrrrr}
M^2 \, L^2 &=& 4.153 \,\,(\times 3) & , &  2.287 \,\,(\times 3) & , & 3.451 \,\,(\times 4) & , & 1.945 \,\,(\times 4) & , \\
& & 0.191 \,\,(\times 8) & , & 0 \,\,(\times 6) & ,
\end{array}
\end{equation}
whereas the normalised scalar masses read
\begin{equation}
\label{N=0&SO4_spectrum_scal}
\begin{array}{ccrrrrrrrr}
M^2 \, L^2 &=& 6.727 \,\,(\times 1) & , &  5.287 \,\,(\times 1) & , & 0.584 \,\,(\times 1) & , & -1.586 \,\,(\times 1) & , \\
& &  -1.588  \,\,(\times 9) & , & -1.751 \,\,(\times 9) & , & 0.630 \,\,(\times 5) & , & -0.983 \,\,(\times 5) & , \\
& &  -0.730  \,\,(\times 4) & , & -1.964 \,\,(\times 4) & , & -1.176 \,\,(\times 8) & , & 0 \,\,(\times 22) & .
\end{array}
\end{equation}

\subsubsection*{$\mathcal{N}=0$ vacuum with $\,\textrm{U}(1)_{\textrm{d}} \times \textrm{SO}(3)_{\textrm{R}}\,$ symmetry and $\, V= -23.456 \,\, g^{2} \left(\tfrac{m}{g}\right) ^{-\frac13}\,$}

The set of normalised vector masses is given by
\begin{equation}
\label{N=0&SO(3)xU(1)_1_spectrum_vec}
\begin{array}{ccrrrrrrrr}
M^2 \, L^2 &=&  
4.295  \,\,(\times 1)  & , &
 3.626 \,\,(\times 2) & , &
 3.312 \,\,(\times 4) & , &
 2.668 \,\,(\times 2) & ,  \\
& &
 2.397 \,\,(\times 4) & , &
 2.294 \,\,(\times 1) & , &
 0.125 \,\,(\times 4) & , &
 0.088 \,\,(\times 2) & ,  \\
 & &
 0.053 \,\,(\times 4) & , &
 0 \,\,(\times 4) & ,
\end{array}
\end{equation}
whereas the normalised scalar masses read
\begin{equation}
\label{N=0&SO(3)xU(1)_1_spectrum_scal}
\begin{array}{ccrrrrrrrr}
M^2 \, L^2 &=& 
6.293 \,\,(\times 1)  & , &
 5.780 \,\,(\times 1) & , &
 -1.714 \,\,(\times 4)  & , &
 -1.620 \,\,(\times 6) & ,  \\
 & &
 -1.582 \,\,(\times 1) & , &
 -1.570 \,\,(\times 3) & , &
 -1.386 \,\,(\times 4)  & , &
 -1.366 \,\,(\times 2) & ,  \\
 & &
 -1.265 \,\,(\times 6) & , &
 -1.169 \,\,(\times 3) & , &
 1.146 \,\,(\times 1) & , &
 -1.136 \,\,(\times 4)  & ,  \\
 & &
 -0.988 \,\,(\times 1) & , &
 -0.920 \,\,(\times 2) & , &
 -0.826 \,\,(\times 4)  & , &
 -0.145 \,\,(\times 1) & ,  \\
 & &
 -0.028 \,\,(\times 2) & , &
 0 \,\,(\times 24) & .
\end{array}
\end{equation}

\subsubsection*{$\mathcal{N}=0$ vacuum with $\,\textrm{U}(1)_{\textrm{d}} \times \textrm{SO}(3)_{\textrm{R}}\,$ symmetry and $\, V= -23.459 \,\, g^{2} \left(\tfrac{m}{g}\right) ^{-\frac13}\,$}

The set of normalised vector masses is given by
\begin{equation}
\label{N=0&SO(3)xU(1)_2_spectrum_vec}
\begin{array}{ccrrrrrrrr}
M^2 \, L^2 &=&  
 4.597 \,\,(\times 1) & , &
 3.267 \,\,(\times 2) & , &
 3.246 \,\,(\times 4) & , &
 2.821 \,\,(\times 2) & ,  \\
 & &
 2.559 \,\,(\times 4) & , &
 2.167 \,\,(\times 1) & , &
 0.148 \,\,(\times 4) & , &
 0.132 \,\,(\times 2) & ,  \\
 & &
 0.012 \,\,(\times 4) & , &
 0 \,\,(\times 4) & ,
\end{array}
\end{equation}
whereas the normalised scalar masses read
\begin{equation}
\label{N=0&SO(3)xU(1)_2_spectrum_scal}
\begin{array}{ccrrrrrrrr}
M^2 \, L^2 &=&
6.223 \,\,(\times 1) & , &
 5.895 \,\,(\times 1) & , &
 -1.707 \,\,(\times 6) & , &
 -1.695 \,\,(\times 4) & ,  \\
 & &
 -1.636 \,\,(\times 3) & , &
 -1.597 \,\,(\times 2) & , &
 -1.586 \,\,(\times 1) & , &
 -1.559 \,\,(\times 4) & ,  \\
 & &
 -1.223 \,\,(\times 1) & , &
 1.161 \,\,(\times 1) & , &
 -1.126 \,\,(\times 4) & , &
 -1.047 \,\,(\times 6) & ,  \\
 & &
 -0.936 \,\,(\times 2) & , &
 -0.805 \,\,(\times 4) & , &
 -0.627 \,\,(\times 3) & , &
 -0.481 \,\,(\times 2) & ,  \\
 & &
 0.085 \,\,(\times 1) & , &
 0 \,\,(\times 24)  & .
\end{array}
\end{equation}

\subsubsection*{$\mathcal{N}=0$ vacuum with $\,\textrm{SO}(3)_{\textrm{R}}\,$ symmetry}

The set of normalised vector masses is given by
\begin{equation}
\label{N=0&SO(3)_spectrum_vec}
\begin{array}{ccrrrrrrrr}
M^2 \, L^2 &=& 
 4.265 \,\,(\times 1) & , &
 3.708 \,\,(\times 1) & , &
 3.593 \,\,(\times 1) & , &
 3.315 \,\,(\times 4) & ,  \\
 & &
 2.679 \,\,(\times 1) & , &
 2.636 \,\,(\times 1) & , &
 2.390 \,\,(\times 4) & , &
 2.307 \,\,(\times 1) & ,  \\
  & &
 0.123 \,\,(\times 4) & , &
 0.095 \,\,(\times 1) & , &
 0.075 \,\,(\times 1) & , &
 0.056 \,\,(\times 4) & ,  \\
  & &
 0.002 \,\,(\times 1) & , &
 0 \,\,(\times 3) & , 
\end{array}
\end{equation}
whereas the normalised scalar masses read
\begin{equation}
\label{N=0&SO(3)_spectrum_scal}
\begin{array}{ccrrrrrrrr}
M^2 \, L^2 &=&
6.298  \,\,(\times 1) & , &
 5.773  \,\,(\times 1) & , &
 -1.717  \,\,(\times 4) & , &
 -1.642 \,\,(\times 3) & ,  \\
 & &
 -1.585  \,\,(\times 1) & , &
 -1.579 \,\,(\times 3) & , &
 -1.563 \,\,(\times 3) & , &
 -1.372 \,\,(\times 4) & ,  \\
 & &
 -1.352  \,\,(\times 1) & , &
 -1.346  \,\,(\times 1) & , &
 -1.283 \,\,(\times 3) & , &
 -1.272 \,\,(\times 3) & ,  \\
 & &
 -1.202 \,\,(\times 3) & , &
 1.142  \,\,(\times 1) & , &
 -1.132 \,\,(\times 4) & , &
 -0.977  \,\,(\times 1) & ,  \\
 & &
 -0.961  \,\,(\times 1) & , &
 -0.874  \,\,(\times 1) & , &
 -0.834 \,\,(\times 4) & , &
 -0.215  \,\,(\times 1) & ,  \\
 & &
 0.040  \,\,(\times 1) & , &
 0  \,\,(\times 25) & .
\end{array}
\end{equation}

\section{Ten-dimensional background geometries} 
\label{app:SO3_R_sector}

\subsection{Uplifting formulae}

The continuous residual symmetry groups $G$ of the solutions specified in table \ref{Table:KnownBFStableCritical points} are all subgroups of SO(7): the compact subgroup of the ISO(7) gauge group from a $D=4$ perspective, and the isometry group of the round metric on S$^6$ from a IIA point of view. With the exception of the $\cN=1$ $\textrm{U}(1)_\textrm{R}$-invariant solutions of \cite{Guarino:2019snw}, all solutions in table \ref{Table:KnownBFStableCritical points} have continuous symmetry groups $G$ that contain the $\textrm{SO}(3)_{\textrm{R}}$ subgroup of SO(7) defined through the alternative embeddings (see (2.1) of \cite{Guarino:2019jef})
\begin{equation}
\label{Embedding_SO3_R}
\begin{array}{cccccccc}
  & & \textrm{G}_{2} & \supset & \textrm{SU}(3)      \\
 \textrm{SO}(7) & \supset & & & & \supset & \textrm{SO}(3)_{\textrm{R}} \; , \\
 & & \textrm{SO}(3)' \times \textrm{SO}(4) & \supset & \textrm{SO}(3)_{\textrm{d}} \times \textrm{SO}(3)_{\textrm{R}} \; .
\end{array}
\end{equation}
In the embedding through the top line, $\textrm{SO}(3)_{\textrm{R}} \sim \textrm{SU}(2)$ is the subgroup of $\textrm{SU}(3)$ such that $\bm{3} \rightarrow \bm{2} + \bm{1}$. In the bottom line, $\,\textrm{SO}(4) \sim \textrm{SO}(3)_{\textrm{L}} \times \textrm{SO}(3)_{\textrm{R}}\,$, and $\,\textrm{SO}(3)_{\textrm{d}}\,$ is the diagonal subgroup of the product $\,\textrm{SO}(3)' \, \times \, \textrm{SO}(3)_{\textrm{L}}\,$. The subgroup $\textrm{SO}(3)_{\textrm{d}} \times \textrm{SO}(3)_{\textrm{R}}$ in the bottom line of (\ref{Embedding_SO3_R}) is also the maximal subgroup of the G$_2$ in the top line. The groups $\textrm{G}_{2}$, $\textrm{SU}(3)$, $\textrm{SO}(3)_{\textrm{d}} \times \textrm{SO}(3)_{\textrm{R}}$ and $\textrm{SO}(3)_{\textrm{R}}$ in (\ref{Embedding_SO3_R}) manifestly appear as possible residual continuous groups $G$ in table \ref{Table:KnownBFStableCritical points}. The other groups $\textrm{U}(1)$, $\textrm{U}(1)_{\textrm{d}}$, $\textrm{U}(1)_{\textrm{R}}$ that appear in the table but not in (\ref{Embedding_SO3_R}) respectively correspond to the U(1) that commutes with SU(3) inside SO(7) and to the Cartan subgroups of $\textrm{SO}(3)_{\textrm{d}}$ and $\textrm{SO}(3)_{\textrm{R}}$.

The sector of $D=4$ $\cN=8$ ISO(7) supergravity invariant under $\textrm{SO}(3)_{\textrm{R}}$ was constructed in \cite{Guarino:2019jef}. As per (\ref{Embedding_SO3_R}), all the non-supersymmetric points of table \ref{Table:KnownBFStableCritical points} (as well as the supersymmetric ones with the exception of the $\cN=1$ $\textrm{U}(1)_\textrm{R}$ points) are contained within the $\textrm{SO}(3)_{\textrm{R}}$-invariant sector. In addition, as explained in appendix \ref{app_vacua}, all these critical points enjoy the additional discrete symmetry $(\mathbb{Z}_2)^3$ that truncates the $\cN=8$ supergravity to its seven-chiral subsector. In conclusion, all relevant points in table \ref{Table:KnownBFStableCritical points} belong to the intersection of the $\textrm{SO}(3)_{\textrm{R}}$ and $(\mathbb{Z}_2)^3 $ invariant subsectors. This is the model described in section 2.1 of \cite{Guarino:2019snw} containing four $\cN=1$ chiral fields. These are respectively composed of four real scalars $\varphi$, $\phi_i$, $i=1,2,3$, and four real pseudoscalars $\chi$, $b_i$, $i=1,2,3$, parameterising the scalar manifold $\left( \textrm{SL}(2)/\textrm{SO}(2) \right)^4$. These are the eight real scalar fields used in section \ref{sec:D2-brane} of the main text. In this language, the relevant critical points of table \ref{Table:KnownBFStableCritical points} are attained at the $D=4$ scalar vevs recorded in table \ref{Table:ScalarVevs} of this appendix.

\begin{table}[t!]
\centering
\footnotesize{
\scalebox{0.68}{
\begin{tabular}{ccccccccccc}
\hline
\\[-2.5mm]
SUSY  	& bos. sym. &  $g^{-2} c^{\frac13} \, V$ &  $c^{-\frac13} \, e^{-\varphi} $ & $c^{-\frac13} \, e^{-\frac{1}{\sqrt{2}} \phi_1} $  & $c^{-\frac13} \, e^{-\frac{1}{\sqrt{2}} \phi_2} $     & $c^{-\frac13} \, e^{-\frac{1}{\sqrt{2}} \phi_3} $   & $ c^{-\frac13} \, \chi$ & $c^{-\frac13} \, b_{1}$ & $ c^{-\frac13} \, b_{2} $ & $ c^{-\frac13} \, b_{3} $
\\[2pt]
\hline
\\[-10pt]
$ \cN=3 $ &	$\textrm{SO}(3)_{\textrm{d}} \times \textrm{SO}(3)_{\textrm{R}}$ &  $-\frac{2^{16/3}}{3^{1/2}} $   &	$2^{-\frac13} \sqrt{3}$ & $2^{-\frac43} \sqrt{3}$  & $2^{-\frac43} \sqrt{3}$  & $2^{-\frac43} \sqrt{3}$ & $-2^{-\frac13} $ & $-2^{-\frac56} $  &  $-2^{-\frac56} $ &  $-2^{-\frac56} $
\\[10pt]
$ \cN=2 $ & $\textrm{SU}(3) \times \textrm{U}(1)$  &   $-2^2 \, 3^{3/2} $&	$ \tfrac{\sqrt{3}}{2}$ & $ \tfrac{1}{\sqrt{2}}$  & $ \tfrac{1}{\sqrt{2}}$  & $ \tfrac{\sqrt{3}}{2}$ & $-\tfrac{1}{2}$ & $0$  & $0$ & $\tfrac{1}{\sqrt{2}}$
\\[10pt]
$ \cN=1 $ &	G$_2$ & $- \frac{2^{28/3} \, 3^{1/2}}{5^{5/2}} $   &	$2^{-\frac73} \sqrt{15}$ & $2^{-\frac73} \sqrt{15}$  & $2^{-\frac73} \sqrt{15}$  & $2^{-\frac73} \sqrt{15}$ & $-2^{-\frac73} $ & $2^{-\frac{11}{6}} $  & $2^{-\frac{11}{6}} $ & $2^{-\frac{11}{6}} $
\\[10pt]
$ \cN=1 $ &	SU(3) &   $-\frac{2^{8} \, 3^{3/2}}{5^{5/2}} $  &	$ \tfrac{\sqrt{15}}{4}$ & $ \tfrac{\sqrt{5}}{4} $  & $  \tfrac{\sqrt{5}}{4} $  & $ \tfrac{\sqrt{15}}{4} $ & $\tfrac{1}{4}$ & $ -\tfrac{\sqrt{3}}{2\sqrt{2}} $  & $ -\tfrac{\sqrt{3}}{2\sqrt{2}}  $ & $ -\tfrac{1}{2\sqrt{2}}  $
\\[10pt]
\hline
\\[-2.5mm]
$ \cN=0 $ &	G$_2$   &  $-\frac{2^{16/3}}{3^{1/2}}$ &	$2^{-\frac43} \sqrt{3}$ & $2^{-\frac43} \sqrt{3}$  & $2^{-\frac43} \sqrt{3}$  & $2^{-\frac43} \sqrt{3}$ & $2^{-\frac43} $ & $-2^{-\frac56} $  & $-2^{-\frac56} $ & $-2^{-\frac56} $%
\\[10pt]
$ \cN=0 $ &	SU(3)   & $-23.413628$  &	$0.732929$ & $0.661767$  & $0.661767$  & $0.732929$ & $0.269815$ & $-0.695046$  & $-0.695046$ & $-0.381576$%
\\[10pt]
$ \cN=0 $ &	SU(3)  &  $-23.456779$ &	$0.837649$ & $0.601165$  & $0.601165$  & $0.837649$ & $0.454656$ & $-0.473498$  & $-0.473498$ & $-0.642981$
\\[10pt]
$ \cN=0 $ &	$\textrm{SO}(3)_{\textrm{d}} \times \textrm{SO}(3)_{\textrm{R}}$ &  $-23.512690$  &	$1.146640$ & $0.651159$  & $0.651159$  & $0.651159$ & $0.067819$ & $-0.583082$  & $-0.583082$ & $-0.583082$
\\[10pt]
$ \cN=0 $ &	$\textrm{U}(1)_{\textrm{d}} \times \textrm{SO}(3)_{\textrm{R}}$  & $-23.456053$ &	$0.960882$ & $0.623955$  & $0.623955$  & $0.732823$ & $0.273189$ & $-0.497774$  & $-0.497774$ & $-0.715748$
\\[10pt]
$ \cN=0 $ &	$\textrm{U}(1)_{\textrm{d}} \times \textrm{SO}(3)_{\textrm{R}}$ & $-23.458780$ & 	$0.908165$ & $0.605946$  & $0.605946$  & $0.793681$ & $0.373634$ & $0.478351$  & $0.478351$ & $-0.688614$
\\[10pt]
$ \cN=0 $ &	$\textrm{SO}(3)_{\textrm{R}}$ & $-23.456098$ & $0.962339$ & $0.625559$  & $0.625872$  & $0.728508$ & $0.267119$ & $0.518046$  & $0.481301$ & $-0.716279$
\\[5pt]
\hline
\end{tabular}}
\caption{\footnotesize{$D=4$ scalar vevs on the $\,\textrm{AdS}_{4} \times \textrm{S}^{6}\,$ type IIA backgrounds of section \ref{sec:D2-brane}. \label{Table:ScalarVevs} }}\normalsize }
\end{table}

The uplift of the $\textrm{SO}(3)_{\textrm{R}}$ sector of the $D=4$ supergravity into the type IIA metric and dilaton was presented in appendix D.1 of \cite{Guarino:2019snw}. The external components of the RR three-form potential can also be easily reconstructed with the tensor hierarchy expressions of appendix B of \cite{Guarino:2019jef} by employing (3.43) of \cite{Guarino:2015vca}. These are all the ingredients needed for the D2 BJ-stability analysis of section \ref{sec:D2-brane}. For convenience, we record here the relevant uplifting formulae following \cite{Guarino:2019snw,Guarino:2019jef}. For all the solutions in table \ref{Table:KnownBFStableCritical points} except the $\cN=1$ $\textrm{U}(1)_\textrm{R}$-invariant ones, the relevant type IIA fields are given in equation (\ref{10Dmetric_SO(3)_R}) of the main text with the quantities $\Delta_1$, $\Delta_2$ and $C$ given in terms of the $\mathbb{R}^7$ coordinates $\mu^I$ constrained to the S$^6$ locus (\ref{eq:S6constraint}) and the $D=4$ scalars fixed at the values recorded in table \ref{Table:ScalarVevs}. It is convenient to pack $\phi_i$ and $b_i$, $i=1,2,3$, into the $3 \times 3$ matrices
\begin{equation} \label{eq:defmats}
\bm{m} = \textrm{diag} \, \big( e^{-\sqrt{2} \phi_1} , \, e^{-\sqrt{2} \phi_2} , \, e^{-\sqrt{2} \phi_3} \big) \; ,
\qquad
\bm{b} = \textrm{diag} \, \big( b_{1} , \, b_{2} , \, b_{3} \big) \; ,
\end{equation}
and to split $\mu^I = (\mu^i , \nu^a) $, $i=1,2,3$, $a=4,5,6,7$, while also introducing the vector notation $\bm{\mu}$ and $\bm{\nu}$ for the components $\mu^i$ and $\nu^a$. With these definitions, we have \cite{Guarino:2019snw}
\begin{eqnarray} 
\label{Delta1}
\Delta_1 = e^{\sqrt{2} (\phi_1+\phi_2+\phi_3 ) } \, \bm{\mu}^\textrm{T} \bm{m} \bm{\mu} + e^{\varphi + \frac{1}{\sqrt{2}}  (\phi_1+\phi_2+\phi_3 )} \, \bm{\nu}^\textrm{T} \bm{\nu} \; , 
\end{eqnarray}
and  \cite{Guarino:2019snw}
{\setlength\arraycolsep{2pt}
\begin{eqnarray}
\label{Delta2}
\Delta_2 &=& e^{-\varphi+\sqrt{2} (\phi_1+\phi_2+\phi_3 ) } \, \big(1+ e^{2\varphi} \chi^2 \big) \big[  \bm{\mu}^\textrm{T} \big( \bm{m} + \tfrac12 \bm{b}^\textrm{T} \bm{b} \big)  \bm{\mu} \big]^2 \nonumber \\[4pt]
&& + e^\varphi \Big[ 1 + \tfrac12 \, \textrm{tr} \big( \bm{b}^\textrm{T} \bm{b} \bm{m}^{-1} \big) + \tfrac18 \big[ \textrm{tr} \big( \bm{b}^\textrm{T} \bm{b} \bm{m}^{-1} \big) \big]^2 -
 \tfrac18  \textrm{tr} \big( \bm{b}^\textrm{T} \bm{b} \bm{m}^{-1} \bm{b}^\textrm{T} \bm{b} \bm{m}^{-1} \big) \nonumber \\[4pt]
&& \qquad \; +\tfrac18 \, e^{\sqrt{2} (\phi_1+\phi_2+\phi_3 ) } \big( \textrm{det} \, \bm{b} \big)^2 \Big]  \big( \bm{\nu}^\textrm{T} \bm{\nu} \big)^2 \nonumber \\[4pt]
&& + \Big[ e^{\frac{1}{\sqrt{2}} (\phi_1 +\phi_2 +\phi_3 )} \big[ 2+ e^{2\varphi}\chi^2 + \tfrac12 \textrm{tr}  \big( \bm{b}^\textrm{T} \bm{b} \bm{m}^{-1} \big) \big] \nonumber \\[4pt]
&& \qquad \;  -\tfrac{1}{\sqrt{2}}  \, e^{\varphi+ \sqrt{2} (\phi_1 +\phi_2 +\phi_3 ) } \, \chi  \,  \textrm{det} \, \bm{b}   \Big] \big[  \bm{\mu}^\textrm{T} \big( \bm{m} + \tfrac12 \bm{b}^\textrm{T} \bm{b} \big)  \bm{\mu} \big] \, \bm{\nu}^\textrm{T} \bm{\nu}  \; .
\end{eqnarray}
and
\begin{equation}
\label{C_func_SO(3)_R}
C = - \frac{L}{3} \, \left[  \,  \bm{\mu}^\textrm{T}  \bm{H}_\4  \bm{\mu}  \,\, + \,\,  H^{0}_\4  \,  \bm{\nu}^\textrm{T} \bm{\nu}   \right] \ ,
\end{equation}
where \cite{Guarino:2019jef}
{\setlength\arraycolsep{2pt}
\begin{eqnarray}
\label{H4_elec_1}
 \bm{H}_\4 &= & g \,  \Big[ \Big( 4 \, e^{ \frac{1}{\sqrt{2} } (\phi_1 +\phi_2 +\phi_3 )} +\sqrt{2} \,  \chi \,  e^{\varphi +\sqrt{2} (\phi_1 +\phi_2 +\phi_3 )} \, \textrm{det} \, \bm{b} \Big) \big( \bm{m} +\tfrac12 \, \bm{b}^\textrm{T} \bm{b} \big)    \nonumber \\[2mm]
 && \quad + e^{-\varphi +\sqrt{2} (\phi_1 +\phi_2 +\phi_3 )}  \big(1 + e^{2\varphi} \chi^2 \big) \Big( ( \textrm{tr} \, \bm{m} ) \, \bm{m} -2 \bm{m} \bm{m} -\tfrac12 \, \bm{m} \, \bm{b}^\textrm{T} \bm{b}  -\tfrac12 \, \bm{b}^\textrm{T} \bm{b}  \, \bm{m}  
 \nonumber \\[2mm]
 &&\quad  -\tfrac14 \, \big( \textrm{tr} \,(  \bm{b}^\textrm{T} \bm{b} ) \big) \, \bm{b}^\textrm{T} \bm{b}  \Big) \Big] 
  +  \tfrac12 m \,  \chi \,  e^{\varphi +\sqrt{2} (\phi_1 +\phi_2 +\phi_3 )} \, \bm{b}^\textrm{T} \bm{b}  \, 
\end{eqnarray}
}and\footnote{In (\ref{H4_elec_1}) and (\ref{H4_elec_2}) we have corrected a few typos in the formulae of \cite{Guarino:2019jef}.}   \cite{Guarino:2019jef}
{\setlength\arraycolsep{2pt}
\begin{eqnarray}
\label{H4_elec_2}
H^{0}_\4 & = & g \,  \Big[ 2 \, e^{\varphi} +  \Big( e^{ \frac{1}{\sqrt{2} } (\phi_1 +\phi_2 +\phi_3 )} + \tfrac{1}{2\sqrt{2}}  \,  \chi \,  e^{\varphi +\sqrt{2} (\phi_1 +\phi_2 +\phi_3 )} \, \textrm{det} \, \bm{b} \Big)  \,  \tr \, \big(  {\bm m} + \tfrac12 \, {\bm b}^\textrm{T} {\bm b}  \big) 
\nonumber \\[2mm]
&& \quad -\tfrac14 e^\varphi  \big( \tr \, ( {\bm b}^\textrm{T} {\bm b}  \,  {\bm m}^{-1} ) \big)^2   +  \tfrac14 e^\varphi  \,  \tr  ( {\bm b}^\textrm{T} {\bm b} \,  {\bm m}^{-1}  {\bm b}^\textrm{T} {\bm b}  \, {\bm m}^{-1}  ) - \tfrac12 \, e^{\varphi +\sqrt{2} ( \phi_1+  \phi_2+  \phi_3)} (\det  {\bm b} )^2 \Big] 
 \nonumber \\[2mm]
 && 
-  \tfrac{1}{2\sqrt{2}} \,  m  \,  e^{\varphi +\sqrt{2} (\phi_1 +\phi_2 +\phi_3 )} \, \textrm{det} \, \bm{b}  \,  ,
\end{eqnarray}
and $\,L=\sqrt{-6/V}\,$ with \cite{Guarino:2019jef}
{\setlength\arraycolsep{2pt}
\begin{eqnarray} 
\label{V_SO3}
V &=& g^2 \Big[ -4 \, e^{ \varphi}  - 4 \, e^{\frac{1}{\sqrt{2}} ( \phi_1+  \phi_2+  \phi_3)} \,  \tr \, \big(  {\bm m} + \tfrac12 \, {\bm b}^\textrm{T} {\bm b}  \big)    \nonumber \\[2mm]
&& \quad \;   +\tfrac12 \, e^{-\varphi +\sqrt{2} ( \phi_1+  \phi_2+  \phi_3)} (1+ e^{2\varphi} \chi^2 ) \Big( 2 \,  \tr \, \big(   {\bm m} {\bm m} + \tfrac12 \, {\bm b}^\textrm{T} {\bm b}   {\bm m}  \big)  -  ( \tr \, {\bm m} )^2  + \tfrac14  \big( \tr \, ( {\bm b}^\textrm{T} {\bm b}  ) \big)^2  \Big)    \nonumber \\[2mm]
&&   \quad \; +\tfrac12 e^\varphi  \big( \tr \, ( {\bm b}^\textrm{T} {\bm b}  \,  {\bm m}^{-1} ) \big)^2 - \tfrac12 e^\varphi \,  \tr \, ( {\bm b}^\textrm{T} {\bm b} \,  {\bm m}^{-1}  {\bm b}^\textrm{T} {\bm b}  \, {\bm m}^{-1}  )  +  e^{\varphi +\sqrt{2} ( \phi_1+  \phi_2+  \phi_3)} (\det  {\bm b} )^2   \nonumber \\[2mm]
&& \quad  \; -\sqrt{2} \, \chi \, e^{\varphi +\sqrt{2} ( \phi_1+  \phi_2+  \phi_3)} ( \det  {\bm b} ) \,   \tr \, \big(  {\bm m} + \tfrac12 \, {\bm b}^\textrm{T} {\bm b}  \big)    \Big] \nonumber \\[2mm]
&&
+ \,  g \, m \, e^{\varphi +\sqrt{2} ( \phi_1+  \phi_2+  \phi_3)}   \Big( \sqrt{2} \,  \det  {\bm b} -\tfrac12 \chi \,  \tr \, ( {\bm b}^\textrm{T} {\bm b} )   \Big) + \tfrac12 \, m^2 \, e^{\varphi +\sqrt{2} ( \phi_1+  \phi_2+  \phi_3)}  \  .
\end{eqnarray}
}The lengthy expression for the internal metric $ds_{\textrm{S}^6}^2$ in (\ref{10Dmetric_SO(3)_R}) can be found in \cite{Guarino:2019snw}. It is not needed for the analysis of section \ref{sec:D2-brane}. While the Freund--Rubin term (\ref{C_func_SO(3)_R}) generically depends on the S$^6$ embedding coordinates $\bm{\mu}$, $\bm{\nu}$, at an AdS$_4$ critical point it becomes independent of them and proportional to the scalar potential \ref{V_SO3}: see (3.37) of \cite{Guarino:2015vca}.

\subsection{Solutions with G$_2$ symmetry} \label{sec:D2G2}

The simplest $\,\textrm{AdS}_{4} \times \textrm{S}^{6}\,$ solutions to analyse are those whose internal symmetry is enlarged to G$_2$, as the resulting solutions become homogeneous. These solutions have the $D=4$ scalars restricted as
\begin{equation} \label{eq:G2locus}
\phi_1 = \phi_2 = \phi_3 = \sqrt{2}  \, \varphi \; , \qquad 
b_1 = b_2 = b_3 = - \sqrt{2}  \, \chi \; .
\end{equation}
On (\ref{eq:G2locus}), the dependence on the internal S$^6$ coordinates  $\mu^I = (\mu^i , \nu^a) $ drop out from (\ref{Delta1})--(\ref{C_func_SO(3)_R}) by virtue of (\ref{eq:S6constraint}). The warp factor and dilaton in (\ref{10Dmetric_SO(3)_R}) become constant ({\it i.e.} independent of the S$^6$ coordinates), and (\ref{10Dmetric_SO(3)_R}) reduce to the expressions given in (4.3) of \cite{Guarino:2015vca}. The individual $\,\textrm{AdS}_{4} \times \textrm{S}^{6}\,$ solutions are obtained by further fixing (\ref{eq:G2locus}) to the corresponding vevs in table \ref{Table:ScalarVevs}. In our conventions, the complete type IIA $\,\cN=1\,$ G$_2$ solution can be found in (4.6) of \cite{Varela:2015uca}, and the $\,\cN=0\,$ G$_2$ solution, in (4.9) of that reference. For these solutions, the coefficient $\,\Theta\,$ in (\ref{Theta_D2_SO(3)_R}) for the force experience by spacetime-filling probe D2-branes becomes constant ({\it i.e.}, independent of the $\,\mu^I\,$ angles), see (\ref{ThetaG2N=1}), (\ref{ThetaG2N=0}).

\subsection{Solutions with at least SU(3) symmetry} \label{sec:D2SU3}

Restricting the $\,D=4\,$ scalars as 
\begin{equation} \label{eq:SU3locus}
\phi_1 = \phi_2 \equiv \sqrt{2} \,  \phi \; , \quad   \phi_3  =  \sqrt{2} \, \varphi \; , \quad 
b_1 = b_2 \equiv - \tfrac{1}{\sqrt{2}} \, \zeta  \; , \quad b_3 = - \sqrt{2} \, \chi \; .
\end{equation}
the resulting configuration is at least SU(3)-invariant. Further imposing $\,\zeta = 0 \,$, an extra U(1) is preserved, while imposing $\,\phi = \varphi\,$ and $\,\zeta = 2\chi \,$, (\ref{eq:SU3locus}) reduces to (\ref{eq:G2locus}) and the symmetry is enhanced to $\,\textrm{G}_2\,$. For this class of solutions, it is useful to choose the $\,\mu^I = (\mu^i , \nu^a) \,$ as in (A.1) of \cite{Varela:2015uca} (but adapted to the choice of basis implied by (\ref{eq:SU3locus}) 
\begin{equation} \label{eq:musSU3}
\mu^1 = \sin \alpha \, \tilde{\mu}^1 \; , \quad 
\mu^2 = \sin \alpha \, \tilde{\mu}^2 \; , \quad 
\mu^3 = \cos \alpha  , \quad 
\nu^a = \sin \alpha \, \tilde{\mu}^{a} \; , 
\end{equation}
with $\,\tilde{\mu}^m = ( \tilde{\mu}^1, \tilde{\mu}^2,\tilde{\mu}^a) \,$, $\,m=1, \ldots, 6\,$, defining an S$^5$ via $\,\delta_{mn} \, \tilde{\mu}^m \tilde{\mu}^n = 1\,$, and $\,\alpha\,$ an $\,S^6\,$ angle ranging as 
\begin{equation} \label{alpharange}
0 \leq \alpha \leq \pi \; .
\end{equation}
The choice (\ref{eq:musSU3}) of embedding coordinates is adapted to the topological description of S$^6$ as the sine-cone over S$^5$ (see (\ref{S6=S5xS1_metric}) of the main text). The S$^5$ is equipped with its canonical Sasaki-Einstein structure, see appendix \ref{app:Geometric_structures}.

Away from the G$_2$-invariant sublocus, the resulting type IIA solutions are co\-ho\-mo\-ge\-nei\-ty-one, with all relevant quantities developing a dependence on the angle $\alpha$. Evaluating (\ref{10Dmetric_SO(3)_R}) with (\ref{Delta1})--(\ref{C_func_SO(3)_R}) on (\ref{eq:SU3locus}), (\ref{eq:musSU3}), the relevant expressions in (2.6) of \cite{Varela:2015uca} are recovered (with $\tilde{\zeta} = 0$ there). The actual $\,\textrm{AdS}_{4} \times \textrm{S}^{6}\,$ solutions are obtained by further fixing the $D=4$ scalars to the vevs recorded in table \ref{Table:ScalarVevs}. In our conventions, the full $\cN=2$ $\textrm{SU}(3) \times \textrm{U}(1)$-invariant $\,\textrm{AdS}_{4} \times \textrm{S}^{6}\,$ solution can be found in (12) of \cite{Guarino:2015jca}, the $\cN=1$ SU(3) solution in (4.4) of \cite{Varela:2015uca}, and the two $\cN=0$ SU(3) solutions are obtained by particularising (2.6), (2.9) of \cite{Varela:2015uca} to the numerical $D=4$ vevs given in table 3 of \cite{Guarino:2015qaa}. The coefficient $\Theta$ in (\ref{Theta_D2_SO(3)_R}) for these solutions accordingly acquires an $\alpha$ dependence, see (\ref{ThetaN=2SU3U1})--(\ref{ThetaN=0SU3p2}).

\subsection{Solutions with an explicit factor of $\textrm{SO}(3)_{\textrm{R}}$} \label{sec:SO3R}

For generic values of the $D=4$ scalars $\varphi$, $\chi$, $\phi_i$, $b_i$, the IIA configurations (\ref{10Dmetric_SO(3)_R}) display only an $\textrm{SO}(3)_{\textrm{R}}$ continuous symmetry. If the scalars are restricted as
\begin{equation} \label{eq:U1dSO3R}
\phi_1 = \phi_2  \; , \quad  
b_1 = b_2  \; ,
\end{equation}
an additional $\textrm{U}(1)_{\textrm{d}}$ symmetry is gained on top of $\textrm{SO}(3)_{\textrm{R}}$. Further imposing 
\begin{equation} \label{eq:SO3dSO3R}
\phi_1 = \phi_2= \phi_3  \; , \quad  
b_1 = b_2 = b_3  \; ,
\end{equation}
the symmetry is enhanced to $\textrm{SO}(3)_{\textrm{d}} \times \textrm{SO}(3)_{\textrm{R}}$. If, on top of (\ref{eq:U1dSO3R}), the further restriction (\ref{eq:SU3locus}) is imposed, the symmetry is enhanced to SU(3) as in section \ref{sec:D2SU3}, and it can be further enhanced to G$_2$ as explained there. Alternatively, the symmetry is enhanced directly to G$_2$ by further restricting (\ref{eq:SO3dSO3R}) as in (\ref{eq:G2locus}). These symmetry enhancement patterns simply reflect the embeddings (\ref{Embedding_SO3_R}). 

In this sector, and away from the SU(3)-enhanced locus, it is convenient to choose S$^6$ embedding coordinates $\mu^I = (\mu^i , \nu^a) \equiv ( \bm{\mu} , \bm{\nu}  )  $ as in (A.1) of \cite{DeLuca:2018buk} ,
\begin{equation} \label{musSO3R}
\mu^i = \cos \beta \, \tilde{\mu}^i \; , \; i = 1,2,3 \; , \qquad 
\nu^a = -\sin \beta \, \tilde{\nu}^a \; , \; a = 4,5,6,7 \; .
\end{equation}
These coordinates are adapted to the topological description of S$^6$ as the join of S$^2$ and S$^3$ (see (A.2) of \cite{DeLuca:2018buk}): $\beta$ is an angle on S$^6$ (different from the angle $\alpha$ of section \ref{sec:D2SU3}), that ranges as
\begin{equation} \label{betarange}
0 \leq \beta \leq \tfrac{\pi}{2} \; ,
\end{equation}
while $\tilde{\mu}^i$ and $\tilde{\nu}^a$ define the S$^2$ and S$^3$ via $\delta_{ij} \, \tilde{\mu}^i \, \tilde{\mu}^j =1$ and $\delta_{ab} \, \tilde{\nu}^a \, \tilde{\nu}^b =1$. For the $\tilde{\mu}^i$ we choose, as usual, 
\begin{equation} \label{eq:mutildes}
\tilde{\mu}^1 = \sin \tilde{\theta} \, \cos \tilde{\phi} \; , \qquad 
\tilde{\mu}^2 = \sin \tilde{\theta} \, \sin \tilde{\phi} \; , \qquad 
\tilde{\mu}^3 = \cos \tilde{\theta} \; . \qquad 
\end{equation}
with ranges
\begin{equation} \label{S2ranges}
0 \leq \tilde{\theta} \leq \pi \; , \qquad 
0 \leq \tilde{\phi} < 2 \pi \; .
\end{equation}
We will not need to specify the $\tilde{\nu}^a$. 

On the $\textrm{SO}(3)_{\textrm{d}} \times \textrm{SO}(3)_{\textrm{R}}$-invariant surface (\ref{eq:SO3dSO3R}), the type IIA configurations (\ref{10Dmetric_SO(3)_R}) are cohomogeneity-one: in the $S^6$ coordinates (\ref{musSO3R}), all the relevant quantities acquire a $\beta$ dependence. The expressions (\ref{10Dmetric_SO(3)_R}) reduce to the corresponding ones in (3.5) of \cite{DeLuca:2018buk} with $D=4$ scalars identified as $\varphi_{\textrm{here}} = \phi_{\textrm{there}}$, $\chi_{\textrm{here}} = \rho_{\textrm{there}}$, $(\phi_1 = \phi_2= \phi_3)_{\textrm{here}} = \sqrt{2} \, \varphi_{\textrm{there}}$, $(b_1 = b_2= b_3)_{\textrm{here}} = -\sqrt{2} \, \chi_{\textrm{there}}$, and $S^6$ angles identified as $\beta_{\textrm{here}} = \alpha_{\textrm{there}}$. The $\textrm{U}(1)_{\textrm{d}} \times \textrm{SO}(3)_{\textrm{R}}$ IIA configurations (\ref{10Dmetric_SO(3)_R}) defined by (\ref{eq:U1dSO3R}) are cohomogeneity-two, with all relevant quantities depending on the angles $\beta$ in (\ref{musSO3R}) and $\tilde{\theta}$ in (\ref{eq:mutildes}). Finally, the IIA configurations with only $\textrm{SO}(3)_{\textrm{R}}$ symmetry are cohomogeneity-three, with all their relevant quantities depending on the angles $\beta$, $\tilde{\theta}$ and $\tilde{\phi}$. The actual $\,\textrm{AdS}_{4} \times \textrm{S}^{6}\,$ solutions are obtained by fixing the $D=4$ scalars as in table \ref{Table:ScalarVevs}. In our conventions, the $\cN=3$ $\textrm{SO}(3)_{\textrm{d}} \times \textrm{SO}(3)_{\textrm{R}}$-invariant solution can be found in (5.1) of \cite{DeLuca:2018buk}. The $\cN=0$ $\textrm{SO}(3)_{\textrm{d}} \times \textrm{SO}(3)_{\textrm{R}}$ solution corresponds to (3.5), (3.8) of \cite{DeLuca:2018buk} with the $D=4$ scalar vevs recorded in table 4 of \cite{Guarino:2015qaa}. The IIA solutions with only $\textrm{U}(1)_{\textrm{d}} \times \textrm{SO}(3)_{\textrm{R}}$ or $\textrm{SO}(3)_{\textrm{R}}$ symmetry have only been determined partially in appendix D.1 of \cite{Guarino:2019snw}.

\section{$\textrm{G}_2$-invariant backgrounds: internal fluxes and WZ terms}
\label{app:massive_IIA}

The goal of this appendix is two-fold: on the one hand, we establish our conventions regarding Romans' massive IIA supergravity \cite{Romans:1985tz} and, on the other hand, we present the various WZ terms entering the effective action of the probe D$p$-branes placed in the class of $\,\textrm{G}_{2}$-invariant backgrounds of section~\ref{sec:Dp-branes}.

Following the conventions of \cite{Guarino:2015vca,Lavrinenko:1999xi}, the bosonic action for the massive IIA closed string sector reads
\begin{equation}
\label{SIIA}
\begin{array}{lll}
S_{\textrm{IIA}}  &=& \frac{1}{2 \hat{\kappa}^2} \int  \hat R\,  \hat{\textrm{vol}}_{10} 
+\frac{1}{2} d\hat \phi \wedge {\hat *d\hat \phi}
+ \frac{1}{2} e^{-\hat \phi}\, \hat H_3 \wedge {\hat *\hat H_3} - \frac{1}{2} e^{\frac{3}{2}\hat\phi}\,  \hat F_2 \wedge   {\hat *\hat F_2}
- \frac{1}{2} e^{\frac{1}{2}\hat\phi}\, \hat F_4 \wedge  {\hat *\hat F_4}   \\[2mm]
& & -  \frac{1}{2} (d\hat C_3 )^2  \wedge \hat B_2  - \frac{1}{6} m \,  d\hat C_3 \wedge (\hat B_2)^3
-\frac{1}{40} m^2\, (\hat B_2)^5 
-\frac{1}{2} m^2\, e^{\frac{5}{2}\hat \phi} \,   \hat{\textrm{vol}}_{10} \ ,
\end{array}
\end{equation}
where the ten-dimensional gravitational coupling constant is expressed in terms of the string length $\,\ell_s = \sqrt{\alpha^\prime}\,$ as $\,2\hat\kappa^2 = (2\pi)^7 \ell_s^8\,$. All the $\,\textrm{AdS}_{4} \times \textrm{S}^{6}\,$ backgrounds of massive IIA supergravity discussed in this work solve the equations of motion that follow from the action (\ref{SIIA}). However, in order to evaluate the probe D$p$-brane action (\ref{SDp}), with $\,p >2\,$, it is convenient to use the democratic formulation of the RR sector of the theory \cite{Bergshoeff:2001pv}. Together with the field strengths entering the action (\ref{SIIA}), namely,
\begin{equation}
\label{10D_Field_strengths_action}
\begin{array}{lll}
\hat H_3 &=&  d\hat B_2 \ ,  \\[2mm]
\hat F_4 &=& d\hat C_3 + \hat C_1\wedge d\hat B_2 + \frac{1}{2} \, m \, \hat B_2\wedge \hat B_2 \ , \\[2mm]
\hat F_2 &=& d\hat C_1 + m\, \hat B_2  \ , \\[2mm]
\hat F_0 &=& m \ ,
\end{array}
\end{equation}
where $\,m\,$ is the Romans mass parameter \cite{Romans:1985tz}, this (re)formulation introduces dual RR potentials $\,\hat{C}_{5}\,$, $\,\hat{C}_{7}\,$ and $\,\hat{C}_{9}\,$, which do not carry an independent dynamics, and have associated field strengths
\begin{equation}
\label{dual_potentials}
\begin{array}{lll}
\hat{F}_{6} & \equiv & e^{\frac{1}{2} \hat{\phi}} \, \hat{\ast} \hat{F}_{4} = d\hat{C}_{5} - \hat{B}_{2} \wedge d\hat{C}_{3} - \frac{m}{6} \, \hat{B}_{2}{}^3 \ , \\[2mm]
\hat{F}_{8} & \equiv & e^{\frac{3}{2} \hat{\phi}} \, \hat{\ast} \hat{F}_{2} = d\hat{C}_{7} - \hat{B}_{2} \wedge d\hat{C}_{5} + \frac{1}{2} \, \hat{B}_{2} \wedge \hat{B}_{2} \wedge  d\hat{C}_{3}  +  \frac{m}{24} \, \hat{B}_{2}{}^4 \ , \\[2mm]
\hat{F}_{10} & \equiv & e^{\frac{5}{2} \hat{\phi}} \,\,  m \,\, \hat{\ast} 1 = d\hat{C}_{9} - \hat{B}_{2} \wedge d\hat{C}_{7} + \frac{1}{2} \, \hat{B}_{2} \wedge \hat{B}_{2} \wedge  d\hat{C}_{5}  -  \frac{1}{6} \hat{B}_{2} \wedge \hat{B}_{2} \wedge \hat{B}_{2} \wedge d\hat{C}_{3} -  \frac{m}{120} \, \hat{B}_{2}{}^5 \ .
\end{array}
\end{equation}
Therefore, in order to obtain the dual potentials $\,\hat{C}_{5}\,$, $\,\hat{C}_{7}\,$ and $\,\hat{C}_{9}\,$ that enter the probe D$p$-brane action (\ref{SDp}), one must integrate the dual field strengths $\,\hat{F}_{6}\,$, $\,\hat{F}_{8}\,$ and $\,\hat{F}_{10}\,$ in (\ref{dual_potentials}) recursively.

For the class of $\textrm{G}_2$-invariant $\,\textrm{AdS}_{4} \times \textrm{S}^{6}\,$ backgrounds in (\ref{10D_G2-sector}), the associated field strengths in (\ref{10D_Field_strengths_action}) take the form \cite{Guarino:2015vca}
\begin{equation}
\label{10D_G2-sector_field_strengths}
\begin{array}{lll}
\hat H_{3} &=& 3 \, B \, \textrm{Re} \, \Upomega \  , \\[4mm]
%
\hat F_{4} &=&  -\frac{3}{L} \, C \, \textrm{vol}_4(\textrm{AdS}_{4})  + \left(\,\tfrac{1}{2} \, m \, B^{2}  - 2 \, g^{-3} \,  \chi \,\right) \, {\cal J} \wedge {\cal J} ,   \\[4mm]
%
%
\hat F_{2} &=& m \, B \, {\cal J}  \ , \\[2mm]
\hat F_{0} &=& m \ , 
\end{array}
\end{equation} 
with $\,L\,$ being the radius of AdS$_{4}$. The quantities $\,B\,$ and $\,C\,$ depend on the constants $\,(\varphi,\chi)\,$ in (\ref{eq:G2locus}) and are explicitly given in (\ref{DeltaBC_func}). Finally, the two-form $\,\mathcal{J}\,$ and three-form $\,\Upomega\,$ are $\,\textrm{SU}(3)$-invariant forms specifying the nearly-K\"ahler structure on $\,\textrm{S}^6\,$ (see appendix~\ref{app:Geometric_structures}). It is then a straightforward (but tedious) task to obtain the dual potentials $\,\hat{C}_{5}\,$, $\,\hat{C}_{7}\,$ and $\,\hat{C}_{9}\,$ from the duality relations (\ref{dual_potentials}) upon particularisation to the background fluxes in (\ref{10D_G2-sector_field_strengths}) and using (\ref{10D_G2-sector}).

In terms of the various gauge potentials, the relevant WZ terms entering the probe D$p$-brane effective actions in the class of $\,\textrm{G}_{2}$-invariant backgrounds (\ref{10D_G2-sector})-(\ref{DeltaBC_func}) analysed in section~\ref{sec:Dp-branes} are given subsequently.

\subsubsection*{WZ terms for the D0-brane}

The class of $\,\textrm{G}_{2}$-invariant backgrounds in (\ref{10D_G2-sector}) has 
\begin{equation}
\hat{C}_{1} = 0 \ ,
\end{equation}
so no WZ term is induced on the worldvolume of a probe D$0$-brane. The effective potential density that derives from the action (\ref{SDp})-(\ref{unmagnetised_F}) is purely gravitational and thus of attractive nature.

\subsubsection*{WZ terms for the D2-brane}

The contribution to (\ref{SD2_SO(3)_R}) coming from the potential $\,\hat{C}_{3}\,$ reads
\begin{equation}
\label{WZ_D2}
\begin{array}{lll}
\hat{C}_{3}
& =  &   \Delta^{\frac{3}{2}} \, C \, \, \hat{\textrm{vol}}_{3} \ ,
\end{array}
\end{equation}
with $\,\hat{\textrm{vol}}_{3}\,$ given in (\ref{vol_3}).

\subsubsection*{WZ terms for the D4-brane}

The term $\,\hat{C}_{3} \wedge \hat{B}_2\,$ contributes to (\ref{SD4}) as
\begin{equation}
\label{WZ_D4_1}
\begin{array}{lll}
\hat{C}_{3} \wedge \hat{B}_2 
%
%
& =  & - \,  \Delta^{\frac{7}{6}} \,  C  \,  g^{2} \, B \,  \cos\alpha  \, \, \hat{\textrm{vol}}_{5} \ ,
\end{array}
\end{equation}
whereas, after integrating the dual field strength $\,\hat{F}_{6}\,$ in (\ref{dual_potentials}), the contribution coming from the potential $\,\hat{C}_{5}\,$ reads
\begin{equation}
\label{WZ_D4_2}
\begin{array}{lll}
\hat{C}_{5} 
& =  & \frac{1}{3} \,\, \Delta^{-\frac{7}{6}}  \,\,  g \,\, \cos\alpha   \left[ L \,  e^{\frac{1}{2} \hat{\phi}} \,  \left( m \, g^3 \,  B^2 - 4 \, \chi \right) - 3 \, g\, B \, C \, \Delta^{\frac{7}{3}}\right] \, \hat{\textrm{vol}}_{5}  \ ,
\end{array}
\end{equation}
with $\,\hat{\textrm{vol}}_{5}\,$ given in (\ref{vol_5}).

\subsubsection*{WZ terms for the D6-brane}

The term $\,\hat{C}_{3} \wedge \hat{B}_2{}^2\,$ contributes to (\ref{SD6}) as
\begin{equation}
\label{WZ_D6_1}
\begin{array}{rll}
\hat{C}_{3} \wedge \hat{B}_2{}^2 
& = &  - \, 2 \,\, \Delta^{\frac{5}{6}}  \,\, C  \,  g^4 \,  B^2  \,\, \hat{\textrm{vol}}_{7} \ ,
\end{array}
\end{equation}
whereas, after integrating the dual field strengths $\,\hat{F}_{6}\,$ and $\,\hat{F}_{8}\,$ in (\ref{dual_potentials}), the contributions involving the potentials $\,\hat{C}_{5}\,$ and $\,\hat{C}_{7}\,$ read
\begin{equation}
\label{WZ_D6_2}
\begin{array}{rll}
\hat{C}_{5} \wedge \hat{B}_2 & = &  \frac{2}{3} \, \Delta^{-\frac{3}{2}} \, g^3 \, B   \left[  L \, e^{\frac{1}{2} \hat{\phi}} \left( m \, g^3 \, B^2   - 4  \, \chi \right) - 3 \, g \, B \, C \, \Delta^{\frac{7}{3}} \right] \, \hat{\textrm{vol}}_{7}  \ , \\[4mm]
\hat{C}_{7}  & = & \frac{1}{3} \, \Delta^{-\frac{3}{2}} \,g^2  \, B  \left[ L \, e^{\frac{1}{2} \hat{\phi}} \left( 2 \, m  \, g^4\, B^2  - 8 \, g \, \chi  + \Delta^{\frac{2}{3}} \, m \, e^{\hat{\phi} }  \right)  - \, 3 \, g^2 \, B \, C \, \Delta^{\frac{7}{3}} \right]  \, \hat{\textrm{vol}}_{7} \ ,
\end{array}
\end{equation}
with $\,\hat{\textrm{vol}}_{7}\,$ given in (\ref{vol_7}).

\subsubsection*{WZ terms for the D8-brane}

The term $\,\hat{C}_{3} \wedge \hat{B}_2{}^3\,$ contributes to (\ref{SD8}) as
\begin{equation}
\label{WZ_D8_1}
\begin{array}{rll}
\hat{C}_{3} \wedge \hat{B}_2{}^3 
& = &  - 6 \,\,  \Delta^{\frac{1}{2}}  \, C \,\, g^6  \, B^3 \,\, \hat{\textrm{vol}}_{9} \ ,
\end{array}
\end{equation}
whereas, after integrating the dual field strengths $\,\hat{F}_{6}\,$,  $\,\hat{F}_{8}\,$ and $\,\hat{F}_{10}\,$ in (\ref{dual_potentials}), the contributions involving the potentials $\,\hat{C}_{5}\,$, $\,\hat{C}_{7}\,$ and $\,\hat{C}_{9}\,$ read
\begin{equation}
\label{WZ_D8_2}
\begin{array}{rll}
\hat{C}_{5} \wedge \hat{B}_2{}^{2} & = &   2 \, \Delta^{-\frac{11}{6}} \,   g^5 \, B^2  \, 
\left[  \,  L \, e^{\frac{1}{2} \hat{\phi}} \, (  m \, g^3  \,  B^2    -  4 \, \chi)   -  3 \, g \, B \, C \, \Delta^{\frac{7}{3}}  \, \right] \,\,   \hat{\textrm{vol}}_{9}  \ , \\[3mm]
\hat{C}_{7} \wedge \hat{B}_2 & = &   \Delta ^{-\frac{11}{6}} \, g^4 \, B^2 \,  
\left[  \,  L \, e^{\frac{1}{2} \hat{\phi}}  \left( 2 \,m \, g^4 \, B^2  - 8 \, g \, \chi + \Delta^{\frac{2}{3}} \, m \,  e^{\hat{\phi}}   \right)  - 3 \, g^2 \, B \,  C \,  \Delta^{\frac{7}{3}}    \right] \,\,    \hat{\textrm{vol}}_{9}  \ , \\[3mm]
\hat{C}_9 & = & \Delta^{-\frac{11}{6}} 
\left[ \,  L \,  e^{\frac{1}{2} \hat{\phi}} 
\left(  m \, g^8 \, B^4  
+   g^4 \, B^2 \, \left(\Delta^{\frac{2}{3}} \, m  \, e^{\hat{\phi}} - 4 \, g \, \chi \right) 
+  \frac{1}{3}  \, \Delta^{\frac{4}{3}} \, m  \, e^{2 \hat{\phi} } \right) \right. \\[1mm]
& & \quad\quad\quad  \left. -  g^6 \, B^3 \, C \,  \Delta^{\frac{7}{3}} 
 \right] \,\,  \hat{\textrm{vol}}_{9}  \ ,
\end{array}
\end{equation}
with $\,\hat{\textrm{vol}}_{9}\,$ given in (\ref{vol_9}).

\section{Geometric structures on $\,\textrm{S}^5\,$ and $\,\textrm{S}^6\,$}
\label{app:Geometric_structures}

In this appendix we collect various results on the $\,\textrm{SU}(2)\,$ and $\,\textrm{SU}(3)\,$ geometric structures on $\,\textrm{S}^5\,$ and $\,\textrm{S}^6\,$ which are of relevance in section~\ref{sec:Dp-branes}. 

The Sasaki--Einstein $\,\textrm{SU}(2)$-structure on $\,\textrm{S}^{5}\,$ is specified by a one-form $\,\boldsymbol{\eta} \equiv d\psi + \boldsymbol{A}_{1}\,$ along the $\,\textrm{S}^{5}\,$ Hopf fibre (see below (\ref{S6=S5xS1_metric})),  together with a real $\,(1,1)$-form $\,\boldsymbol{J}\,$ and a complex $\,(2,0)$-form $\,\boldsymbol{\Omega}\,$. The latter is charged under the $\,\textrm{U}(1)_\psi\,$ generated by $\partial_\psi$, see (\ref{SU(2)_differential}) below. Following the conventions in \cite{Varela:2015uca} (except for the sign on the rightmost term of the second relation), these forms satisfy the algebraic 
\begin{equation}
\label{SU(2)_algebraic}
\boldsymbol{J} \wedge \boldsymbol{\Omega} = 0 
\hspace{8mm} , \hspace{8mm}
\boldsymbol{\Omega} \wedge \bar{\boldsymbol{\Omega}} = 2 \boldsymbol{J} \wedge \boldsymbol{J} = - 4 \, \textrm{vol}_{4}(\mathbb{CP}_{2})  \ ,
\end{equation}
and differential relations
\begin{equation}
\label{SU(2)_differential}
d \boldsymbol{\eta} = 2 \, \boldsymbol{J}
\hspace{5mm} , \hspace{5mm} 
d \boldsymbol{J} = 0
\hspace{5mm} , \hspace{5mm} 
d \boldsymbol{\Omega} = 3  \, i  \, \boldsymbol{\Omega} \wedge \boldsymbol{\eta} \ .
\end{equation}
Choosing the $\,\textrm{S}^5\,$ embedding coordinates $\,\tilde{\mu}^m\,$, $\,m=1, \ldots, 6\,$, with $\,\delta_{mn} \, \tilde{\mu}^m \tilde{\mu}^n = 1\,$, as
\begin{eqnarray} \label{eq:musS5}
& \tilde{\mu}^1 +i \tilde{\mu}^2 = \cos \gamma  \, e^{i \psi}  \; , \; \nonumber \\[5pt]
& \tilde{\mu}^3 +i \tilde{\mu}^4 = - \sin \gamma \, \cos \tfrac{\theta}{2} \, e^{\frac{i}{2} ( 2\psi + \tau +\phi)}  \; , \qquad
\tilde{\mu}^5 +i \tilde{\mu}^6 =  - \sin \gamma \, \sin \tfrac{\theta}{2} \, e^{\frac{i}{2} ( 2\psi + \tau -\phi)}   \; ,
\end{eqnarray}
with angular ranges $\,\gamma \in \left[ 0 , \frac{\pi}{2} \right]\,$, $\,\tau \in \left[ 0 , 4 \pi \right]\,$, $\,\theta \in \left[ 0 , \pi \right]\,$, $\,\phi \in \left[ 0 , 2 \pi \right]\,$, and $\,\psi \in \left[ 0 , 2\pi \right]\,$,  the round metric on $\,\textrm{S}^5\,$ inherited from the flat metric $\,ds^2 = \delta_{mn} \, d \tilde{\mu}^m d \tilde{\mu}^n\,$ on $\,\mathbb{R}^6\,$ can be written as 
\begin{equation}
\label{S5_metric}
ds^{2}_{\textrm{S}^5} = ds^{2}_{\mathbb{CP}_{2}} + \bm{\eta}^2  \ ,
\end{equation}
with $\,ds^{2}_{\mathbb{CP}_{2}}\,$ the Fubini-Study metric on $\,\mathbb{CP}_{2}\,$ normalised so that the Ricci tensor equals six times the metric:
\begin{equation}
\label{CP2_metric}
ds^{2}_{\mathbb{CP}_{2}} = d \gamma^2 + \frac{1}{4} \, \sin^2 \gamma \left( \sigma_{1}^2 + \sigma_{2}^2 +  \cos^2\gamma \, \sigma_{3}^2 \right) \ .
\end{equation}
Here, we have employed a set of $\,\textrm{SU}(2)\,$ left-invariant forms 
\begin{equation}
\sigma_{1} = -\sin\tau \, d\theta + \cos \tau \, \sin \theta \, d \phi \ , \quad
\sigma_{2} = \cos \tau \, d\theta + \sin \tau \, \sin \theta \, d \phi \ , \quad
\sigma_{3} = d\tau + \cos \theta \, d \phi \ .
\end{equation}
The volume form associated to (\ref{CP2_metric}) is
\begin{equation}
\textrm{vol}_{4}(\mathbb{CP}_{2}) = - \left(  \frac{\sin\gamma}{2} \right)^3  \cos\gamma  \,\,   d\gamma \wedge \sigma_{1} \wedge \sigma_{2} \wedge  \sigma_{3} \ .
\end{equation}

With the above definitions, the Sasaki-Einstein SU(2)-structure on $\,\textrm{S}^5\,$ can be written 
\begin{equation}
\label{SU(2)_structure_forms}
\begin{array}{lll}
\boldsymbol{\eta} &=& d\psi + \boldsymbol{A}_{1}\,\,=\,\, d\psi +  \frac{\sin\gamma}{2} \, \sin\gamma \,\, \sigma_{3} \ ,  \\[2mm] 
\boldsymbol{J} &=& \frac{\sin\gamma}{2} \, \cos\gamma \,\,  d\gamma \wedge \sigma_{3} + \left(\frac{\sin\gamma}{2}\right)^2 \,\, \sigma_{1} \wedge \sigma_{2}   \ ,  \\[2mm] 
\boldsymbol{\Omega} &=& e^{3 i \psi} \,  \frac{\sin\gamma}{2}  \, (d\gamma + i \, \frac{\sin\gamma}{2}  \cos\gamma \,\, \sigma_{3} ) \wedge (\sigma_{1} + i \, \sigma_{2})  \ . 
\end{array}
\end{equation}
An obvious $\,\textrm{S}^2 \sim \mathbb{CP}_{1} \subset \mathbb{CP}_{2}\,$ occurs in the metric (\ref{CP2_metric}) at $\,\gamma=\frac{\pi}{2}\,$. The normalised (unit radius) metric on $\,\textrm{S}^2\,$ reads
\begin{equation}
\label{CP1_metric}
ds^{2}_{\textrm{S}^2} = \sigma_{1}^2 + \sigma_{2}^2 = d\theta^2 + \sin^2\theta \, d\phi^2 \ ,
\end{equation}
with $\,\sigma_{1} \wedge \sigma_{2} = -\sin\theta \,\, d\theta  \wedge d\phi  = -\textrm{vol}_{2}(\textrm{S}^2)\,$. The form $\,\boldsymbol{J}\,$ in (\ref{SU(2)_structure_forms}) consistently reduces, up to a normalisation factor, to 
\begin{equation}
\label{J_S2}
\boldsymbol{J}_{\mathbb{CP}_{1}} =  \sigma_{1} \wedge \sigma_{2}=  -\textrm{vol}_{2}(\textrm{S}^2) \ .
\end{equation}

Finally, the SU(2)-structure on $\,\textrm{S}^5\,$ can be used to reconstruct the homogeneous nearly-K\"ahler $\,\textrm{SU}(3)\,$-structure on $\,\textrm{S}^6\,$ specified by a real two-form $\,\mathcal{J}\,$ and a holomorphic three-form $\,\Upomega\,$. Bringing the S$^5$ embedding coordinates (\ref{eq:musS5}) to their S$^6$ counterparts (\ref{eq:musSU3}), one has (see appendix~A of \cite{Varela:2015uca})
\begin{equation}
\label{SU(2)/SU(3)_correspondence}
\begin{array}{rll}
\mathcal{J} &=& \sin^2 \alpha \, \cos\alpha  \,  \boldsymbol{J} + \sin^3 \alpha \, \textrm{Re} \, \boldsymbol{\Omega} + \sin\alpha \, d\alpha \wedge  \boldsymbol{\eta} \ , \\[2mm]
\textrm{Re} \, \Upomega & = & - \sin^3 \alpha  \, \boldsymbol{J} \wedge d\alpha + \sin^2 \alpha \, \cos\alpha \,  \textrm{Re} \, \boldsymbol{\Omega} \wedge d\alpha - \sin^3 \alpha \, \textrm{Im} \, \boldsymbol{\Omega} \wedge \boldsymbol{\eta} \ , \\[2mm]
\textrm{Im} \, \Upomega & = & - \sin^4 \alpha  \, \boldsymbol{J} \wedge \boldsymbol{\eta}  + \sin^3 \alpha \, \cos\alpha \,  \textrm{Re} \, \boldsymbol{\Omega} \wedge \boldsymbol{\eta}  + \sin^2 \alpha \, \textrm{Im} \, \boldsymbol{\Omega} \wedge d \alpha \ .
\end{array}
\end{equation}
Using (\ref{SU(2)/SU(3)_correspondence}), the algebraic relations in (\ref{SU(2)_algebraic}) imply
\begin{equation}
\label{SU(3)_algebraic}
{\cal J} \wedge \Upomega = 0
\hspace{8mm} , \hspace{8mm}
\Upomega \wedge \bar \Upomega = -\tfrac{4i}{3} {\cal J} \wedge {\cal J} \wedge {\cal J} =  8 \, i \, \textrm{vol}_{6}(\textrm{S}^6)  \ ,
\end{equation} 
as appropriate to an SU(3)-structure, whereas the differential relations (\ref{SU(2)_differential}) imply the nearly-K\"ahler conditions
\begin{equation}
\label{SU(3)_differential}
d {\cal J} = 3 \,  \textrm{Re} \,  \Upomega  
\hspace{8mm} , \hspace{8mm}
d \, \textrm{Im} \, \Upomega =  -2 \,  {\cal J} \wedge {\cal J} \ . 
\end{equation}

\bibliography{references}

\end{document}